\documentclass[12pt,a4paper]{article}
\usepackage{tikz}
\usepackage{pgf}

\usepackage{amsmath}
\usepackage{booktabs}
\usepackage{cite}
\usepackage[colorlinks=true,linkcolor=black, citecolor=black,urlcolor=black]{hyperref}

\usepackage{multirow,graphics}
\usepackage{enumerate}
\usepackage{amstext}
\usepackage{amssymb}
\usepackage{amsmath}

\usepackage{graphicx}
\usepackage{subcaption}
\usepackage{color}
\usepackage[nodisplayskipstretch]{setspace}
\usepackage{tcolorbox}
\usepackage{float}
\usepackage{wasysym}
\usepackage{ytableau}
\usepackage{textcomp}

\usetikzlibrary{decorations.pathmorphing}
\usetikzlibrary{decorations.markings}
\usetikzlibrary{intersections}
\usetikzlibrary{calc}
  \usetikzlibrary{positioning}
\usetikzlibrary{shapes}
\usetikzlibrary{fit}
\usetikzlibrary{backgrounds}

\tikzset{
photon/.style={decorate, decoration={snake}},
particle/.style={postaction={decorate},
    decoration={markings,mark=at position .5 with {\arrow{>}}}},
antiparticle/.style={postaction={decorate},
    decoration={markings,mark=at position .5 with {\arrow{<}}}},
gluon/.style={decorate, decoration={coil,amplitude=2pt, segment length=4pt},color=purple},
wilson/.style={color=blue, thick},
scalarZ/.style={postaction={decorate},decoration={markings, mark=at position .5 with{\arrow[scale=1]{stealth}}}},
scalarX/.style={postaction={decorate}, dashed, dash pattern = on 4pt off 2pt, dash phase = 2pt, decoration={markings, mark=at position .53 with{\arrow[scale=1]{stealth}}}},
scalarZw/.style={postaction={decorate},decoration={markings, mark=at position .75 with{\arrow[scale=1]{stealth}}}},
scalarXw/.style={postaction={decorate}, dashed, dash pattern = on 4pt off 2pt, dash phase = 2pt, decoration={markings, mark=at position .60 with{\arrow[scale=1]{stealth}}}},
frozen/.style={inner sep=0.7mm, rectangle,draw},
frozenblue/.style={rectangle, draw, fill=blue!20, inner sep=0.7mm},
blue/.style={rectangle, rounded corners, fill=blue!20, inner sep=0.7mm},
red/.style={rectangle, rounded corners, fill=red!20, inner sep=0.7mm},
>=stealth,
norm/.style={->, draw, shorten <=2pt, shorten >=2pt},
diag/.style={->, draw, shorten <=5pt, shorten >=3pt},
every node/.style={inner sep=0.5mm}
}

\newcommand\ns[1]{\text{ns}[#1]}
\newcommand\hns[1]{\text{hns}[#1]}

\newcommand\hnsqbar[1]{\text{hns}_{\bar Q}[#1]}

\newcommand\be{\begin{equation}}
\newcommand\ee{\end{equation}}

\newcommand{\cA}{\mathcal{A}}
\newcommand{\la}{\langle}
\newcommand{\ra}{\rangle}
\newcommand{\ab}[1]{\la #1 \ra}
\newcommand{\dab}[1]{\la \la #1 \ra \ra}
\newcommand{\quadd}[4]{\la #1 (#2)(#3)(#4) \ra}

\normalsize
\makeatletter
\divide\@tempdima\baselineskip
\@tempcnta=\@tempdima
\skip\@mpfootins = \skip\footins
\renewcommand{\@dotsep}{10000}
\makeatother

\begin{document}
\numberwithin{equation}{section}
\begin{center}
\phantom{vv}

\vspace{3cm}
\bigskip

{\Large \bf    Cluster adjacency beyond MHV}

\bigskip
{\mbox {\bf James Drummond, Jack Foster,  \"Omer G\"urdo\u gan}}%
\footnote{ {\sffamily \{\tt j.a.foster, j.m.drummond, o.c.gurdogan\}@soton.ac.uk }}
\bigskip

{\em School of Physics \& Astronomy, University of Southampton,\\
  Highfield, Southampton, SO17 1BJ, United Kingdom.}

\vspace{3cm}
\bigskip
\vspace{30pt} {\bf Abstract}
\end{center}

\noindent We explore further the notion of cluster adjacency,
focussing on non-MHV amplitudes. We extend the notion of adjacency to
the BCFW decomposition of tree-level amplitudes. Adjacency controls
the appearance of poles, both physical and spurious, in individual
BCFW terms. We then discuss how this notion of adjacency is connected
to the adjacency already observed at the level of symbols of
scattering amplitudes which controls the appearance of branch cut
singularities. Poles and symbols become intertwined by cluster
adjacency and we discuss the relation of this property to the
$\bar{Q}$-equation which imposes constraints on the derivatives of the
transcendental functions appearing in loop amplitudes.

\newpage
\phantom{vv}
\vspace{1cm}
\hrule
\tableofcontents

\bigskip
\medskip

\hrule
\newpage
\section{Introduction}

The analytic behaviour of scattering amplitudes has been a subject of great interest for decades \cite{Eden:1966dnq}. Recent developments in the theory of amplitudes have led to the application of an array of mathematical ideas to their calculation. The study of poles in tree-level amplitudes led to the BCFW recursion relations \cite{Britto:2005fq}, that of cuts of integrals to the unitarity approach \cite{Bern:1994zx}. The combination of these ideas has fed into new constructions of loop integrands for many amplitudes \cite{ArkaniHamed:2012nw,Arkani-Hamed:2013jha}. The study of polylogarithmic iterated integrals \cite{Chen:1977oja,Remiddi:1999ew,Goncharov:2005sla,Brown:2009qja} has led to a much greater understanding of loop integrals and motivated a greater push to classify and understand more general functions of elliptic type and beyond \cite{BrownLevin,CaronHuot:2012ab,Bloch:2013tra,Bourjaily:2017bsb,Bourjaily:2018ycu,Broedel:2018qkq}. These developments have inspired recent advances \cite{Henn:2013pwa} in the well-studied subject of differential equations for loop integrals \cite{Chetyrkin:1981qh,Kotikov:1990kg,Bern:1993kr,Remiddi:1997ny} which have been applied to processes of interest for QCD or gauge theories in general. It is clear that the greater understanding we have of the role of singularities in field theory amplitudes the greater our ability is to calculate them and the deeper our understanding of field theory becomes.

Here we will focus on the study of poles and branch cuts in perturbative amplitudes and the algebraic and geometrical structures which govern their appearance. A very helpful toy model in this regard is the planar limit of $\mathcal{N}=4$ super Yang-Mills theory where many approaches can be taken to calculate amplitudes. In perturbation theory an analytic bootstrap programme has been employed for certain amplitudes, allowing the construction of explicit data for many loop orders \cite{Dixon:2011pw,Dixon:2011nj,Dixon:2013eka,Dixon:2014voa,Dixon:2014iba,Drummond:2014ffa,Dixon:2015iva,Caron-Huot:2016owq,Dixon:2016nkn}. A different technique relying on the relation of the planar amplitudes with light-like Wilson loops \cite{Alday:2007hr,Drummond:2007aua,Brandhuber:2007yx,Drummond:2007cf,Bern:2008ap,Drummond:2008aq} is based on multiple expansions in a near-collinear OPE limit \cite{Alday:2010ku,Basso:2013vsa,Basso:2013aha,Basso:2014koa,Basso:2014nra}, much like correlation functions of local operators in conformal field theories. The interplay of these techniques has revealed surprising structures at the heart of scattering amplitudes.

An important observation about the perturbative amplitudes came with the work of \cite{Golden:2013xva} where a link was made between the locations of branch point singularities in scattering amplitudes and certain coordinates (`$\mathcal{A}$-coordinates') of cluster algebras \cite{1021.16017,1054.17024}. In \cite{Drummond:2017ssj} we extended this connection to the interplay of such singularities with each other. Specifically we noticed that the cluster algebras also control the possible sequences of such branch cut singularities; a non-trivial analytic continuation around a given singularity may only be followed by certain others. The set of which singularities are visible on any given Riemann sheet is dictated by the clusters themselves. We refer to this property of amplitudes as `cluster adjacency'. The adjacency relations we find encompass the Steinmann relations \cite{Steinmann,Steinmann2} which place constraints on consecutive discontinuities of amplitudes \cite{Bartels:2008ce}. Such relations can be made manifest on appropriately defined infrared finite quantities and then become a powerful constraint in the analytic bootstrap programme \cite{Caron-Huot:2016owq}.

We will develop the connection between singularities and cluster algebras further. We emphasise that, although the connection to cluster algebras is phrased in algebraic terms, there is also a very geometric picture to the structure of relations between branch point singularities. The geometry in question is that of cluster polytopes and in particular the intricate structure of their boundaries, which captures the possible nested sequences of cluster subalgebras. The picture which emerges is different from, but shares many features with, the positive geometry arising in the description of integrands in \cite{ArkaniHamed:2012nw,Arkani-Hamed:2013jha}.

Since the monodromies of analytic functions in general and amplitudes in particular are typically non-abelian in nature, the cluster adjacency controlling their appearance has a non-abelian character; the order in which $\mathcal{A}$-coordinates appear in the symbol is important. Here we also develop an abelian version of adjacency which controls the poles of individual terms in tree-level amplitudes. We find that precisely the same notion of adjacency holds for individual BCFW terms for NMHV amplitudes and beyond. Since poles multiply in a commutative fashion the adjacency constraints apply to all poles in a given term.

When considering NMHV loop amplitudes we have expressions which simultaneously exhibit non-trivial poles and branch cuts. We find that the cluster structure also imposes relations between the two. Specifically we find that the derivatives of  individual terms in NMHV loop amplitudes are constrained in such a way that they are compatible with the poles of the multiplying rational function. The cluster adjacency we find actually comprises a subset of the constraints which follow from dual superconformal symmetry \cite{Drummond:2008vq}. At loop level these constraints are expressed through the $\bar{Q}$ equation of \cite{CaronHuot:2011kk,Bullimore:2011kg}. So the cluster adjacency structure simultaneously implies both branch cut relations, e.g. the Steinmann relations, and derivative relations such as those following from dual superconformal symmetry .

\section{Amplitudes in planar $\mathcal{N}=4$ super Yang-Mills theory}
Here we recall a few basic properties of scattering amplitudes which are necessary for the discussion of singularities and the link to cluster algebras.

\subsection{Kinematics and symmetries}

The $\mathcal{N}=4$ super Yang-Mills on-shell multiplet may be organised into an on-shell superfield $\Phi$, a function of an on-shell momentum $p^{\alpha \dot{\alpha}} = \lambda^\alpha \tilde{\lambda}^{\dot{\alpha}}$ and a Grassmann variable $\eta^A$ transforming in the $su(4)$ fundamental representation,
\be
\Phi=G^+ +\eta^A\Gamma_A+\tfrac{1}{2!}\eta^A\eta^B S_{AB}+\tfrac{1}{3!}\eta^A\eta^B\eta^C\epsilon_{ABCD}\bar \Gamma^D+\tfrac{1}{4!}\eta^A\eta^B\eta^C\eta^D\epsilon_{ABCD}G^-\,,
\ee
The colour-ordered partial amplitudes in planar $\mathcal{N}=4$ super Yang-Mills exhibit dual superconformal symmetry \cite{Drummond:2008vq} which motivates the introduction of dual variables as follows,
\be
p_i^{\alpha \dot\alpha} =
\lambda_i^\alpha \tilde{\lambda}_i^{\dot\alpha}
= x_{i+1}^{\alpha \dot\alpha} - x_{i}^{\alpha \dot\alpha}\,, \qquad
q_i^{\alpha A} = \lambda_i^{\alpha} \eta_i^A
= \theta_{i+1}^{\alpha A} - \theta_{i}^{\alpha A}\,.
\label{xthetadef}
\ee
The dual symmetries act as superconformal transformations in the $(x,\theta)$ space. The fact that the momenta are null means that the geometry in the dual space is associated with null lines, for which Penrose's (super)twistor variables are most appropriate \cite{Hodges:2009hk},
\be
\mathcal{Z}_i = (Z_i \, | \, \chi_i)\,, \qquad
Z_i^{\alpha,\dot\alpha} =
(\lambda_i^\alpha , x_i^{\beta \dot\alpha}\lambda_{i\beta})\,,
\qquad
\chi_i^A= \theta_i^{\alpha A}\lambda_{i \alpha} \,.
\ee
Here the $Z_i$ variables correspond to points in $\mathbb{P}^3$.

When considering amplitudes we should take care of the structure of infrared divergences and the associated dual conformal anomaly \cite{Drummond:2007cf,Drummond:2007au}. For our considerations here the appropriate way to do this will be to extract from the amplitude the so-called `BDS-like' form of the MHV superamplitude (denoted $\tilde{A}_n$),
\begin{equation}
\label{}
A_{n} = \tilde{A}_{n} \mathcal{E}_n\,.
\end{equation}
The precise form of $\tilde{A}_n$ can be found in \cite{Alday:2009yn} and is not of great relevance here. The important point is that the remaining factor $\mathcal{E}_n$ is dual conformally invariant and comprises all of the non-trivial information about the scattering amplitudes, once the dual conformal Ward identity of \cite{Drummond:2007cf,Drummond:2007au} is taken into account. The function $\mathcal{E}_n$ can be written purely in terms of the supertwistors $\mathcal{Z}_i$ and has an expansion in Grassmann degree which encompasses the decomposition of different amplitudes into MHV, NMHV and so on,
\be
\label{MHVexp}
\mathcal{E}_n  = \mathcal{E}_{n,{\rm MHV}} + \mathcal{E}_{n,{\rm NMHV}} + \ldots\,.
\ee

The MHV term in (\ref{MHVexp}) is of degree zero in the Grassmann $\chi_i$ variables and hence is just a function of the $Z_i$. Dual conformal symmetry implies it is a function of the four-brackets $\langle ijkl \rangle$. It is homogeneous of degree zero in each $Z_i$ and so is a function on the space ${\rm Conf}_n(\mathbb{P}^3)$ (the configuration space of $n$ points in $\mathbb{P}^3$).

The NMHV term in (\ref{MHVexp}) is of Grassmann degree four and can be written in terms of the Yangian invariants (called \emph{R-invariants}),
\be
[ijklm] = \frac{(\langle\langle ijklm \rangle\rangle}{\langle ijkl \rangle \langle jklm \rangle \langle klmi \rangle \langle lmij \rangle \langle mijk \rangle}\,,
\ee
multiplied by dual conformally invariant functions $E_{ijklm}$ on ${\rm Conf}_n(\mathbb{P}^3)$,
\be
\mathcal{E}_{n,{\rm NMHV}} = \sum [ijklm] E_{ijklm}(Z_1,\ldots,Z_n)\,,
\ee
where $\la \la i j k l m \ra \ra = (\chi_i \la j k l m \ra + \text{cyclic})^4$. In what follows the functions $\mathcal{E}_{n,{\rm MHV}}$ and $\mathcal{E}_{n,{\rm NMHV}}$ (and hence the functions $E_{ijklm}$) admit perturbative expansions of the form
\be\label{eq:gLoopExpansion}
F=\sum_{L=0}^\infty g^{2L} F^{(L)}\,.
\ee
For the hexagon and heptagon amplitudes that we focus on here we need only consider MHV and NMHV terms in the expansion (\ref{MHVexp}) since other amplitudes are obtained by parity conjugation of these ones. 

\subsection{Analytic structure in perturbation theory}

In perturbation theory the functions appearing in hexagon and heptagon amplitudes are (according to all current evidence) polylogarithms of degree $2L$ where $L$ is the loop order. Polylogarithms are a class of iterated integrals over logarithmic singularities. Here we will define them in a recursive fashion. We declare that polylogarithms come with a grading and that a polylogarithm $f^{(k)}$ of degree (or \emph{weight}) $k$ obeys
\be
\label{dlogpolys}
d f^{(k)} = \sum_{a \in \mathcal{A}} f_{[a]}^{(k-1)} d \log a\,,
\ee
where the $a$ are some rational (or algebraic) functions of some number of variables (called \emph{letters}) and the sum runs over a finite set $\mathcal{A}$ of such functions (an \emph{alphabet}). The space of functions of degree one is spanned by the set of logarithms of the letters $a$ themselves. The choice of the set $\mathcal{A}$ then determines a class of polylogarithmic functions recursively in the degree. For example, in the case of functions of a single variable $x$, the choice $\mathcal{A} = \{ x, 1-x \}$ yields the class of harmonic polylogarithms \cite{Remiddi:1999ew} with indices $0$ or $1$. In particular this example includes the classical polylogarithms ${\rm Li}_n(x)$.

The formula (\ref{dlogpolys}) encodes the $(k-1,1)$ part of the coproduct of the function $f^{(k)}$. We write this as
\be
f^{(k-1,1)}  = \sum_{a \in \mathcal{A}} [f_{[a]}^{(k-1)} \otimes a]\,,
\ee
where by convention we just record the argument of the $d \log$ in the second argument of the tensor product. The arguments of the $(k-1,1)$ coproduct must obey the integrability relation
\be
\label{intcond}
\sum_{a \in \mathcal{A}} d f_{[a]}^{(k-1)} \wedge d \log a = 0\,,
\ee
which follows from $d^2 f^{(k)}=0$.

If we continue applying the definition of the $(n,1)$ coproduct iteratively to each of the functions $f^{(k-1)}_{[a]}$ all the way down to weight zero we obtain the \emph{symbol}, an element of the $k$-fold tensor product of the space of one-forms spanned by the $d \log a$ for $a \in \mathcal{A}$ (or more compactly a \emph{word} in the alphabet $\mathcal{A}$),
\be
\label{fksymbol}
S[f^{(k)}] = f^{(1,\dots,1)} = \sum_{(a_1,\ldots,a_k)} \!\! c_{a_1,\ldots,a_k} \, [a_1 \otimes a_2 \otimes \ldots \otimes a_k]\,, \quad c_{a_1,\ldots,a_k} \in \mathbb{Q}\,, \quad a_i \in \mathcal{A}\,.
\ee
Note that by common convention we write the letters $a$ rather than $d \log a$ in the arguments of the tensor product. This leads to the property that symbols with products of functions in their arguments decompose as follows,
\be
\label{symmult}
[a \otimes b\, b ' \otimes c] = [a \otimes b \otimes c] + [a \otimes b' \otimes c]\,,
\ee
and similarly that symbols with powers of functions in their arguments obey
\be
\label{sympower}
[a \otimes b^{\, p} \otimes c] = p\, [a \otimes b \otimes c]\, \qquad p \in \mathbb{Q}\,.
\ee
In the following we will discuss examples where the alphabet $\mathcal{A}$ is given by the set of $\mathcal{A}$-coordinates associated to a cluster algebra.

The symbol $S[f^{(k)}]$ displays both the branch cut structure and the differential structure of the function $f^{(k)}$. From the definition of the symbol (\ref{fksymbol}) and the behaviour of polylogarithms under derivative action, (\ref{dlogpolys}) we see derivatives act on the symbol by action on the rightmost element of the tensor product,
\be
d \, [a_1 \otimes \ldots \otimes a_k] = [a_1 \otimes \ldots \otimes a_{k-1}]\, d \log a_k\,.
\ee

The symbol (\ref{fksymbol}) obeys integrability relations,
\be
\sum_{\vec{a}} c_{\vec{a}}\, [a_1 \otimes \ldots \otimes a_{i-1} \otimes a_{i+2} \otimes \ldots \otimes a_k]\, (d \log a_{i} \wedge d \log a_{i+1}) = 0\,, \quad i=1,\ldots,k-1
\ee
which follow from the fact that $d^2 f  = 0$ for all the functions of all weights and encode the commutativity of partial derivatives.

Similarly a logarithmic branch cut discontinuity around a singularity at $a=0$ is obtained from terms beginning with the letters $a$, assuming the alphabet is chosen so that no other letter vanishes at $a=0$,
\be
{\rm disc}_{a=0} [a_1 \otimes a_2 \otimes \ldots \otimes a_k] = (2 \pi i)[ a_2 \otimes  \ldots \otimes a_k]\,.
\ee

The symbol is an efficient tool for simplifying polylogarithmic expressions, as demonstrated in the derivation of the simple formula of \cite{Goncharov:2010jf} for the two-loop MHV hexagon amplitude \cite{DelDuca:2009au}.
A first step in the bootstrap calculations of \cite{Dixon:2011pw,Dixon:2011nj,Dixon:2013eka,Dixon:2014voa,Dixon:2014iba,Drummond:2014ffa,Dixon:2015iva,Caron-Huot:2016owq,Dixon:2016nkn} is to build integrable words in a given alphabet. We quickly review here the method described in \cite{Dixon:2016nkn} for performing this task. The construction of integrable words can be done iteratively in the weight. We suppose that we have a basis $\{ f^{(k)}_i \}$ of integrable words up to weight $k$. This means that we know how to decompose integrable words of weight $k$ into their $(k-1,1)$ coproducts
\be
f^{(k)}_i = \sum_{a,j} M^{(k)}_{ija} [f^{(k-1)}_j \otimes a]\,.
\ee

Now we would like to construct integrable words of weight $(k+1)$. We build an ansatz for the $(k,1)$ coproduct with constants $c_{ai}$,
\be
f^{(k,1)} = \sum_{a,i} c_{ai} [f^{(k)}_i \otimes a]\,.
\ee
The constraints we have to solve come from the integrability condition (\ref{intcond}),
\be
\label{bootstrapint}
\sum_{a,i} c_{ai} df^{(k)} \wedge d\log a  = \sum_{a,i} c_{ai} \sum_{b,j} M^{(k)}_{ijb} f^{(k-1)}_j d \log b \wedge d \log a = 0\,.
\ee
where the first equality expresses $df^{(k)}$ using (\ref{dlogpolys}).

The two-forms $d\log a \wedge d\log b$ are not generally all linearly independent. They satisfy linear relations known as Arnold relations which essentially come from partial fraction identities. We suppose that $\{ \omega^{(2)}_m \}$ form a basis for the space of independent two forms. Then there exists a tensor $Y$ which expresses each two-form $d\log a \wedge d \log b$ in terms of the independent basis
\be
d \log a \wedge d \log b = \sum_m Y_{ab,m}\, \omega^{(2)}_m\,.
\ee
It follows that the condition (\ref{bootstrapint}) becomes
\be
\sum_{a,i} c_{ai} \sum_{b,j} M^{(k)}_{ijb} f^{(k-1)}_j Y_{ab,m} \, \omega^{(2)}_m = 0\,.
\ee
Since the $\omega^{(2)}_m$ form a basis for the independent two-forms and the $f^{(k-1)}_j$ form a basis for the integrable words of weight $(k-1)$ the condition becomes
\be
\sum_{a,i} c_{ai} \sum_{b} M^{(k)}_{ijb} Y_{ab,m} = 0\,.
\ee
In other words we need to compute the kernel of the matrix
\be
\label{linalg}
\mathcal{M}_{AB} = \sum_b M^{(k)}_{ijb} Y_{ab,m}\,,\qquad A=(jm)\,,\, B=(ai)\,,
\ee
where we grouped indices into multi-indices $A,B$. 

To obtain a solution to (\ref{linalg}) is a linear algebra problem that can be helpfully addressed with available packages. The package SpaSM \cite{spasm} for sparse modular linear algebra operations is particularly helpful as the matrices involved are typically sparse and all quantities involved can be chosen to be integer-valued. 
However it is solved, one obtains a basis for the kernel of $\mathcal{M}$, i.e. a set of linearly independent null vectors $\{ v_{A, l} \}$ where $l=1,\ldots,{\rm dim} (\ker \mathcal{M})$. Expanding the multi-index $A=(ai)$ we obtain the desired basis of weight $(k+1)$ words,
\be
f^{(k+1)}_l = \sum_{a,i} M^{(k+1)}_{lia} [f^{(k)}_i \otimes a]\,, \qquad M^{(k+1)}_{lia} = v_{ai,l}\,.
\ee

The above procedure has been used extensively in several works as a first step in the analytic bootstrap programme for amplitudes.

\section{Cluster Algebras and Grassmannians}

\sloppy In \cite{Golden:2013xva} the important observation was made that the symbols of the two-loop MHV remainder functions constructed in \cite{CaronHuot:2011ky} were written in terms of alphabets that exclusively contained $\mathcal{A}$-coordinates of cluster algebras associated to Grassmannians ${\rm Gr}(4,n)$, or more precisely, the $(3n-15)$-dimensional spaces $\allowbreak{\rm Conf}_n(\mathbb{P}^3) \allowbreak= {\rm Gr}(4,n)/(\mathbb{C}^*)^{n-1}$. Beyond two loops, great progress has been made in understanding the hexagon ($n=6$) and heptagon ($n=7$) amplitudes via the analytic bootstrap programme. All current evidence is compatible with the hypothesis that the hexagon and heptagon amplitudes are polylogarithmic at all orders in perturbation theory and moreover that their symbol alphabets are given by the set of $\mathcal{A}$-coordinates for the cases ${\rm Conf}_6(\mathbb{P}^3)$ and ${\rm Conf}_7(\mathbb{P}^3)$ respectively. The associated cluster algebras are isomorphic to the ones based on $A_3$ and $E_6$ respectively. Here we will review some of the important aspects of cluster algebras. Many of the points we recall here are covered already in \cite{Golden:2013xva} but we review them as we will need many of the ideas to explain the notion of adjacency for the cluster polylogarithms appearing in the expressions for scattering amplitudes.

Cluster algebras are commutative associative algebras with generators referred to as cluster coordinates which arise in families called clusters. They can be specified by giving an initial cluster with a set of $\mathcal{A}$-coordinates together with a mutation rule which allows the generation of further clusters and cluster coordinates. To each cluster can be associated a quiver diagram with $\mathcal{A}$-coordinates associated to the nodes. Such a quiver is described by the adjacency matrix $b_{ij}$ defined via
\be
b_{ij} = (\text{no. of arrows } i \rightarrow j) - (\text{no. of arrows } j \rightarrow i)\,.
\ee
The adjacency matrix specifies how the cluster changes under a mutation. If one performs a mutation on a node labelled by $\mathcal{A}$-coordinate $a_k$ then the adjacency matrix of the new cluster is given by 
\begin{equation}
\label{mutationb}
  b'_{ij} =
  \begin{cases}
    -b_{ij}                  &   k \in \{i,j\}\,,\\
    b_{ij}                   &   b_{ik}b_{kj} \leq 0\,,\\
    b_{ij}  + b_{ik}b_{kj}    &   b_{ik}, b_{kj} > 0\,,\\
    b_{ij}  - b_{ik}b_{kj}    &   b_{ik}, b_{kj} < 0\,.\\
  \end{cases}
\end{equation}
and the $\mathcal{A}$-coordinate $a_k$ associated to that node is replaced by
\be
\label{mutationa}
a_k' = \frac{1}{a_k}\biggl[\prod_{i|b_{ik}>0} a_i^{b_{ik}} +\prod_{i|b_{ik}<0} a_i^{-b_{ik}} \biggr]\,.
\ee

\begin{figure}
\begin{center}
{\footnotesize
\begin{tikzpicture}
\pgfmathsetmacro{\nw}{1.3}
\pgfmathsetmacro{\vvwnw}{2.5}
\pgfmathsetmacro{\vvvwnw}{2.85}
\pgfmathsetmacro{\nh}{0.6}
\pgfmathsetmacro{\aa}{0.6}
\pgfmathsetmacro{\ep}{0.1}
\node at (-0.5*\nw -\aa,\aa+0.5*\nh) {$\langle 1\,2\,3\,4 \rangle$};
\draw[] (-\aa,\aa) -- (-\aa -\nw,\aa) -- (-\aa -\nw, \aa+\nh) -- (-\aa,\aa+\nh) -- cycle;
\node at (0.5*\nw +0*\aa,-0*\aa-0.5*\nh) {$\langle 1\,2\,3\,5 \rangle$};
\node at (0.5*\nw +0*\aa,-1*\aa-1.5*\nh) {$\langle 1\,2\,4\,5 \rangle$};
\node at (0.5*\nw +0*\aa,-2*\aa-2.5*\nh) {$\langle 1\,3\,4\,5 \rangle$};
\node at (0.5*\nw +0*\aa,-3*\aa-3.5*\nh) {$\langle 2\,3\,4\,5 \rangle$};
\draw[] (0,-3*\aa-3*\nh) -- (0,-3*\aa-4*\nh) -- (\nw,-3*\aa-4*\nh) -- (\nw,-3*\aa-3*\nh) -- cycle;
\node at (1.5*\nw +1*\aa,-0*\aa-0.5*\nh) {$\langle 1\,2\,3\,6 \rangle$};
\node at (1.5*\nw +1*\aa,-1*\aa-1.5*\nh) {$\langle 1\,2\,5\,6 \rangle$};
\node at (1.5*\nw +1*\aa,-2*\aa-2.5*\nh) {$\langle 1\,4\,5\,6 \rangle$};
\node at (1.5*\nw +1*\aa,-3*\aa-3.5*\nh) {$\langle 3\,4\,5\,6 \rangle$};
\draw[] (\nw+\aa,-3*\aa-3*\nh) -- (\nw+\aa,-3*\aa-4*\nh) -- (2*\nw+\aa,-3*\aa-4*\nh) -- (2*\nw+\aa,-3*\aa-3*\nh) -- cycle;
\node at (3*\nw + 2*\aa+0.5*\vvvwnw,-0*\aa-0.5*\nh) {$\langle 1\,2\,3\,n \scalebox{0.65}[1.0]{\( - \)} 1 \rangle$};
\node at (3*\nw + 2*\aa+0.5*\vvvwnw,-1*\aa-1.5*\nh) {$\langle 1\,2\,n \scalebox{0.65}[1.0]{\( - \)} 2\,n \scalebox{0.65}[1.0]{\( - \)} 1 \rangle$};
\node at (3*\nw + 2*\aa+0.5*\vvvwnw,-2*\aa-2.5*\nh) {$\langle 1\,n \scalebox{0.65}[1.0]{\( - \)} 3\,n \scalebox{0.65}[1.0]{\( - \)} 2\,n \scalebox{0.65}[1.0]{\( - \)} 1 \rangle$};
\node at (3*\nw + 2*\aa+0.5*\vvvwnw,-3*\aa-3.5*\nh) {$\langle n \scalebox{0.65}[1.0]{\( - \)} 4\,n \scalebox{0.65}[1.0]{\( - \)} 3\,n \scalebox{0.65}[1.0]{\( - \)} 2\,n \scalebox{0.65}[1.0]{\( - \)} 1 \rangle$};
\draw[] (3*\nw+2*\aa,-3*\aa-3*\nh) -- (3*\nw+2*\aa,-3*\aa-4*\nh) -- (3*\nw+2*\aa+\vvvwnw,-3*\aa-4*\nh) -- (3*\nw+2*\aa+\vvvwnw,-3*\aa-3*\nh) -- cycle;
\node at (3*\nw + 3*\aa+1*\vvvwnw+0.5*\vvwnw,-0*\aa-0.5*\nh) {$\langle 1\,2\,3\,n \rangle$};
\draw[] (3*\nw+3*\aa+1*\vvvwnw,-0*\aa-0*\nh) -- (3*\nw+3*\aa+1*\vvvwnw,-0*\aa-1*\nh) -- (3*\nw+3*\aa+1*\vvvwnw+1*\vvwnw,-0*\aa-1*\nh) -- (3*\nw+3*\aa+1*\vvvwnw+1*\vvwnw,-0*\aa-0*\nh) -- cycle;
\node at (3*\nw + 3*\aa+1*\vvvwnw+0.5*\vvwnw,-1*\aa-1.5*\nh) {$\langle 1\,2\,n \scalebox{0.65}[1.0]{\( - \)} 1\,n \rangle$};
\draw[] (3*\nw+3*\aa+1*\vvvwnw,-1*\aa-1*\nh) -- (3*\nw+3*\aa+1*\vvvwnw,-1*\aa-2*\nh) -- (3*\nw+3*\aa+1*\vvvwnw+1*\vvwnw,-1*\aa-2*\nh) -- (3*\nw+3*\aa+1*\vvvwnw+1*\vvwnw,-1*\aa-1*\nh) -- cycle;
\node at (3*\nw + 3*\aa+1*\vvvwnw+0.5*\vvwnw,-2*\aa-2.5*\nh) {$\langle 1\,n \scalebox{0.65}[1.0]{\( - \)} 2\,n \scalebox{0.65}[1.0]{\( - \)} 1\,n \rangle$};
\draw[] (3*\nw+3*\aa+1*\vvvwnw,-2*\aa-2*\nh) -- (3*\nw+3*\aa+1*\vvvwnw,-2*\aa-3*\nh) -- (3*\nw+3*\aa+1*\vvvwnw+1*\vvwnw,-2*\aa-3*\nh) -- (3*\nw+3*\aa+1*\vvvwnw+1*\vvwnw,-2*\aa-2*\nh) -- cycle;
\draw[] (3*\nw+3*\aa+1*\vvvwnw,-3*\aa-3*\nh) -- (3*\nw+3*\aa+1*\vvvwnw,-3*\aa-4*\nh) -- (3*\nw+3*\aa+1*\vvvwnw+1*\vvwnw,-3*\aa-4*\nh) -- (3*\nw+3*\aa+1*\vvvwnw+1*\vvwnw,-3*\aa-3*\nh) -- cycle;
\node at (3*\nw + 3*\aa+1*\vvvwnw+0.5*\vvwnw,-3*\aa-3.5*\nh) {$\langle n \scalebox{0.65}[1.0]{\( - \)} 3\,n \scalebox{0.65}[1.0]{\( - \)} 2\,n \scalebox{0.65}[1.0]{\( - \)} 1\,n \rangle$};
\node at (2.25*\nw + 2*\aa,-0.5*\nh) {$\ldots$};
\node at (2.25*\nw + 2*\aa,-1.5*\nh-\aa) {$\ldots$};
\node at (2.25*\nw + 2*\aa,-2.5*\nh-2*\aa) {$\ldots$};
\node at (2.25*\nw + 2*\aa,-3.5*\nh-3*\aa) {$\ldots$};
\draw[->] (-\aa+0*\ep,\aa-\ep) -- (0-0*\ep,0+\ep);
\draw[->] (0.5*\nw,-\nh-\ep) -- (0.5*\nw,-\nh-\aa+\ep);
\draw[->] (0.5*\nw,-2*\nh-\aa-\ep) -- (0.5*\nw,-2*\nh-2*\aa+\ep);
\draw[->] (0.5*\nw,-3*\nh-2*\aa-\ep) -- (0.5*\nw,-3*\nh-3*\aa+\ep);
\draw[->] (1*\nw+\ep,-0.5*\nh) -- (1*\nw+\aa-\ep,-0.5*\nh);
\draw[->] (1*\nw+\ep,-1.5*\nh-\aa) -- (1*\nw+\aa-\ep,-1.5*\nh-\aa);
\draw[->] (1*\nw+\ep,-2.5*\nh-2*\aa) -- (1*\nw+\aa-\ep,-2.5*\nh-2*\aa);
\draw[->] (1*\nw+\aa-0*\ep,-\nh-\aa+\ep) -- (1*\nw+0*\ep,-\nh-\ep);
\draw[->] (1*\nw+\aa-0*\ep,-2*\nh-2*\aa+\ep) -- (1*\nw+0*\ep,-2*\nh-1*\aa-\ep);
\draw[->] (1*\nw+\aa-0*\ep,-3*\nh-3*\aa+\ep) -- (1*\nw+0*\ep,-3*\nh-2*\aa-\ep);
\draw[->] (1.5*\nw+1*\aa,-\nh-\ep) -- (1.5*\nw+1*\aa,-\nh-\aa+\ep);
\draw[->] (1.5*\nw+1*\aa,-2*\nh-1*\aa-\ep) -- (1.5*\nw+1*\aa,-2*\nh-2*\aa+\ep);
\draw[->] (1.5*\nw+1*\aa,-3*\nh-2*\aa-\ep) -- (1.5*\nw+1*\aa,-3*\nh-3*\aa+\ep);
\draw[->] (3*\nw+0.5*\vvvwnw+2*\aa,-\nh-\ep) -- (3*\nw+0.5*\vvvwnw+2*\aa,-\nh-\aa+\ep);
\draw[->] (3*\nw+0.5*\vvvwnw+2*\aa,-2*\nh-\aa-\ep) -- (3*\nw+0.5*\vvvwnw+2*\aa,-2*\nh-2*\aa+\ep);
\draw[->] (3*\nw+0.5*\vvvwnw+2*\aa,-3*\nh-2*\aa-\ep) -- (3*\nw+0.5*\vvvwnw+2*\aa,-3*\nh-3*\aa+\ep);
\draw[->] (3*\nw+2*\aa+1*\vvvwnw+\ep,-0.5*\nh) -- (3*\nw+3*\aa+1*\vvvwnw-\ep,-0.5*\nh);
\draw[->] (3*\nw+2*\aa+1*\vvvwnw+\ep,-1.5*\nh-1*\aa) -- (3*\nw+3*\aa+1*\vvvwnw-\ep,-1.5*\nh-1*\aa);
\draw[->] (3*\nw+2*\aa+1*\vvvwnw+\ep,-2.5*\nh-2*\aa) -- (3*\nw+3*\aa+1*\vvvwnw-\ep,-2.5*\nh-2*\aa);
\draw[->] (3*\nw+3*\aa+1*\vvvwnw-0*\ep,-\nh-\aa+\ep) -- (3*\nw+2*\aa+1*\vvvwnw+0*\ep,-\nh-\ep);
\draw[->] (3*\nw+3*\aa+1*\vvvwnw-0*\ep,-2*\nh-2*\aa+\ep) -- (3*\nw+2*\aa+1*\vvvwnw+0*\ep,-2*\nh-1*\aa-\ep);
\draw[->] (3*\nw+3*\aa+1*\vvvwnw-0*\ep,-3*\nh-3*\aa+\ep) -- (3*\nw+2*\aa+1*\vvvwnw+0*\ep,-3*\nh-2*\aa-\ep);
\end{tikzpicture}
}
\end{center}
\caption{The initial cluster of the Grassmannian series ${\rm Gr}(4,n)$.}
\label{Gr4ninitial}
\end{figure}
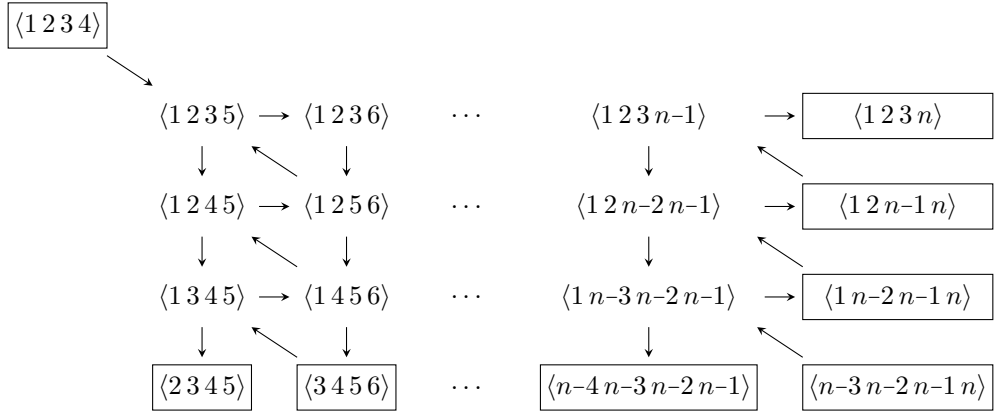

For the set of cluster algebras associated to ${\rm Conf}_n(\mathbb{P}^3)$ we take the initial cluster depicted in Fig. \ref{Gr4ninitial}.
The boxed nodes are referred to as \emph{frozen} nodes and the remainder are \emph{unfrozen}. Other clusters (and hence other $\mathcal{A}$-coordinates) are obtained by mutating on the unfrozen nodes according to the above rules. In the cases of $n=6,7$ the number of distinct clusters obtained is finite. For $n=6$ all $\mathcal{A}$-coordinates are Pl\"ucker coordinates of the form $\langle ijkl \rangle$ while $n=7$ some $\mathcal{A}$-coordinates are quadratic in Pl\"uckers.

If we pick a particular $\mathcal{A}$-coordinate $a$ and look at all clusters containing $a$ we obtain a cluster subalgebra. Such clusters may be generated by starting in one cluster containing $a$ and performing all possible combinations of mutations on the other nodes. In this way, to each $\mathcal{A}$-coordinate we associate a codimension-one subalgebra. Similarly we may pick a pair of coordinates $\{ a, b \}$ and, as long as there is at least one cluster where they both appear, associate to them a codimension-two subalgebra by performing all possible mutations on the other nodes. If there is no cluster where $a$ and $b$ appear together then there is no such subalgebra. The fact that some pairs can be found together (we call them `admissible' or `adjacent') while other pairs cannot is at the heart of the cluster adjacency property describing the behaviour of singularities of scattering amplitudes. Note that frozen nodes are present in every cluster and hence are always admissible with any other $\mathcal{A}$-coordinate.

We can continue further and associate codimension-three subalgebras with admissible triplets $\{a,b,c\}$ where $a$, $b$ and $c$ can all be found together in some cluster and so on. Finally when we have fixed an admissible set of $(3n-15)$ $\mathcal{A}$-coordinates we uniquely specify a cluster which we could alternatively describe as a dimension-zero subalgebra.

Note that while $\mathcal{A}$-coordinates are called `coordinates' they are not strictly coordinates on ${\rm Conf}_n(\mathbb{P}^3)$ because they are not homogeneous under rescalings of the twistors. A natural set of homogeneous coordinates for ${{\rm Conf}_n(\mathbb{P}^3)}$  are the cluster $\mathcal{X}$-coordinates. They are defined with respect to a given cluster for each unfrozen node $j$ and are related to the $\mathcal{A}$-coordinates and the adjacency matrix of the cluster via
\be
x_j = \prod_i a_i^{b_{ij}}\,,
\ee
where the product runs over all nodes (frozen and unfrozen) labelled by $i$.
Under mutation on an unfrozen node $k$ the $\mathcal{X}$-coordinates change according to 
\be
x_i' = \begin{cases}
    1/x_i              &   k = i\,,\\
    x_i\bigl(1+x_k^{{\rm sgn}(b_{ik})}\bigr)^{b_{ik}} & k \neq i\,.
  \end{cases}
\ee
Note that if node $i$ is not connected to node $k$ then $b_{ik}=0$ and $x_i'=x_i$.

The adjacency matrix $b_{ij}$ actually defines a Poisson structure on the space ${\rm Conf}_n(\mathbb{P})$ via the formula
\be
\label{poisson}
\{ x_i , x_j \} = b_{ij} x_i x_j.
\ee
The choice of cluster is irrelevant since the formula (\ref{poisson}) is preserved under mutation. Note that only the restriction of the adjacency matrix to the unfrozen nodes actually appears in (\ref{poisson}). We recall that a Poisson structure can be described in terms of a bivector $b$ such that $b(d f , d g) = \{f,g\}$. The adjacency matrix of a cluster then gives the components of the Poisson bivector in the coordinate system given by the (logarithms of the) cluster $\mathcal{X}$-coordinates for that cluster. 

If we restrict attention to the real case then the condition that all $\mathcal{X}$-coordinates obey $0<x<\infty$ defines a region inside ${\rm Conf}_n(\mathbb{RP}^3)$. The region should be visualised as a polytope with a boundary that is made of facets corresponding to codimension-one sub-algebras of the original cluster algebra (which, as we described before are associated to individual $\mathcal{A}$-coordinates). Each facet has boundaries corresponding to codimension-two subalgebras (associated to admissible pairs of $\mathcal{A}$-coordinates) and so on. If we continue all the way down we arrive at dimension-zero subalgebras given by the clusters themselves and corresponding to corners of the polytope in the sense that the corner is the origin in the associated set of $\mathcal{X}$-coordinates.

The cluster $\mathcal{X}$-coordinates are edge coordinates in that they can be associated to the one-dimensional edges (axes) which meet at the vertex corresponding to the cluster. On each edge the associated $\mathcal{X}$-coordinate runs over $0<x<\infty$, in correspondence with the fact that the $\mathcal{X}$-coordinate associated to a given edge inverts under the mutation along that edge.

\subsection{Hexagons and the $A_3$ associahedron}
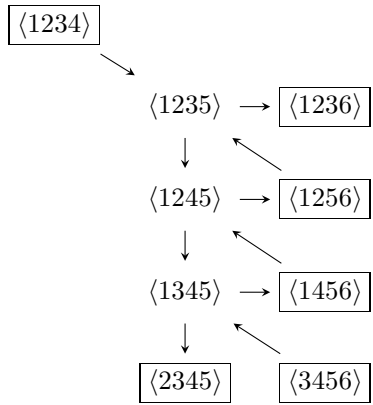
\begin{figure}
{\footnotesize
\begin{center}
\makeatletter  
\newcommand{\phantombox}[1]{%
  \setbox0=\hbox{#1}%
  \begin{tcolorbox}[colframe=white,colback=white,boxrule=0.4pt,
    left=2pt,right=2pt,top=3pt,bottom=3pt,boxsep=0pt,width=1.2cm, valign = center,  halign=center, sharp corners = all]
    #1
  \end{tcolorbox}
}
\newcommand{\frozenbox}[1]{%
  \setbox0=\hbox{#1}%
  \begin{tcolorbox}[colframe=black,colback=white,boxrule=0.5pt,
      left=2pt,right=2pt,top=2pt,bottom=2pt,boxsep=0pt,width=1.2cm, halign=center, sharp corners = all]
    #1
  \end{tcolorbox}
}
\makeatother
\begin{tikzpicture}%
    [
    unfrozen/.style={},
    frozen/.style={inner sep=1.2mm,outer sep=0mm,yshift=0},
    node distance = 0.4cm
    ]
    \node[frozen]        (f0) at (0,5) {$\frozenbox{$\langle 1234 \rangle$}$};
    \node[frozen, below right = of f0]        (t1)  {$\phantombox{$\langle1235 \rangle$}$};
    \node[frozen, right = of t1]        (t2)  {$\frozenbox{$\langle 1236 \rangle$}$};
    \node[frozen, below = of t1]        (m1)  {$\phantombox{$\langle 1245 \rangle$}$};
    \node[frozen, right = of m1]        (m2)  {$\frozenbox{$\langle 1256 \rangle$}$};
    \node[frozen, below = of m1]        (b1)  {$\phantombox{$\langle1345 \rangle$}$};
    \node[frozen, right = of b1]        (b2)  {$\frozenbox{$\langle1456 \rangle$}$};
    %
    %
    \node[frozen, below = of b1]        (f4)  {$\frozenbox{$\langle 2345 \rangle$}$};
    \node[frozen, right = of f4]        (f5)  {$\frozenbox{$\langle 3456 \rangle$}$};
    \draw[->] (f0) -- (t1);
    \draw[->] (t1) -- (t2);    \draw[->] (t1) -- (m1) ;   
    \draw[->] (m1) -- (m2);  \draw[->] (m1) -- (b1) ; \draw[->] (m2) -- (t1);     
    \draw[->] (b1) -- (b2);  
    \draw[->] (b1) -- (f4) ; 
    \draw[->] (b2) -- (m1);     
    \draw[->] (f5) -- (b1);
\end{tikzpicture}
\end{center}}
\caption{The quiver diagram for the initial cluster for the algebra associated to ${\rm Conf}_6(\mathbb{P}^3)$.}
\label{hexinitial}
\end{figure}

For $\mathrm{Conf}_6(\mathbb{P}^3)$, the initial cluster is represented by the quiver diagram given in Fig. \ref{hexinitial} with Pl\"ucker coordinates at each of the nodes. The unfrozen $\mathcal{A}$-coordinates of this cluster are
\be
a_1 = \langle 1235 \rangle\,, \qquad a_2 = \langle 1245 \rangle\,, \qquad a_3 = \langle 1345 \rangle\,.
\ee

\begin{figure}
{\footnotesize
\begin{center}
\begin{tikzpicture}[scale=0.46]
 \draw[join=bevel,thick,gray,dashed]  (5,-2.75) -- (5,-5.25);
 \draw[join=bevel,->,gray,dashed] (1.25,3.75) -- (2.5,2.5);
\node[gray] at (0.25,4.25){$\langle 1345 \rangle$};
\draw[join=bevel,->,gray,dashed] (8.75,3.75) -- (7.5,2.5);
\node[gray] at (9.75,4.25){$\langle 1356 \rangle$};
\draw[join=bevel,->,gray,dashed] (8.75,-4.25) -- (7.5,-3);
\node[gray] at (9.75,-4.75){$\langle 2346 \rangle$};
\draw[join=bevel,->,gray,dashed] (1.25,-4.25) -- (2.5,-3);
\node[gray] at (0.25,-4.75){$\langle 1246 \rangle$};
 \draw[join=bevel,thick,gray,dashed] (0,0) -- (2.75,0.25) ;
  \draw[join=bevel,thick,gray,dashed]  (7.25,0.25) -- (10,0);
  \draw[join=bevel,thick,gray,dashed]  (2.75,0.25) -- (5,3.25) -- (7.25,0.25) -- (5,-2.75) -- cycle;
  \draw[join=bevel,thick,gray,dashed]  (5,3.25) -- (5,5);
  \draw[join=bevel,thick,fill=none] (0,0) -- (1,3) -- (2,-0.25) -- (1,-3.5)  -- cycle;
  \draw[join=bevel,thick,fill=none] (2,-0.25) -- (8,-0.25);
  \draw[join=bevel,thick,fill=none] (10,0) -- (9,3) -- (8,-0.25) -- (9,-3.5)  -- cycle;
  \draw[join=bevel,thick,fill=none] (1,3) -- (5,5) -- (9,3);
  \draw[join=bevel,thick,fill=none] (1,-3.5) -- (5,-5.25) -- (9,-3.5);
    \draw[thick,fill=none] (0,0) -- (1,3) -- (5,5) -- (9,3) -- (10,0) -- (9,-3.5) -- (5,-5.25) -- (1,-3.5) -- cycle;
\draw[join=bevel,->] (-0.75,-0.125) -- (1,-0.125);
\node at (-2,-0.125){$\langle 1245 \rangle$};
\draw[join=bevel,->] (10.75,-0.125) -- (9,-0.125);
\node at (12,-0.125){$\langle 2356 \rangle$};
\draw[join=bevel,->] (5,5.75) arc (-200:-160:5.75);
\node at (5,6.5){$\langle 1235 \rangle$};
\draw[join=bevel,->] (5,-6) arc (-20:20:6.25);
\node at (5,6.5){$\langle 1235 \rangle$};
\node at (5,-6.75){$\langle 2456 \rangle$};
\node[gray] at (5,0.25){$\langle 1346 \rangle$};
\draw[join=bevel,thick,->] (1,3) -- (1.5,1.375);
\draw[join=bevel,thick,->] (2,-0.25) -- (5,-0.25);
\draw[join=bevel,thick,->] (8,-0.25) -- (8.5,-1.875);
 \end{tikzpicture}
 \end{center}
 }
 \caption{The $A_3$ Stasheff polytope with six pentagonal faces and three square faces, each labelled with the corresponding $\mathcal{A}$-coordinate. The initial cluster corresponds to the vertex at the top left corner at the intersection of the faces labelled by $\langle 1235 \rangle$, $\langle 1245 \rangle$, $\langle 1345\rangle$. The three-step path leads from the initial cluster to one obtained by a cyclic rotation by one unit.}
 \label{Stasheff}
 \end{figure}
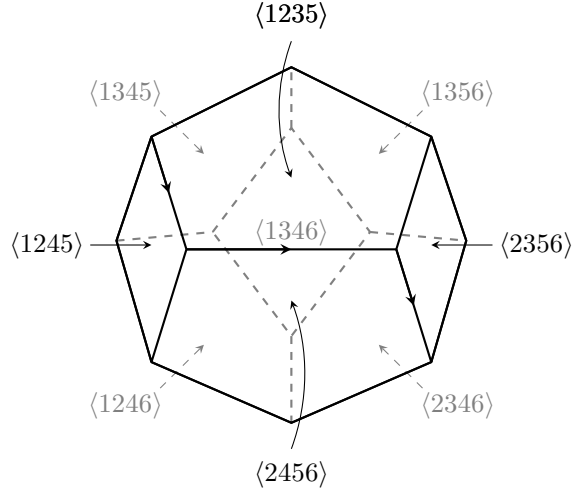

By repeated mutation of the above data according to (\ref{mutationb}) and (\ref{mutationa}) one obtains 14 distinct clusters arranged in the topology of the Stasheff polytope or associahedron illustrated in Fig. \ref{Stasheff}. In total nine distinct unfrozen $\mathcal{A}$-coordinates are obtained, corresponding to the nine faces of the polytope, in addition to the six frozen ones present in every cluster. Three are square faces and six are pentagonal. Each cluster corresponds to a vertex, with the unfrozen $\mathcal{A}$-coordinates of the cluster corresponding to the faces of the polytope which meet at the vertex. The frozen $\mathcal{A}$-coordinates $\langle i\, i+1\,i+2\,i+3 \rangle$, being present in every cluster, are not shown in Fig. \ref{Stasheff}. The initial cluster drawn in Fig. \ref{hexinitial} corresponds to the cluster in the top left of Fig. \ref{Stasheff}. The edges between clusters correspond to mutation operations.

Fig. \ref{Stasheff} also makes manifest the discrete symmetries of the ${\rm Conf}_6(\mathbb{P}^3)$ cluster algebra. A cyclic rotation of the initial cluster can be generated by a threefold sequence of mutations, as indicated by the arrows. This corresponds to mutating on the three unfrozen nodes in Fig. \ref{hexinitial} in turn, starting at the bottom and moving to the top. A threefold cyclic rotation corresponds to a reflection in the equatorial plane of Fig. \ref{Stasheff} and also corresponds to the parity transformation $Z_i \mapsto Z_{i-1} \wedge Z_i \wedge Z_{i+1}$ when applied to homogeneous quantities. Finally, the reflection $Z_i \mapsto Z_{7-i}$ corresponds to a left-right reflection of Fig. \ref{Stasheff} together with a reflection in the equatorial plane.

\begin{figure}
{\footnotesize
\begin{center}
\begin{tikzpicture}[scale=0.8]
   \foreach \i in {1,...,6}
   {
     \coordinate (a\i) at (60+60*\i:1.6);
   }
   \draw[thick]
     (a1)
     \foreach \i in {2,...,6}
   {
     -- (a\i)
   } -- cycle;

   \draw[thick] (a6) -- (a2);
   \draw[thick] (a6) -- (a3);
   \draw[thick] (a6) -- (a4);
    \foreach \i in {1,...,6}
   {
     \node at (60+60*\i:1.9) {$\i$};
   }
\end{tikzpicture}
\end{center}
}
\caption{The two-brackets $(ij)$ can be identified with chords on a hexagon between the vertices $i$ and $j$. A triangulation of the hexagon then corresponds to a cluster of the $A_3$ or ${\rm Conf}_6(\mathbb{P}^3)$ polytope. Above is shown the triangulation corresponding to the initial cluster of Fig. \ref{hexinitial} comprised of the chords $(26) = \langle 1345 \rangle$, $(36) = \langle 1245 \rangle$ and $(46) = \langle 1235 \rangle$ together with the six edges which correspond to the frozen nodes.}
\label{hexagonchords}
\end{figure}
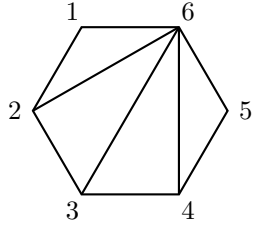

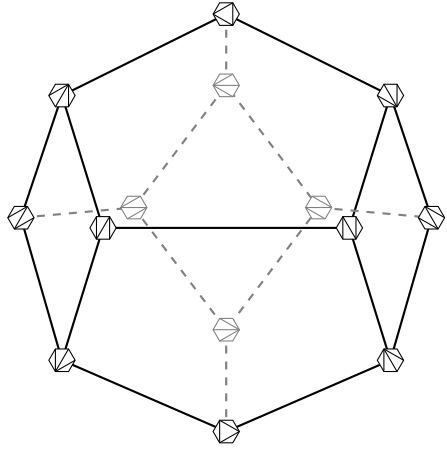
\begin{figure}
\begin{center}
\begin{tikzpicture}[scale=0.54]
  \draw[thick,fill=none] (0,0) -- (1,3) -- (2,-0.25) -- (1,-3.5)  -- cycle;
  \draw[thick,fill=none] (2,-0.25) -- (8,-0.25);
  \draw[thick,fill=none] (10,0) -- (9,3) -- (8,-0.25) -- (9,-3.5)  -- cycle;
  \draw[thick,fill=none] (1,3) -- (5,5) -- (9,3);
  \draw[thick,fill=none] (1,-3.5) -- (5,-5.25) -- (9,-3.5);
  \draw[thick,gray,dashed] (0,0) -- (2.75,0.25) ;
  \draw[thick,gray,dashed]  (7.25,0.25) -- (10,0);
  \draw[thick,gray,dashed]  (2.75,0.25) -- (5,3.25) -- (7.25,0.25) -- (5,-2.75) -- cycle;
  \draw[thick,gray,dashed]  (5,3.25) -- (5,5);
  \draw[thick,gray,dashed]  (5,-2.75) -- (5,-5.25);

\foreach \i in {1,...,6}
   {
     \coordinate (a\i) at ($(60+60*\i:0.32)+(0,0)$);
   }
   \draw[fill=white]
     (a1)
     \foreach \i in {2,...,6}
   {
     -- (a\i)
   } -- cycle;

   \draw (a6) -- (a2);
   \draw (a6) -- (a3);
   \draw (a3) -- (a5);

 \foreach \i in {1,...,6}
   {
     \coordinate (a\i) at ($(60+60*\i:0.32)+(1,3)$);
   }
   \draw[fill=white]
     (a1)
     \foreach \i in {2,...,6}
   {
     -- (a\i)
   } -- cycle;

   \draw (a6) -- (a2);
   \draw (a6) -- (a3);
   \draw (a6) -- (a4);

 \foreach \i in {1,...,6}
   {
     \coordinate (a\i) at ($(60+60*\i:0.32)+(2,-0.25)$);
   }
   \draw[fill=white]
     (a1)
     \foreach \i in {2,...,6}
   {
     -- (a\i)
   } -- cycle;

   \draw (a1) -- (a3);
   \draw (a6) -- (a3);
   \draw (a6) -- (a4);

 \foreach \i in {1,...,6}
   {
     \coordinate (a\i) at ($(60+60*\i:0.32)+(1,-3.5)$);
   }
   \draw[fill=white]
     (a1)
     \foreach \i in {2,...,6}
   {
     -- (a\i)
   } -- cycle;

   \draw (a1) -- (a3);
   \draw (a6) -- (a3);
   \draw (a3) -- (a5);

 \foreach \i in {1,...,6}
   {
     \coordinate (a\i) at ($(60+60*\i:0.32)+(5,5)$);
   }
   \draw[fill=white]
     (a1)
     \foreach \i in {2,...,6}
   {
     -- (a\i)
   } -- cycle;

   \draw (a2) -- (a4);
   \draw (a4) -- (a6);
   \draw (a6) -- (a2);

\foreach \i in {1,...,6}
   {
     \coordinate (a\i) at ($(60+60*\i:0.32)+(5,-5.25)$);
   }
   \draw[fill=white]
     (a1)
     \foreach \i in {2,...,6}
   {
     -- (a\i)
   } -- cycle;

   \draw (a1) -- (a3);
   \draw (a3) -- (a5);
   \draw (a5) -- (a1);

\foreach \i in {1,...,6}
   {
     \coordinate (a\i) at ($(60+60*\i:0.32)+(8,-0.25)$);
   }
   \draw[fill=white]
     (a1)
     \foreach \i in {2,...,6}
   {
     -- (a\i)
   } -- cycle;

   \draw (a1) -- (a3);
   \draw (a1) -- (a4);
   \draw (a4) -- (a6);

\foreach \i in {1,...,6}
   {
     \coordinate (a\i) at ($(60+60*\i:0.32)+(9,3)$);
   }
   \draw[fill=white]
     (a1)
     \foreach \i in {2,...,6}
   {
     -- (a\i)
   } -- cycle;

   \draw (a1) -- (a4);
   \draw (a2) -- (a4);
   \draw (a4) -- (a6);
   
\foreach \i in {1,...,6}
   {
     \coordinate (a\i) at ($(60+60*\i:0.32)+(10,0)$);
   }
   \draw[fill=white]
     (a1)
     \foreach \i in {2,...,6}
   {
     -- (a\i)
   } -- cycle;

   \draw (a1) -- (a4);
   \draw (a1) -- (a5);
   \draw (a2) -- (a4);
   
   \foreach \i in {1,...,6}
   {
     \coordinate (a\i) at ($(60+60*\i:0.32)+(9,-3.5)$);
   }
   \draw[fill=white]
     (a1)
     \foreach \i in {2,...,6}
   {
     -- (a\i)
   } -- cycle;

   \draw (a1) -- (a3);
   \draw (a1) -- (a4);
   \draw (a1) -- (a5);

   \foreach \i in {1,...,6}
   {
     \coordinate (a\i) at ($(60+60*\i:0.32)+(2.75,0.25)$);
   }
   \draw[gray,fill=white]
     (a1)
     \foreach \i in {2,...,6}
   {
     -- (a\i)
   } -- cycle;

   \draw[gray] (a2) -- (a5);
   \draw[gray] (a2) -- (a6);
   \draw[gray] (a3) -- (a5);

   \foreach \i in {1,...,6}
   {
     \coordinate (a\i) at ($(60+60*\i:0.32)+(5,3.25)$);
   }
   \draw[gray,fill=white]
     (a1)
     \foreach \i in {2,...,6}
   {
     -- (a\i)
   } -- cycle;

   \draw[gray] (a2) -- (a4);
   \draw[gray] (a2) -- (a5);
   \draw[gray] (a2) -- (a6);
   
      \foreach \i in {1,...,6}
   {
     \coordinate (a\i) at ($(60+60*\i:0.32)+(7.25,0.25)$);
   }
   \draw[gray,fill=white]
     (a1)
     \foreach \i in {2,...,6}
   {
     -- (a\i)
   } -- cycle;

   \draw[gray] (a1) -- (a5);
   \draw[gray] (a2) -- (a5);
   \draw[gray] (a2) -- (a4);
   
      \foreach \i in {1,...,6}
   {
     \coordinate (a\i) at ($(60+60*\i:0.32)+(5,-2.75)$);
   }
   \draw[gray,fill=white]
     (a1)
     \foreach \i in {2,...,6}
   {
     -- (a\i)
   } -- cycle;

   \draw[gray] (a1) -- (a5);
   \draw[gray] (a2) -- (a5);
   \draw[gray] (a3) -- (a5);   

 \end{tikzpicture}
 \end{center}
 \caption{The Stasheff polytope for ${\rm Conf}_6(\mathbb{P}^3) \cong \mathcal{M}_{0,6}$ with the clusters labelled by the different triangulations of a hexagon.}
 \label{Stasheff_triang}
 \end{figure}

The space ${\rm Conf}_6(\mathbb{P}^3)$ can be identified with the space ${\rm Conf}_6(\mathbb{P}^1) \cong \mathcal{M}_{0,6}$, that is the moduli space of six points on the Riemann sphere modulo $sl_2$ transformations. At the level of Pl\"ucker coordinates this can be achieved by identifying an ordered four-bracket $\langle ijkl \rangle$ (such that $i<j<k<l$) with an ordered two-bracket $(mn)$ (with  $m<n$) made of the absent labels from the set $\{1, \ldots , 6\}$. In other words we make the identifications $\langle 1345 \rangle = (26)$, $\langle 1245 \rangle = (36)$, $\langle 1235 \rangle = (46)$ and so on. In this way the nodes of each quiver diagram can be identified with chords of a hexagon. The edges of the hexagon correspond to adjacent two-brackets, e.g. $(12) = \langle 3456 \rangle$. Such an identification is described in Fig. \ref{hexagonchords}. With the triangulation labelling of clusters to hand we may illustrate all triangulations on the Stasheff polytope, as shown in Fig. \ref{Stasheff_triang}. 

It is important to stress again that Fig. \ref{Stasheff} is not just a pictorial representation of a set of topological relations between clusters. If we restrict attention to the case of real twistors then ${\rm Conf}_6(\mathbb{RP}^3) \cong \mathcal{M}_{0,6}(\mathbb{R})$ is a three-dimensional space. The interior of the polytope in Fig. \ref{Stasheff} is precisely the region inside ${\rm Conf}_6(\mathbb{RP}^3)$ where all the cluster $\mathcal{X}$-coordinates obey $0<x<\infty$. Each corner is the origin in the set of $\mathcal{X}$-coordinates defined by the corresponding cluster. As an example, the $\mathcal{X}$-coordinates of the initial cluster for the ${\rm Conf}_6(\mathbb{P}^3)$ polytope are
\begin{align}
x_1 &= \frac{\langle 1234 \rangle \langle 1256 \rangle}{\langle 1236 \rangle \langle 1245 \rangle} = \frac{(56)(34)}{(45)(36)}\,, \notag \\
x_2 &= \frac{\langle 1235 \rangle \langle 1456 \rangle}{\langle 1256 \rangle \langle 1345 \rangle} = \frac{(46)(23)}{(34)(26)}\,, \notag \\
x_3 &= \frac{\langle 1245 \rangle \langle 3456 \rangle}{\langle 2345 \rangle \langle 1456 \rangle} = \frac{(36)(12)}{(16)(23)}\,.
\end{align}
The vertex corresponding to the initial cluster is the origin $x_1=x_2=x_3=0$ in this coordinate system. The cluster coordinates run from $0$ to $\infty$ along the three one-dimensional edges which meet at the vertex. This is in accord with the fact that under a mutation (which corresponds to moving along an edge to an adjacent vertex) the associated $\mathcal{X}$-coordinate inverts.

The adjacency matrix for the unfrozen nodes of the initial cluster is 
\be
\label{A3bmatrix}
b = 
\left(
\begin{tabular}{ccc}
0 & 1 & 0 \\
-1 & 0 & 1 \\
0 &-1&0
\end{tabular}
\right)\,.
\ee
We recall that the $\mathcal{X}$-coordinates are (log) canonical coordinates for the Poisson bracket. The adjacency matrix $b$ in (\ref{A3bmatrix}) is singular and has rank two. This means that there is a coordinate $\Delta$ which Poisson commutes with every function $\{ \Delta , f \} = 0$. It is the product of two of the $x_i$ above,
\be
\label{Delta}
\Delta = x_1 x_3 = \frac{(12)(34)(56)}{(16)(23)(45)} = \frac{\langle 1234 \rangle \langle 1256 \rangle \langle 3456 \rangle}{\langle 1236 \rangle \langle 1456 \rangle \langle 2345 \rangle}\,.
\ee
Equivalently, there is a canonical (up to a constant rescaling) one-form $d \log \Delta$ which is null under the action of the Poisson bivector,
\be
b(d\log \Delta , \cdot) = 0\,.
\ee
Note that $\Delta$ is built purely from frozen $\mathcal{A}$-coordinates. 

Another natural set of coordinates are \emph{dihedral} coordinates \cite{Brown:2009qja} which can be defined (here with respect to the trivial ordering $\{1,\ldots,n\}$) for all the moduli spaces $\mathcal{M}_{0,n}$ (or $A_{n-3}$ cluster algebras) via
\be
u_{ij} = \frac{(i\,j+1)(i+1\,j)}{(i\,j)(i+1\,j+1)}\,.
\ee
We require that the labels $i$ and $j$ are separated by at least two (as for unfrozen two-brackets). For the case $n=6$ this implies that there are nine such dihedral coordinates, each labelled by the chords of the hexagon.

The interior of the Stasheff polytope is the region where all nine $u_{ij}$ obey $0< u_{ij} < 1$. A face of the polytope is the locus defined by $u_{ij}=0$ where $(ij)$ is the chord associated to that face. When a particular $u_{ij}=0$ then all the $u_{kl}$ such that the chord $(kl)$ intersects the chord $(ij)$ take the value $1$. The vertices of the polytope are then the origin in the coordinate system defined by taking the dihedral coordinates associated to the triangulation of the corresponding cluster. For example, the initial cluster is associated to the origin in the coordinates $\{u_{26},u_{36},u_{46}\}$ and the equations $u_{26}=0$, $u_{36}=0$ and $u_{46}=0$ define the faces labelled by $\langle 1345 \rangle = (26)$, $\langle 1245 \rangle = (36)$ and $\langle 1235 \rangle = (46)$ respectively in Fig. \ref{Stasheff}. These dihedral coordinates are related to the $\mathcal{X}$-coordinates above via
\begin{align}
u_{26} = \frac{x_3}{1+x_3}\,, \qquad
u_{36} = \frac{x_2(1+x_3)}{1+x_2+x_2 x_3}\,, \qquad
u_{46} = \frac{x_1(1+x_2+x_2 x_3)}{1+x_1+x_1x_2+x_1x_2x_3} \,. 
\end{align}
From the above relations it is clear that the three faces meeting at the vertex are equivalently defined either by vanishing of dihedral coordinates or by vanishing of cluster $\mathcal{X}$-coordinates.

The dihedral coordinates form a complete set of nine multiplicatively independent homogeneous combinations of the $\mathcal{A}$-coordinates. They can therefore be taken as an alphabet for the construction of polylogarithms on ${\rm Conf}_6(\mathbb{P}^3) = \mathcal{M}_{0,6}$. They are related to the nine letters taken in e.g. \cite{Dixon:2011nj} for the construction of hexagon functions as follows,
\begin{eqnarray}
\label{hexusual}
u = u_{26} u_{35} u_{25} u_{36}\,, \qquad & 1-u = u_{14}\,, \qquad &y_u = \frac{u_{35}}{u_{26}}\,,\notag \\
v = u_{13} u_{46} u_{36} u_{14}\,, \qquad & 1-v = u_{25}\,, \qquad &y_v = \frac{u_{13}}{u_{46}}\,,\notag \\
w = u_{24} u_{15} u_{14} u_{25}\,, \qquad & 1-w = u_{36}\,, \qquad &y_w = \frac{u_{15}}{u_{24}}\,.
\end{eqnarray}
The null Poisson coordinate $\Delta$ is given by
\be
\label{Delta}
\Delta = \frac{u_{24} u_{26} u_{46}}{u_{13} u_{15} u_{35}} = \frac{1}{y_u y_v y_w}\,.
\ee

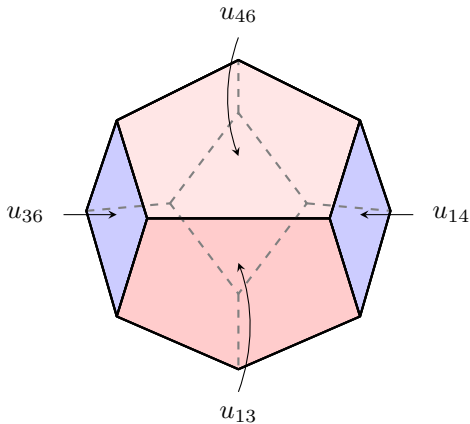
\begin{figure}
{\footnotesize
\begin{center}
\begin{tikzpicture}[scale=0.4]
  \draw[join=bevel,thick,fill=blue!20] (0,0) -- (1,3) -- (2,-0.25) -- (1,-3.5)  -- cycle;
  \draw[join=bevel,thick,fill=red!10] (2,-0.25) -- (8,-0.25) -- (9,3) -- (5,5) -- (1,3) -- cycle ;
  \draw[join=bevel,thick,fill=blue!20] (10,0) -- (9,3) -- (8,-0.25) -- (9,-3.5)  -- cycle;
  \draw[thick,fill=none] (1,3) -- (5,5) -- (9,3);
  \draw[join=bevel,thick,fill=red!20] (1,-3.5) -- (5,-5.25) -- (9,-3.5) -- (8,-0.25) --  (2,-0.25) -- cycle;
   \draw[thick,gray,dashed] (0,0) -- (2.75,0.25) ;
  \draw[thick,gray,dashed]  (7.25,0.25) -- (10,0);
  \draw[thick,gray,dashed]  (2.75,0.25) -- (5,3.25) -- (7.25,0.25) -- (5,-2.75) -- cycle;
  \draw[thick,gray,dashed]  (5,3.25) -- (5,5);
  \draw[thick,gray,dashed]  (5,-2.75) -- (5,-5.25);
   \draw[join=bevel,thick,fill=none] (0,0) -- (1,3) -- (2,-0.25) -- (1,-3.5)  -- cycle;
    \draw[join=bevel,thick,fill=none] (2,-0.25) -- (8,-0.25) -- (9,3) -- (5,5) -- (1,3) -- cycle ;
  \draw[join=bevel,thick,fill=none] (10,0) -- (9,3) -- (8,-0.25) -- (9,-3.5)  -- cycle;
   \draw[join=bevel,thick,fill=none] (1,-3.5) -- (5,-5.25) -- (9,-3.5) -- (8,-0.25) --  (2,-0.25) -- cycle;
     \draw[thick,fill=none] (0,0) -- (1,3) -- (5,5) -- (9,3) -- (10,0) -- (9,-3.5) -- (5,-5.25) -- (1,-3.5) -- cycle;
\draw[->] (-0.75,-0.125) -- (1,-0.125);
\node at (-2,-0.125){$u_{36}$};
\draw[->] (10.75,-0.125) -- (9,-0.125);
\node at (12,-0.125){$u_{14}$};
\draw[->] (5,5.75) arc (-200:-160:5.75);
\node at (5,6.5){$u_{46}$};
\draw[->] (5,-6) arc (-20:20:6.25);
\node at (5,-6.75){$u_{13} $};
 \end{tikzpicture}
 \end{center}
 }
 \caption{The $A_3$ polytope with four faces labelled by their dihedral coordinates. The double scaling limits $u_{46}\rightarrow0$ and its parity conjugate version $u_{13} \rightarrow 0$ are are the highlighted red pentagons. The soft limits $u_{36} \rightarrow 0$ and $u_{14} \rightarrow 0$ are the blue squares. The line joining the two squares corresponds to the collinear limit $u_{13} = u_{46} = 0$.}
 \label{Stasheffhighlights}
 \end{figure}

There are two distinct types of codimension-one subalgebras in the $A_3$ polytope. Each pentagonal face of Fig. \ref{Stasheff} corresponds to an $A_2$ subalgebra. For example, freezing the node labelled by $\langle 1235 \rangle = (46)$ in the initial cluster, and mutating the other nodes generates the pentagon of clusters around the edge of the corresponding face of the polytope. The condition $u_{46}=0$ corresponds to restricting to the pentagonal boundary. Physically, taking the limit $u_{46} \rightarrow 0$ corresponds to taking the double scaling limit where $v\rightarrow 0$ on the branch where $y_v \rightarrow \infty$. Its parity conjugate version is the limit $u_{13} \rightarrow 0$ which corresponds to $v \rightarrow 0$ on the branch where $y_v \rightarrow 0$. These double-scaling limits are highlighted in red in Fig. \ref{Stasheffhighlights}.

The other type of codimension-one subalgebra is $A_1 \times A_1$, corresponding to a square face, as can be obtained from freezing the node $\langle 1245\rangle = (36)$ in the initial cluster and mutating the others. The condition $u_{36}=0$ defines this face and taking the limit $u_{36} \rightarrow 0 $ corresponds to taking the soft limit where $u \rightarrow 0$, $v\rightarrow 0$, $w \rightarrow 1$. Note that this limit is a limit to a codimension one (i.e. dimension two) subspace. This is important because, although the soft limit itself (of the remainder function) is independent of the location approached on the face, after analytic continuation the same limit corresponds to a Regge limit which is not independent of where on the face is being approached. The remaining transverse kinematic dependence of the amplitude in the Regge limit is precisely parametrised by the two-dimensional square face. The limit $u_{36} \rightarrow 0$ and a cyclically rotated one $u_{14} \rightarrow 0$ are highlighted as blue squares in Fig. \ref{Stasheffhighlights}.

Admissible pairs of unfrozen nodes are pairs of faces which intersect on the boundary, e.g. the pair $\{\langle 1235 \rangle , \langle 2456 \rangle\} = \{(46),(13)\}$ is admissible and intersects in a codimension-two (i.e. dimension-one) $A_1$ subalgebra corresponding to the shared edge of those two faces. The edge in question is defined by $u_{46} = u_{13} = 0$ and corresponds to taking the collinear limit of the hexagon amplitudes. Note that the collinear limit indeed interpolates between two soft limits corresponding to the square faces labelled by $(36)$ and $(14)$.

The pair $\{\langle 1245 \rangle  , \langle 2356 \rangle \} = \{(36),(14)\}$ on the other hand is not admissible as the corresponding faces do not intersect on the boundary of Fig. \ref{Stasheff}. The absence of such an intersection is directly related to the Steinmann relations obeyed by scattering amplitudes, or even more basically, to the absence of overlapping factorisation poles in tree-level amplitudes. In general we can describe admissible pairs as non-intersecting chords $(ij)$ of the polygon while intersecting chords give non-admissible pairs. Frozen $\mathcal{A}$-coordinates correspond to the edges of the polygon and therefore do not intersect any chord and hence are admissible with every other $\mathcal{A}$-coordinate.

Finally, admissible triples correspond to corners of Fig. \ref{Stasheff}, i.e. to clusters themselves. They are codimension-three or dimension-zero subalgebras and as an example we could take the triplet $\{ \langle 1235 \rangle , \langle 1245 \rangle , \langle 1345 \rangle\}$ which defines the initial cluster.

The full space ${\rm Conf}_6(\mathbb{RP}^3) \cong \mathcal{M}_{0,6}(\mathbb{R})$ is tiled by 60 regions identical to the Stasheff polytope of Fig. \ref{Stasheff}. In general \cite{Brown:2009qja}, the moduli spaces $\mathcal{M}_{0,n}(\mathbb{R})$ are tiled by $n!/(2n)$ regions which are $(n-3)$-dimensional polytopes, each corresponding to a choice of dihedral structure (i.e. an ordering modulo cyclic  transformations and reflections) on the $n$ points in $\mathbb{RP}^1$. 

Each vertex of the polytope provides a natural base point for the contour of integration over which a symbol made of homogeneous combinations of the $\mathcal{A}$-coordinates can be iteratively integrated to produce a polylogarithmic function \cite{Brown:2009qja}.

\subsection{Heptagons and the $E_6$ polytope}

\begin{figure}
{\footnotesize
\begin{center}
\makeatletter  
\newcommand{\phantombox}[1]{%
  \setbox0=\hbox{#1}%
  \begin{tcolorbox}[colframe=white,colback=white,boxrule=0.4pt,
    left=2pt,right=2pt,top=3pt,bottom=3pt,boxsep=0pt,width=1.2cm, valign = center,  halign=center, sharp corners = all]
    #1
  \end{tcolorbox}
}
\newcommand{\frozenbox}[1]{%
  \setbox0=\hbox{#1}%
  \begin{tcolorbox}[colframe=black,colback=white,boxrule=0.5pt,
      left=2pt,right=2pt,top=2pt,bottom=2pt,boxsep=0pt,width=1.2cm, halign=center, sharp corners = all]
    #1
  \end{tcolorbox}
}
\makeatother
\begin{tikzpicture}%
    [
    unfrozen/.style={},
    frozen/.style={inner sep=1.2mm,outer sep=0mm,yshift=0},
    node distance = 0.5cm
    ]
    \node[frozen]        (f0) at (0,5) {$\frozenbox{$\langle 1234 \rangle$}$};
    \node[frozen, below right = of f0]        (t1)  {$\phantombox{$\langle1235 \rangle$}$};
    \node[frozen, right = of t1]        (t2)  {$\phantombox{$\langle1236 \rangle$}$};
    \node[frozen, below = of t1]        (m1)  {$\phantombox{$\langle1245 \rangle$}$};
    \node[frozen, below = of t2]        (m2)  {$\phantombox{$\langle1256 \rangle$}$};
    \node[frozen, below = of m1]        (b1)  {$\phantombox{$\langle1345 \rangle$}$};
    \node[frozen, below = of m2]        (b2)  {$\phantombox{$\langle1456 \rangle$}$};
    \node[frozen, right = of t2]        (f1)  {$\frozenbox{$\langle1237 \rangle$}$};
    \node[frozen, right = of m2]        (f2)  {$\frozenbox{$\langle1267 \rangle$}$};
    \node[frozen, right = of b2]        (f3)  {$\frozenbox{$\langle1567 \rangle$}$};
    \node[frozen, below = of b1]        (f4)  {$\frozenbox{$\langle2345 \rangle$}$};
    \node[frozen, below = of b2]        (f5)  {$\frozenbox{$\langle3456 \rangle$}$};
    \node[frozen, right= of f5]         (f6)  {$\frozenbox{$\langle4567 \rangle$}$};
    
    \draw[->] (f0) -- (t1);
    \draw[->] (t1) -- (t2);    \draw[->] (t1) -- (m1) ;    \draw[->] (t2) -- (f1) ;  \draw[->] (t2) -- (m2) ;
    \draw[->] (m1) -- (m2);  \draw[->] (m1) -- (b1) ; \draw[->] (m2) -- (t1);      \draw[->] (m2) -- (f2) ;  \draw[->] (m2) -- (b2) ; \draw[->] (f2) -- (t2);
    \draw[->] (b1) -- (b2);  \draw[->] (b1) -- (f4) ; \draw[->] (b2) -- (m1);      \draw[->] (b2) -- (f3) ;  \draw[->] (b2) -- (f5) ; \draw[->] (f3) -- (m2);
    \draw[->] (f6) -- (b2);
    \draw[->] (f5) -- (b1);
\end{tikzpicture}
\end{center}}
\caption{The initial cluster of the ${\rm Conf}_7(\mathbb{P}^3)$ cluster algebra, relevant for heptagon amplitudes.}
\label{heptinitial}
\end{figure}
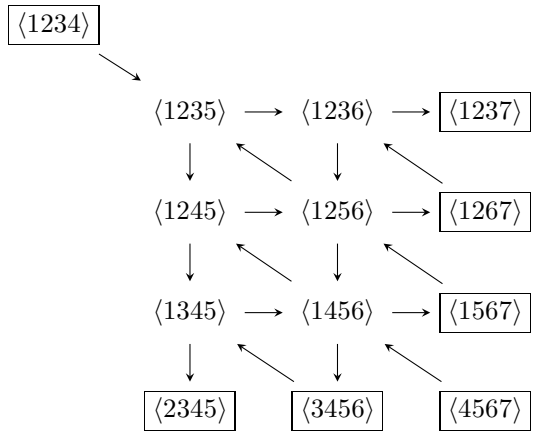

For $\mathrm{Gr}(4,7)$, the initial cluster is represented by the quiver diagram of Fig. \ref{heptinitial}. Each cluster contains six unfrozen nodes as well as the seven frozen ones labelled by the adjacent four-brackets $\langle i\, i+1\,i+2\,i+3\rangle$. Repeated mutation generates a total of 833 distinct clusters containing a total of 42 distinct unfrozen $\mathcal{A}$-coordinates in addition to the 7 frozen ones. 

\begin{figure}
\begin{center}
\begin{tikzpicture}%
    [
    unfrozen/.style={},
    frozen/.style={inner sep=1.5mm,outer sep=0mm,yshift=0},
    node distance = 0.6cm
    ]
    \node[frozen]  (t1)  at (1,4) {$a_{24}$};
    \node[frozen, right = of t1]        (t2)  {$a_{37}$};
    \node[frozen, below = of t1]        (m1)  {$a_{13}$};
    \node[frozen, below = of t2]        (m2)  {$a_{17}$};
    \node[frozen, below = of m1]        (b1)  {$a_{32}$};
    \node[frozen, below = of m2]        (b2)  {$a_{27}$};
    
    \draw[->] (t1) -- (t2); \draw[->] (t1) -- (m1); \draw[->] (t2) -- (m2);
    \draw[->] (m1) -- (m2); \draw[->] (m1) -- (b1); \draw[->] (m2) -- (t1); \draw[->] (m2) -- (b2) ;
    \draw[->] (b1) -- (b2); \draw[->] (b2) -- (m1);      

  \end{tikzpicture}\,.
\end{center}
\caption{The initial cluster for ${\rm Conf}_7(\mathbb{P}^3)$ labelled by homogenised $\mathcal{A}$-coordinates.}
\label{heptinitialhom}
\end{figure}
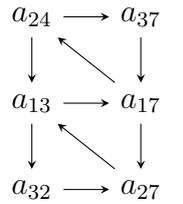

A useful feature of cases of ${\rm Gr}(k,n)$ where the pair $(k,n)$ is coprime (such as the heptagon case) is that one may use the frozen $\mathcal{A}$-coordinates to render the unfrozen ones homogeneous \cite{Drummond:2014ffa}. In this way one can make a natural set of 42 homogeneous letters labelled in one-to-one correspondence with the 42 unfrozen $\mathcal{A}$-coordinates. They are given by the following six quantities together with their cyclic rotations,
\begin{equation}
\begin{aligned}[b]
  a_{11} &= \frac{\langle 1234\rangle\langle1567\rangle\langle2367\rangle}{\langle1237\rangle\langle1267\rangle\langle3456\rangle}\\
  a_{31} &= \frac{\langle1567\rangle\langle2347\rangle}{\langle1237\rangle\langle4567\rangle}\\
  a_{51} &= \frac{\langle1(23)(45)(67)\rangle}{\langle1234\rangle\langle1567\rangle}
\end{aligned}
\,\,\,
\begin{aligned}[b]
  a_{21} &= \frac{\langle1234\rangle\langle2567\rangle}{\langle1267\rangle\langle2345\rangle}\\
  a_{41} &= \frac{\langle2457\rangle\langle3456\rangle}{\langle2345\rangle\langle4567\rangle}  \\
  a_{61} &= \frac{\langle1(34)(56)(72)\rangle}{\langle1234\rangle\langle1567\rangle}\,,
\end{aligned}\,
\label{heptletters}
\end{equation}
Here we use the notation 
\be
\langle 1 (23) (45) (67) \rangle = \langle 1234 \rangle \langle 5671 \rangle - \langle 1235 \rangle \langle 4671\rangle\,.
\ee
By labelling the nodes of the quiver diagram with the homogenised $\mathcal{A}$-coordinates, the initial cluster can be illustrated as in Fig. \ref{heptinitialhom}.

Just as in the hexagon case we should try to visualise the 833 clusters being connected together in a polytope (the $E_6$ polytope). The polytope is a six-dimensional space with 42 codimension one (i.e dimension five) boundary faces, corresponding to the 42 unfrozen $\mathcal{A}$-coordinates. Considering the dimension and the number of vertices it is not as visually instructive to plot the full polytope as a graph. Nevertheless similar general features are present as in the hexagon case. 

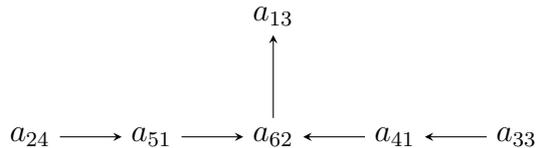
\begin{figure}
\begin{center}
\begin{tikzpicture}[scale=0.8]


\node[] (a51) at (-2, 0) {$a_{51}$}; 
\node[] (a24) at (-4, 0)   {$a_{24}$};           
\node[] (a62) at (0, 0)    {$a_{62}$};          
\node[] (a41) at (2, 0)  {$a_{41}$};          
\node[] (a33) at (4, 0)    {$a_{33}$};           
\node[] (a13) at (0, 2)    {$a_{13}$};           

\draw[norm] (a33) -- (a41);
\draw[norm] (a41) -- (a62) ;
\draw[norm] (a62) -- (a13) ;
\draw[norm] (a24) -- (a51) ;
\draw[norm] (a51) -- (a62) ;

\end{tikzpicture}

\end{center}
\caption{The initial cluster of ${\rm Conf}_7(\mathbb{P}^3)$ does not have the
  topology of an $E_6$ Dynkin diagram but it is possible to mutate it
  to one which does. This cluster contains homogenised $\mathcal{A}$-coordinates of all six types given in  (\ref{heptletters}). 
  }
\label{howtonshep}
\end{figure}

To illustrate the structure of possible subalgebras it is helpful to bring the initial cluster to a cluster with the topology of an $E_6$ Dynkin diagram by a sequence of mutations as shown in Fig. \ref{howtonshep}.
A helpful feature of the $E_6$-shaped cluster is that its homogenised $\mathcal{A}$-coordinates contain one representative of each of the six cyclically related classes given in eq. (\ref{heptletters}). The codimension-one subalgebras obtained by freezing any given letter are then obvious. Freezing $a_{13}$ and mutating on the other nodes generates an $A_5$ subalgebra. Freezing $a_{25}$ or $a_{33}$ will generate a $D_5$ subalgebra. Freezing $a_{41}$ or $a_{51}$ generates an $A_4 \times A_1$ subalgebra. Finally freezing $a_{62}$ generates an $A_2 \times A_2 \times A_1$ subalgebra. The $E_6$-shaped cluster is special in this regard. For example, the initial cluster contains only $a_{1i}$, $a_{2i}$ and $a_{3i}$ types of coordinates and therefore is at the intersection only of $D_5$ and $A_5$ type subalgebras.

Admissible pairs in the $E_6$ case correspond to codimension two subalgebras, i.e. dimension four subalgebras. For example the admissible pair $\{ a_{13}, a_{62} \}$ corresponds to an $A_2 \times A_2$ subalgebra while the pair $\{ a_{51} , a_{41} \}$ corresponds to an $A_2 \times A_1 \times A_1$ subalgebra. Admissible triplets correspond to dimension three subalgebras and so on.

Each cluster (or dimension zero subalgbera) corresponds to a vertex on the boundary of the $E_6$ polytope and the six associated cluster $\mathcal{X}$-coordinates define a local coordinate system such that the vertex is the origin. Once again the $\mathcal{X}$-coordinates can be associated to the one-dimensional edges of the polytope and the interior of the polytope is the region where all $\mathcal{X}$-coordinates obey $0<x<\infty$. The six $\mathcal{X}$-coordinates for the $E_6$-shaped cluster are shown in Fig. \ref{E6xcoords}. The five-dimensional face corresponding to the $A_5$ subalgebra is the boundary component defined by $x_6 =0$ with all other $x_i$ obeying $0<x_i<\infty$. The condition $x_4 = 0$ defines a face corresponding to an $A_4 \times A_1$ subalgebra and so on.

As in the $A_3$ case we may define another set of coordinates $u_{ij}$ such that the $u_{ij}=0$ defines the codimension one face labelled by $a_{ij}$. In terms of the cluster $\mathcal{X}$-coordinates of the $E_6$ shaped cluster we have the following six face coordinates\footnote{Such variables have already been derived by Arkani-Hamed and collaborators \cite{Nimaslides} for finite cluster algebras from a different perspective. Here we obtain them from the cluster $\mathcal{X}$-coordinates. We would like to thank Nima Arkani-Hamed for discussions of this point.},
\begin{align}
\label{usfromxs}
u_{13} &= \frac{x_1}{1+x_1}         &  u_{62} &= \frac{x_6(1+x_1)}{1+x_6+x_1 x_6}             \\
u_{51} &= \frac{x_5(1+x_6+x_1x_6)}{1+x_5+x_5x_6+x_1x_5x_6}         &   u_{41} &=\frac{x_4(1+x_6+x_1x_6)}{1+x_4+x_4x_6+x_1x_4x_6}  \notag \\
u_{24} &= \frac{x_2(1+x_5+x_5x_6+x_1x_5x_6)}{1+x_2+x_2x_5+x_2x_5x_6+x_1x_2x_5x_6}   &  u_{33}&= \frac{x_3(1+x_4+x_4x_6+x_1x_4x_6)}{1+x_3+x_3x_4+x_3x_4x_6+x_1x_3x_4x_6} \,.\notag
\end{align}
Again the origin in the cluster $\mathcal{X}$-coordinates coincides with the origin in the face coordinates.
In terms of the homogenised $\mathcal{A}$-coordinates we have
\begin{align}
u_{13} &= \frac{a_{62}}{a_{11}a_{13}}         &  u_{62} &= \frac{a_{11}a_{41}a_{51}}{a_{62}a_{67}}         \notag    \\
u_{51} &= \frac{a_{24}a_{67}}{a_{46}a_{51}}         &   u_{41} &=\frac{a_{33}a_{67}}{a_{41}a_{56}}  \notag \\
u_{24} &= \frac{a_{46}}{a_{24}a_{31}}   &  u_{33}&= \frac{a_{56}}{a_{22}a_{33}}\,.
\label{usfromas}
\end{align}
Again we clearly have $0<u_{ij}<1$ in the interior of the polytope from (\ref{usfromxs}).
From the equations (\ref{usfromas}) and cyclically related equations one can define a complete set of 42 homogeneous coordinates $u_{ij}$ which makes an alternative multiplicatively independent set to the $a_{ij}$. The variables $u_{ij}$ have the property that $u_{ij}=0$ implies $u_{kl}=1$ if the face labelled by $a_{kl}$ is not adjacent to the face labelled by $a_{ij}$. In other words, setting one $u_{ij}$ to zero for a given face means that all the $u_{kl}$ corresponding to non-adjacent faces go to 1.

Just as in the $A_3$ case there are specific sequences of mutations which generate a cyclic transformation of the $\mathcal{A}$-coordinates in a given cluster. Rather than describe it here for $E_6$ we give a general discussion for ${\rm Conf}_n(\mathbb{P}^{k-1})$ in the next section.

\begin{figure}
\begin{center}
\begin{tikzpicture}[scale=0.7]

\node[] (x51) at (-2, 0) {$x_5$}; 
\node[] (x24) at (-4, 0)   {$x_2$};           
\node[] (x62) at (0, 0)    {$x_6$};          
\node[] (x41) at (2, 0)  {$x_4$};          
\node[] (x33) at (4, 0)    {$x_3$};           
\node[] (x13) at (0, 2)    {{$x_1$}};           
\node[] (a51) at (9, 0) {$\frac{a_{24}}{a_{62}} $}; 
\node[] (a24) at (7, 0)   {$\frac{1}{a_{51}}$};           
\node[] (a62) at (11, 0)    {$\frac{a_{41} a_{51}}{a_{13}}$};          
\node[] (a41) at (13, 0)  {$\frac{a_{33}}{a_{62}}$};          
\node[] (a33) at (15, 0)    {$\frac{1}{a_{41}}$};           
\node[] (a13) at (11, 2)    {{\footnotesize $a_{62}$}};           
\node[] (eq) at (5.5, 1)    {{$=$}};
\draw[norm] (a33) -- (a41);
\draw[norm] (a41) -- (a62) ;
\draw[norm] (a62) -- (a13) ;
\draw[norm] (a24) -- (a51) ;
\draw[norm] (a51) -- (a62) ;
\draw[norm] (x33) -- (x41);
\draw[norm] (x41) -- (x62) ;
\draw[norm] (x62) -- (x13) ;
\draw[norm] (x24) -- (x51) ;
\draw[norm] (x51) -- (x62) ;

\end{tikzpicture}

\end{center}
\caption{The $E_6$-shaped cluster with $\mathcal{X}$-coordinates shown at each of the nodes.}
\label{E6xcoords}
\end{figure}

\subsection{General cyclic mutations for $n>7$}
\label{gencyc}

\begin{figure}
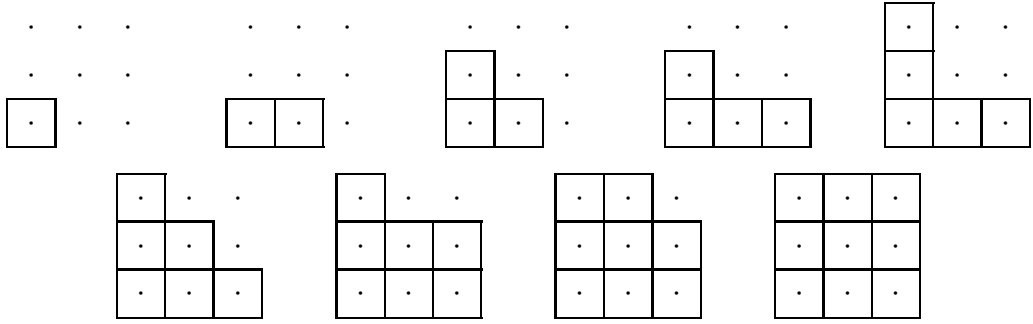

\begin{center}
\begin{ytableau}
\none[\cdot] & \none[\cdot] & \none[\cdot] \\
\none[\cdot] & \none[\cdot] & \none[\cdot] \\
\cdot & \none[\cdot] & \none[\cdot] \\
\end{ytableau}
\qquad
\begin{ytableau}
\none[\cdot] & \none[\cdot] & \none[\cdot] \\
\none[\cdot] & \none[\cdot] & \none[\cdot] \\
\cdot & \cdot & \none[\cdot] \\
\end{ytableau}
\qquad
\begin{ytableau}
\none[\cdot] & \none[\cdot] & \none[\cdot] \\
\cdot & \none[\cdot] & \none[\cdot] \\
\cdot & \cdot & \none[\cdot] \\
\end{ytableau}
\qquad
\begin{ytableau}
\none[\cdot] & \none[\cdot] & \none[\cdot] \\
\cdot & \none[\cdot] & \none[\cdot] \\
\cdot & \cdot & \cdot \\
\end{ytableau}
\qquad
\begin{ytableau}
\cdot & \none[\cdot] & \none[\cdot] \\
\cdot & \none[\cdot] & \none[\cdot] \\
\cdot & \cdot & \cdot \\
\end{ytableau}
\\ \vspace{3mm}
\begin{ytableau}
\cdot & \none[\cdot] & \none[\cdot] \\
\cdot & \cdot & \none[\cdot] \\
\cdot & \cdot & \cdot \\
\end{ytableau}
\qquad
\begin{ytableau}
\cdot & \none[\cdot] & \none[\cdot] \\
\cdot & \cdot & \cdot \\
\cdot & \cdot & \cdot \\
\end{ytableau}
\qquad
\begin{ytableau}
\cdot & \cdot & \none[\cdot] \\
\cdot & \cdot & \cdot \\
\cdot & \cdot & \cdot \\
\end{ytableau}
\qquad
\begin{ytableau}
\cdot & \cdot & \cdot \\
\cdot & \cdot & \cdot \\
\cdot & \cdot & \cdot \\
\end{ytableau}
\end{center}
\caption{A series of mutations which result in a rotation of the Gr$(4,8)$ initial cluster by one unit. The dots represent unfrozen nodes (arrows have been removed for clarity) and the squares represent the mutated nodes. Note there are no gaps between mutated nodes and we always mutate from the bottom up and from left to right. }
\label{fig:YoungTableaux}
\end{figure}

For $n>7$, the ${\rm Conf}_n(\mathbb{P}^3)$ cluster algebra is infinite. We can still define a positive region where all $\mathcal{X}$-coordinates are positive but the structure of its boundary is much less clear. We can still, however, understand certain finite aspects of these infinite algebras. For instance we can mutate from the initial cluster in Fig. \ref{Gr4ninitial} to another one in which all the $\cA$-coordinate labels have been rotated by one unit. We do this by mutating in a manner that mirrors building Young tableaux, instead building from the bottom-left to the top-right (as opposed from top-left to bottom-right) as demonstrated in Fig. \ref{fig:YoungTableaux}.

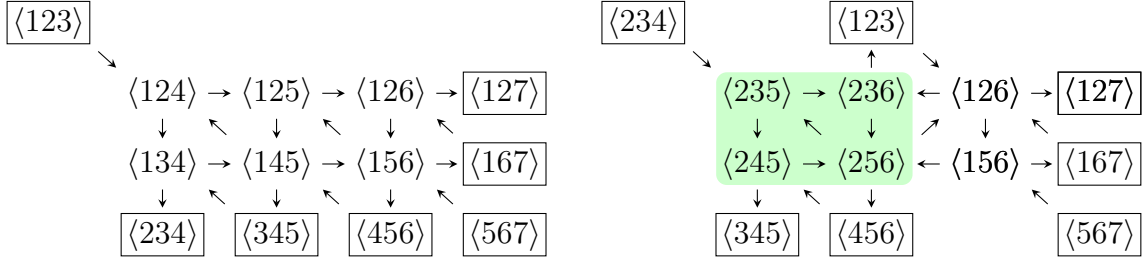
\begin{figure}
	\begin{subfigure}{0.4\textwidth}
		\begin{tikzpicture}[scale=0.75]

		\node[frozen] (123) at (-2, 1.25) {$\ab{123}$};
		\node (124) at (0, 0) {$\ab{124}$};
		\node (125) at (2, 0) {$\ab{125}$};
		\node (126) at (4,0) {$\ab{126}$};
		\node[frozen] (127) at (6,0) {$\ab{127}$};
		\node (134) at (0,-1.25) {$\ab{134}$};
		\node (145) at (2,-1.25) {$\ab{145}$};
		\node (156) at (4, -1.25) {$\ab{156}$};
		\node[frozen] (167) at (6,-1.25) {$\ab{167}$};
		\node[frozen] (234) at (0,-2.5) {$\ab{234}$};
		\node[frozen] (345) at (2,-2.5) {$\ab{345}$};
		\node[frozen] (456) at (4, -2.5) {$\ab{456}$};
		\node[frozen] (567) at (6, -2.5) {$\ab{567}$};

		\draw[->,shorten <=4pt, shorten >=3pt] (123.south east) -- (124.north 		west);
		\draw[norm] (124) -- (125);
		\draw[norm] (125) -- (126);
		\draw[norm] (126) -- (127);
		\draw[norm] (124) -- (134);
		\draw[norm] (125) -- (145);
		\draw[norm] (126) -- (156);
		\draw[norm] (134) -- (145);
		\draw[norm] (145) -- (156);
		\draw[norm] (156) -- (167);
		\draw[norm] (134) -- (234);
		\draw[norm] (145) -- (345);
		\draw[norm] (156) -- (456);
		\draw[diag] (145.north west) -- (124.south east);
		\draw[diag] (156.north west) -- (125.south east);
		\draw[diag] (167.north west) -- (126.south east);
		\draw[diag] (345.north west) -- (134.south east);
		\draw[diag] (456.north west) -- (145.south east);
		\draw[diag] (567.north west) -- (156.south east);

		\end{tikzpicture}
	\end{subfigure}
\hspace*{1.5cm}
\begin{subfigure}{0.4\textwidth}
		\begin{tikzpicture}[scale=0.75]

		\node[frozen] (234) at (-2, 1.25) {$\ab{234}$};
		\node[frozen] (123) at (2, 1.25) {$\ab{123}$};
		\node (235) at (0, 0) {$\ab{235}$};
		\node (236) at (2, 0) {$\ab{236}$};
		\node (126) at (4,0) {$\ab{126}$};
		\node[frozen] (127) at (6,0) {$\ab{127}$};
		\node (245) at (0,-1.25) {$\ab{245}$};
		\node (256) at (2,-1.25) {$\ab{256}$};
		\node (156) at (4, -1.25) {$\ab{156}$};
		\node[frozen] (167) at (6,-1.25) {$\ab{167}$};
		\node[frozen] (345) at (0,-2.5) {$\ab{345}$};
		\node[frozen] (456) at (2,-2.5) {$\ab{456}$};
		\node[frozen] (567) at (6, -2.5) {$\ab{567}$};

		\begin{scope}[blend mode=overlay,overlay]
          \node[rectangle,fill=green!20, rounded corners, fit=(235)(236)(245)(256), inner sep=0.3pt] {};
        \end{scope}
		\node (235) at (0, 0) {$\ab{235}$};
		\node (236) at (2, 0) {$\ab{236}$};
		\node (126) at (4,0) {$\ab{126}$};
		\node[frozen] (127) at (6,0) {$\ab{127}$};
		\node (245) at (0,-1.25) {$\ab{245}$};
		\node (256) at (2,-1.25) {$\ab{256}$};
		\node (156) at (4, -1.25) {$\ab{156}$};

		\draw[->,shorten <=4pt, shorten >=3pt] (234.south east) -- (235.north west);
		\draw[norm] (235) -- (236);
		\draw[norm] (126) -- (236);
		\draw[norm] (126) -- (127);
		\draw[norm] (235) -- (245);
		\draw[norm] (236) -- (256);
		\draw[norm] (236) -- (123);
		\draw[norm] (126) -- (156);
		\draw[norm] (245) -- (256);
		\draw[norm] (156) -- (256);
		\draw[norm] (156) -- (167);
		\draw[norm] (245) -- (345);
		\draw[norm] (256) -- (456);
		\draw[diag] (123.south east) -- (126.north west);
		\draw[diag] (256.north west) -- (235.south east);
		\draw[diag] (167.north west) -- (126.south east);
		\draw[diag] (456.north west) -- (245.south east);
		\draw[diag] (567.north west) -- (156.south east);
		\draw[diag] (256.north east) -- (126.south west);

		\end{tikzpicture}
	\end{subfigure}
\caption{The ${\rm Conf}_7(\mathbb{P}^2)$ initial cluster (left) and the cluster resulting from a cyclic mutation of a ${\rm Conf}_6(\mathbb{P}^2)$ subalgebra, highlighted in green (right). ${\rm Conf}_7(\mathbb{P}^2) \sim {\rm Conf}_7(\mathbb{P}^3)$ but we have given this example to demonstrate this procedure is valid for ${\rm Conf}_n(\mathbb{P}^{k-1}) \text{ } \forall \text{ } k,n$.}
\label{cycexample}
\end{figure}

We can use this method to rotate initial-type sub-algebras within a cluster in order to search for clusters with specific Pl\"uckers. In fact we will use this method later to prove that all R-invariants are cluster adjacent. An example is given in Fig. \ref{cycexample}.
As we can see, the ${\rm Conf}_6(\mathbb{P}^2)$ sub-topology remains unchanged but the labels have all been rotated by one unit. The other nodes have rearranged themselves such that the frozen nodes connected to the sub-algebra have shifted round the cluster. Mutating on $\ab{156}$ followed by $\ab{126}$ will result in the same topology as the left cluster but with each label rotated by one unit. We can repeat this process any number of times to achieve the desired number of rotations.

\section{Cluster adjacent polylogarithms}

In \cite{Drummond:2017ssj} the notion of cluster adjacency for symbols
and polylogarithms was introduced. It extends the role
of cluster algebras in describing the analytic structure of the
scattering amplitudes, at least in the hexagon and heptagon cases for planar $\mathcal{N}=4$ super Yang-Mills theory. The
structure of the cluster algebra restricts the way given
$\mathcal{A}$-coordinates may appear next to each other in the
symbol of appropriately defined IR finite quantities. In particular,
\emph{for two $\mathcal{A}$-coordinates to appear next
to each other in the symbol they must appear together in some cluster}.
In other words they must either be a repeat of the same $\mathcal{A}$-coordinate or be an admissible pair.

The property of cluster adjacency is closely related to the Steinmann relations whose role in constraining the analytic structure of scattering amplitudes was stressed in \cite{Bartels:2008ce}. In \cite{Caron-Huot:2016owq} it was realised that the Steinmann relations were employed to greatly increase the power of the hexagon bootstrap programme and in \cite{Dixon:2016nkn} the same conditions were extended to the heptagon case. In fact the Steinmann conditions can be extended to hold on all adjacent pairs in the symbol \cite{DP,Yorgosslides}, not only in the first two entries. The cluster adjacency property outlined above implies the Steinmann conditions, including the extended ones. In the hexagon (or $A_3$) case this is simply the statement that the square faces of the associahedron in Fig. \ref{Stasheff} are not adjacent to each other. In the heptagon ($E_6$) case it follows from the fact that the face labelled by $a_{11}$ only intersects those labelled by $a_{14}$ and $a_{15}$ but not those labelled by the other $a_{1i}$. What is less obvious but nevertheless appears to hold for the hexagon and heptagon symbols is that the extended Steinmann relations \emph{together with the physical initial entry conditions} actually imply cluster adjacency.

Note that the property of cluster adjacency is described in terms of the inhomogeneous $\mathcal{A}$-coordinates. The polylogarithms describing the known dual conformal invariant amplitudes are functions on the space ${\rm Conf}_n(\mathbb{P}^3)$ and their symbols are normally described in terms of homogeneous multiplicative combinations of $\mathcal{A}$-coordinates. Such combinations can be expanded out into non-manifestly homogeneous combinations by the identities (\ref{symmult}) and (\ref{sympower}). The resulting expressions are the ones which obey the adjacency criterion. 

In the heptagon case we may take the homogenised $\mathcal{A}$-coordinates (\ref{heptletters}) as our symbol alphabet and the statement of adjacency becomes very direct. In the hexagon case this is not possible, essentially due to the existence of the purely frozen homogeneous combination $\Delta$ defined eq. (\ref{Delta}).

In general, beyond the hexagon and heptagon amplitudes we discuss here, we expect a number of new features whose interplay with cluster adjacency is not yet clear. Firstly there will exist algebraic symbol letters with square roots which are not immediately related to $\mathcal{A}$-coordinates which are all polynomials in the Pl\"ucker coordinates. These already appear in the N${}^2$MHV octagon at one loop in the four-mass box contributions. Moreover at high enough multiplicity and loop order there will appear non-polylogarithmic functions, e.g. in the ten-point N${}^3$MHV amplitude at two loops \cite{CaronHuot:2012ab}. Nevertheless we believe that some suitably extended notion of cluster adjacency will also hold beyond the hexagon and heptagon amplitudes.

\subsection{Neighbour sets}
\label{sec:neighbour-sets}

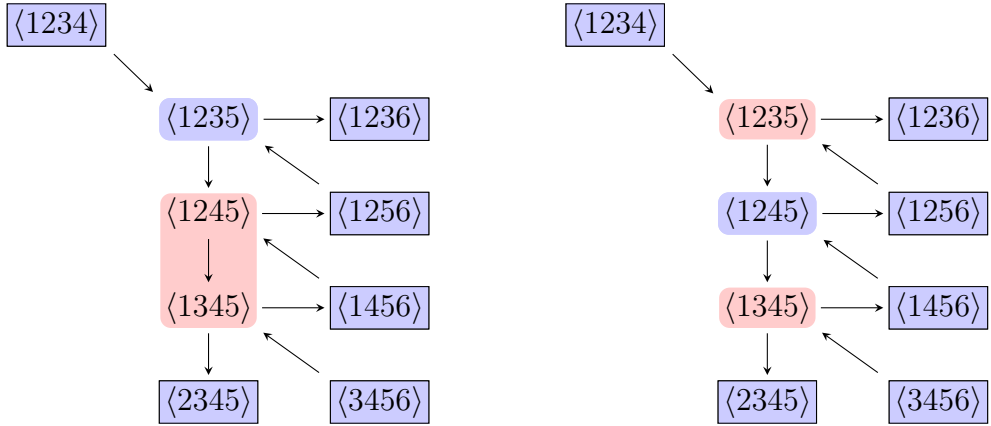
\begin{figure}
\begin{center}
\begin{tikzpicture}

\node[frozenblue] (1234) at (-2, 1.25) {$\ab{1234}$};
\node[blue] (1235) at (0, 0) {$\ab{1235}$};
\node[frozenblue] (1236) at (2.25, 0) {$\ab{1236}$};
\node (1245) at (0,-1.25) {$\ab{1245}$};
\node[frozenblue] (1256) at (2.25,-1.25) {$\ab{1256}$};
\node (1345) at (0,-2.5) {$\ab{1345}$};
\node[frozenblue] (1456) at (2.25,-2.5) {$\ab{1456}$};
\node[frozenblue] (2345) at (0,-3.75) {$\ab{2345}$};
\node[frozenblue] (3456) at (2.25,-3.75) {$\ab{3456}$};
\begin{scope}[blend mode=overlay,overlay]
          \node[rectangle, rounded corners, fit=(1245)(1345),fill=red!20, inner sep=0pt] {};
        \end{scope}
\node[frozenblue] (1234) at (-2, 1.25) {$\ab{1234}$};
\node[blue] (1235) at (0, 0) {$\ab{1235}$};
\node[frozenblue] (1236) at (2.25, 0) {$\ab{1236}$};
\node (1245) at (0,-1.25) {$\ab{1245}$};
\node[frozenblue] (1256) at (2.25,-1.25) {$\ab{1256}$};
\node (1345) at (0,-2.5) {$\ab{1345}$};
\node[frozenblue] (1456) at (2.25,-2.5) {$\ab{1456}$};
\node[frozenblue] (2345) at (0,-3.75) {$\ab{2345}$};
\node[frozenblue] (3456) at (2.25,-3.75) {$\ab{3456}$};

\draw[->,shorten <=4pt, shorten >=3pt] (1234.south east) -- (1235.north west);
\draw[norm] (1235) -- (1236);
\draw[norm] (1245) -- (1256);
\draw[norm] (1345) -- (1456);
\draw[norm] (1235) -- (1245);
\draw[norm] (1245) -- (1345);
\draw[norm] (1345) -- (2345);
\draw[diag] (1256.north west) -- (1235.south east);
\draw[diag] (1456.north west) -- (1245.south east);
\draw[diag] (3456.north west) -- (1345.south east);

\end{tikzpicture}
\qquad\qquad
\begin{tikzpicture}

\node[frozenblue] (1234) at (-2, 1.25) {$\ab{1234}$};
\node[] (1235) at (0, 0) {$\ab{1235}$};
\node[frozenblue] (1236) at (2.25, 0) {$\ab{1236}$};
\node[blue] (1245) at (0,-1.25) {$\ab{1245}$};
\node[frozenblue] (1256) at (2.25,-1.25) {$\ab{1256}$};
\node (1345) at (0,-2.5) {$\ab{1345}$};
\node[frozenblue] (1456) at (2.25,-2.5) {$\ab{1456}$};
\node[frozenblue] (2345) at (0,-3.75) {$\ab{2345}$};
\node[frozenblue] (3456) at (2.25,-3.75) {$\ab{3456}$};

\begin{scope}[blend mode=overlay,overlay]
          \node[rectangle, rounded corners, fit=(1235),fill=red!20, inner sep=0pt] {};
        \end{scope}
\begin{scope}[blend mode=overlay,overlay]
          \node[rectangle, rounded corners, fit=(1345),fill=red!20, inner sep=0pt] {};
        \end{scope}

\node[frozenblue] (1234) at (-2, 1.25) {$\ab{1234}$};
\node[] (1235) at (0, 0) {$\ab{1235}$};
\node[frozenblue] (1236) at (2.25, 0) {$\ab{1236}$};
\node[blue] (1245) at (0,-1.25) {$\ab{1245}$};
\node[frozenblue] (1256) at (2.25,-1.25) {$\ab{1256}$};
\node (1345) at (0,-2.5) {$\ab{1345}$};
\node[frozenblue] (1456) at (2.25,-2.5) {$\ab{1456}$};
\node[frozenblue] (2345) at (0,-3.75) {$\ab{2345}$};
\node[frozenblue] (3456) at (2.25,-3.75) {$\ab{3456}$};

\draw[->,shorten <=4pt, shorten >=3pt] (1234.south east) -- (1235.north west);
\draw[norm] (1235) -- (1236);
\draw[norm] (1245) -- (1256);
\draw[norm] (1345) -- (1456);
\draw[norm] (1235) -- (1245);
\draw[norm] (1245) -- (1345);
\draw[norm] (1345) -- (2345);
\draw[diag] (1256.north west) -- (1235.south east);
\draw[diag] (1456.north west) -- (1245.south east);
\draw[diag] (3456.north west) -- (1345.south east);

\end{tikzpicture}

\end{center}
\caption{The initial cluster of ${\rm Conf}_6(\mathbb{P}^3)$ has the topology of
  an $A_3$ Dynkin diagram. Freezing $\ab{1235}=(46)$ results in a
  $A_2$ subalgebra whereas freezing $\ab{1245}=(36)$ results in a
  $A_1 \times A_1$ subalgebra. These subalgebras generate the letters
  in $\ns{\ab{1235}}$ and $\ns{\ab{1245}}$, respectively.}
\label{howtonshex}
\end{figure}

We define the \emph{neighbour set} $\ns{a}$ of a given ${\cal A}$-coordinate $a$ as the set of $\mathcal{A}$-coordinates $b$ such that $\{a,b\}$ form an admissible pair together with $a$ itself. This set automatically includes all the frozen $\mathcal{A}$-coordinates. In terms of the polytope the unfrozen nodes in the neighbour set correspond to all faces that share a codimension-two boundary with the face labelled by $a$ (i.e. are adjacent to $a$) together with the face labelled by $a$ itself. One way of systematically
constructing neighbour sets is to go to a convenient cluster and freeze the ${\cal A}$-coordinate whose neighbour set is being considered. The neighbour set then consists of all unfrozen ${\cal A}$-coordinates generated in this codimension-one subalgebra, the frozen
coordinates and the coordinate $a$ itself. This is demonstrated in Figure
\ref{howtonshex}. Note that the notion of a neighbour set depends on
the cluster algebra in question, as well as the choice of $\mathcal{A}$-coordinate $a$. 

Through this procedure we find the following neighbour sets for the unfrozen 
hexagon $\mathcal{A}$-coordinates:
\begin{equation}
  \begin{aligned}[t]
    \ns{\ab{1235}} &= \{\ab{1235}, \ab{2456}, \ab{2356}, \ab{1356}, \ab{1345}, \ab{1245}, \,\text{\& frozen coordinates.}\}\\
    \ns{\ab{1245}} &= \{\ab{1245}, \ab{2456}, \ab{1345}, \ab{1246},  \ab{1235},\, \text{\& frozen coordinates.}\}\,.\\
\end{aligned}
\end{equation}
As stated above, apart from $a$ itself, the unfrozen elements of the neighbour set of $a$ are associated with the faces of the Stasheff
polytope which neighbour the face associated with $a$. The edges
where these faces intersect correspond to the remaining $A_1$ algebra
in a cluster containing the two letters associated with the two
faces, cf. Figure~\ref{Stasheff}.

\begin{figure}
{\footnotesize
\begin{center}
\begin{tikzpicture}[scale=0.8]
   \foreach \i in {1,...,6}
   {
     \coordinate (a\i) at (60+60*\i:1.6);
   }
   \draw[thick]
     (a1)
     \foreach \i in {2,...,6}
   {
     -- (a\i)
   } -- cycle;

   \draw[] (a6) -- (a2);
   \draw[] (a1) -- (a5);
    \foreach \i in {1,...,6}
   {
     \node at (60+60*\i:1.9) {$\i$};
   }
  \end{tikzpicture}
  \qquad
  \begin{tikzpicture}[scale=0.8]
   \foreach \i in {1,...,6}
   {
     \coordinate (a\i) at (60+60*\i:1.6);
   }
   \draw[thick]
     (a1)
     \foreach \i in {2,...,6}
   {
     -- (a\i)
   } -- cycle;

   \draw[] (a6) -- (a2);
   \draw[] (a1) -- (a4);
    \foreach \i in {1,...,6}
   {
     \node at (60+60*\i:1.9) {$\i$};
   }
  \end{tikzpicture}
  \qquad
  \begin{tikzpicture}[scale=0.8]
   \foreach \i in {1,...,6}
   {
     \coordinate (a\i) at (60+60*\i:1.6);
   }
   \draw[thick]
     (a1)
     \foreach \i in {2,...,6}
   {
     -- (a\i)
   } -- cycle;

   \draw[] (a6) -- (a3);
   \draw[] (a1) -- (a4);
    \foreach \i in {1,...,6}
   {
     \node at (60+60*\i:1.9) {$\i$};
   }
  \end{tikzpicture}
  
\end{center}
}
\caption{Forbidden pairs correspond to crossing chords of the hexagon.}
\label{hexagonchordscrossing}
\end{figure}
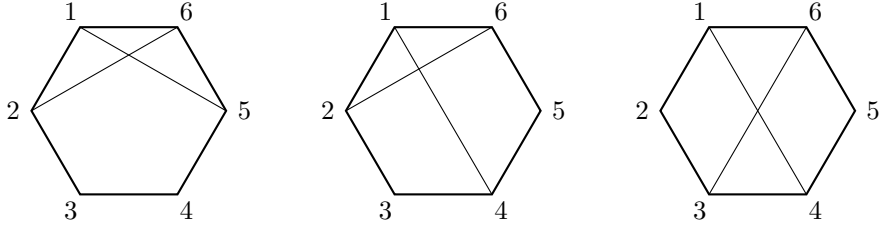

An equivalent way to state the neighbouring principle for the $A_3$ case (and more generally for the $A_n$ case) is that $\mathcal{A}$-coordinates corresponding to chords on the hexagon which cross are non-neighbouring, i.e. are forbidden to appear next to each other in the symbol. Examples are shown in Fig. \ref{hexagonchordscrossing}.

There are 12 coordinates in the
neighbour set of the ${\cal A}$-coordinate $\langle 2456 \rangle = (13)$ including itself
and the 6 frozen coordinates. When writing down homogeneous functions,
it convenient to work with a homogeneous alphabet and there are 6
homogeneous combinations that can be constructed using the allowed
neighbours of $\langle 1235 \rangle = (46)$. Such a \emph{homogeneous neighbour set} can be chosen as the five $\mathcal{X}$-coordinates associated to the edges of the pentagonal face labelled by $(46)$ together with $\Delta$ from eq. (\ref{Delta}) as follows:
\begin{equation}
  \begin{minipage}[c]{2cm}
    \small
    ${\rm  hns}[(46)] = $
  \end{minipage}\left\{
  \frac{(13)(46)}{(16)(34)},
  \frac{(24)(16)}{(12)(46)},
  \frac{(36)(12)}{(23)(16)},
  \frac{(14)(23)}{(12)(34)},
  \frac{(26)(34)}{(23)(46)},
  \frac{(12)(34)(56)}{(23)(45)(16)}\right\}\,.
\end{equation}
Similarly, there are five homogeneous combinations that are
made out of the 11 allowed neighbours of $\langle 1245 \rangle = (36)$. They may be taken as the two $\mathcal{X}$-coordinates associated to the square (opposite edges on a square have the same $\mathcal{X}$-coordinate) as well as any three of the four $\mathcal{X}$-coordinates which are associated to the edges which lead away from the square face. A choice is as follows:
\begin{equation}
  \begin{minipage}[c]{2cm}
    \small
     ${\rm  hns}[(36)] =$
  \end{minipage}
  \left\{
  \frac{(14)(23)}{(12)(34)},
  \frac{(14)(56)}{(16)(45)},
  \frac{(13)(24)}{(12)(34)},
  \frac{(15)(46)}{(16)(45)},
  \frac{(13)(45)}{(34)(15)}\right\}\,.\hfill
\end{equation}

For the cases of the cluster algebras associated to ${\rm Conf}_n(\mathbb{P}^{k-1})$ with $(k,n)$ coprime, one has the advantage of using frozen
coordinates to homogenise all remaining letters to construct a
homogeneous alphabet. Since frozen coordinates appear in every cluster
by definition, they cannot spoil cluster adjacency. Hence for $(k,n)$ coprime,
it is possible to talk about the cluster adjacency directly in terms of homogeneous
letters such as those in equation (\ref{heptletters}) for seven-particle
scattering and ignore the frozen coordinates altogether.

The heptagon alphabet (\ref{heptletters}) consists of 42 letters $a_{ij}$
grouped into six types. The neighbour sets of
these letters can be worked out in the same way as in the hexagon
case, for example starting with the $E_6$-shaped cluster in Fig.
\ref{howtonshep}, freezing the letter one is interested in and performing all possible mutations on the others. One finds
the following homogeneous neighbour sets for the letters $a_{11}$, $a_{21}$,
$a_{41}$ and $a_{61}$:
\begin{equation}
\label{heptns}
  \begin{aligned}[t]
    \hns{a_{11}} =\{ &a_{11}, a_{14}, a_{15}, a_{21}, a_{22}, a_{24}, a_{25}, a_{26}, a_{31}, a_{33}, a_{34}, a_{35}, a_{37}, a_{41},a_{43}, a_{46}, a_{51},\\
     \qquad        &a_{53}, a_{56}, a_{62}, a_{67}\}\\
    \hns{a_{21}} =\{&a_{11}, a_{13}, a_{14}, a_{15}, a_{17}, a_{21}, a_{23}, a_{24}, a_{25}, a_{26},a_{31}, a_{33}, a_{34}, a_{36},a_{37}, a_{41}, a_{43},\\
    \qquad        &a_{45}, a_{46}, a_{52}, a_{53}, a_{55}, a_{57}, a_{62}, a_{64}, a_{66}\}\\
    \hns{a_{41}} =\{&a_{11}, a_{13}, a_{16}, a_{21}, a_{23}, a_{24}, a_{26}, a_{31}, a_{33}, a_{35}, a_{36}, a_{41}, a_{43}, a_{46},a_{51}, a_{62}, a_{67}\}\\
    \hns{a_{61}} =\{&a_{12}, a_{17}, a_{23}, a_{25}, a_{27}, a_{32}, a_{34}, a_{36}, a_{42}, a_{47}, a_{52}, a_{57}, a_{61}\}\,.
\end{aligned}
\end{equation}
All other homogeneous neighbour sets for $\text{Conf}_7(\mathbb{P}^3)$ can be obtained as cyclic
rotations, reflections or parity conjugates of these.

\subsection{Definition of cluster adjacent polylogarithms}
\label{CApolysdef}

We recall a polylogarithm of weight $k$ obeys
\be
d f^{(k)} = \sum_{a \in \mathcal{A}} f_{[a]}^{(k-1)} d \log a\,,
\ee
where for us $\mathcal{A}$ is the set of all $\mathcal{A}$-coordinates of our cluster algebra. A cluster adjacent polylogarithm is one where the $f_{[a]}^{(k-1)}$ above additionally obey
\be
\label{adjacentdef}
d f_{[a]}^{(k-1)}  = \sum_{b \in \ns{a}} f^{(k-2)}_{[b],a} d \log b\,,
\ee
where the sum is only over $b$ in the neighbour set of $a$. We also insist that the $f_{[a]}^{(k-1)}$ are themselves cluster adjacent polylogarithms in the same sense, i.e.
\be
\label{adjacency2nd}
d f_{[b],a}^{(k-2)} = \sum_{c \in \ns{b}} f_{[c],ba}^{(k-3)} d \log c\,,
\ee
and so on all the way down to weight zero. It follows from the above that all adjacent pairs in the symbol of a cluster adjacent polylogarithm  $[\ldots \otimes a \otimes b \otimes \ldots]$ are such that $a \in \ns{b}$ or equivalently $b \in \ns{a}$. 

Note that the above discussion is phrased in terms of the inhomogeneous $\mathcal{A}$-coordinates, even though we are always interested in homogeneous functions $f^{(k)}$. This simply means that all the $d f^{(k)}$ above can be rewritten purely in terms of homogeneous combinations of $\mathcal{A}$-coordinates and the sum in (\ref{adjacentdef}) could be taken over the homogeneous neighbour set of $a$. In general, not all the cluster adjacency properties will be manifest in such a homogeneous representation, as happens in the hexagon case. In particular if we choose to write take sum in (\ref{adjacentdef}) over the homogeneous neighbour set of $a$, then each homogeneous $b$ should be expanded in terms of the inhomogeneous $\mathcal{A}$-coordinates in order to then reveal the cluster adjacent nature of the expression (\ref{adjacency2nd}).

In the heptagon case one can phrase the whole discussion in terms of the homogenised unfrozen coordinates and the sum in (\ref{adjacentdef}) can be taken over the homogeneous neighbour sets given in (\ref{heptns}). Since the frozen factors play no role in cluster adjacency this property can be made manifest at the same time as homogeneity.


\subsection{Neighbour-set functions}
\label{sec:neisets}

When constructing integrable cluster-adjacent functions, it is natural
to introduce the concept of \emph{neighbour-set functions}. They are
defined as polylogarithms which satisfy
\be
d f^{(k)} = \sum_{b \in \ns{a}} f_{[b]}^{(k-1)} d \log b
\ee
for a given choice of $\mathcal{A}$-coordinate $a$. The final entries of the symbols of such functions are selected only from the
neighbour set of a given $\mathcal{A}$-coordinate.  As can be seen from (\ref{adjacentdef}) above, any cluster adjacent weight-$k$
function only requires neighbour set functions in its $(k-1,1)$ coproduct. 
Hence, when constructing cluster adjacent functions of weight $k$ one can use a reduced ansatz for the $(k-1,1)$ coproduct
\begin{equation}
\label{CAans}
  f^{(k-1,1)}
  =
  \sum_{a \in {\cal A}}
  \sum_{i=1}^{d_{[a]}^{(k-1)}}
  \, c_{ai}\,\bigl[f^{(k-1)}_{[a], i} \otimes a\bigr] \, ,
\end{equation}
where $f^{(k-1)}_{[a], i}$ are elements of a basis for 
the space of homogeneous weight-($k-1$) functions whose final entries are in the neighbour-set of $a$ and
$d_{[a]}^{(k-1)}$ is the dimension of this
space. If the $\mathcal{A}$-coordinates $a$ in (\ref{CAans}) above cannot be chosen as unfrozen ones homogenised purely in terms of frozen ones, then the coefficients $c_{ai}$ are assumed to be constrained to ensure homogeneity of the resulting expression. Eliminating any cluster-adjacency violation in the ansatz reduces the size of the resulting linear algebra problem. The notion of a neighbour set function is compatible with any possible choices of constraints in the initial entries, for example when constructing hexagon symbols to describe six-point amplitudes in planar $\mathcal{N}=4$ super Yang-Mills theory.

We now illustrate neighbour set functions for $\text{Conf}_6(\mathbb{P}^3)$. In this case, there are two types of unfrozen
${\cal A}$-coordinates with neighbour set functions:
$(13)$ \& cyclic and $(14)$ \& cyclic. 
The neighbour-set functions for the hexagon are then defined as
homogeneous, cluster-adjacent functions that obey the initial entry
condition, i.e. begin with the three-cross ratios of the hexagon ($u$, $v$ or $w$ from eq. \ref{hexusual}), and
end with aforementioned homogeneous combinations that are
cluster-adjacent to $(13)$ or $(14)$. The dimensions of such spaces
for a few weights are compared to the full space of
cluster-adjacent hexagon symbols is given in Table \ref{neidimshex}.
{
  \renewcommand{\arraystretch}{1.2}
\begin{table}
  \centering
  {\small
    \begin{tabular}{@{}lllllllllllllll@{}}
      \toprule
    Weight&2&3&4&5&6&7&8&9&10&11&12&13&14\\
    \midrule
    ${\rm hns}[(13)]$&3&6&11&21&39&73&132&237&415&717&1216&2036&3358\\

    ${\rm hns}[(14)]$&3&5&10&19&36&66&120&213&374&644&1096&1835&3041\\

      Full $A_3$&6&13&26&51&98&184&340&613&1085&1887&3224&5431&9014\\
      \bottomrule
  \end{tabular}
  }
  \caption{Dimensions of the spaces of integrable words in the
    hexagon alphabet with hexagon initial entries $\{u,v,w\}$ only and final entries drawn from the neighbour sets ${\rm hns}[(13)]$, ${\rm hns}[(14)]$ or from the full nine-letter $A_3$ alphabet.}
  \label{neidimshex}
\end{table}
}

We have also computed the neighbour-set functions of the heptagon
letters up to weight seven. The dimensions of the neighbour-set
function spaces depend on the letter and they are summarised in Table
\ref{neisets}. For weights 2-7 we find the span of all $a_{2i}$ and
$a_{3i}$ neighbour-set function spaces covers the entire
cluster-adjacent function space of the corresponding weight.

{
  \renewcommand{\arraystretch}{1.2}
\begin{table}
  \centering
  \begin{tabular}{lllllll}
    \toprule
    Weight&2&3&4&5&6&7\\
    \midrule
    ${\rm hns}[a_{1i}]$&10&29&83&229&612&1577\\
    ${\rm hns}[a_{2i}]$ &15&43&117&311&804&2025\\
    ${\rm hns}[a_{4i}]$ &6&14&34&87&224&570\\
    ${\rm hns}[a_{6i}]$&4&11&29&76&193&476\\
    Full $E_6$&28&97&308&911&2555&6826\\
    \bottomrule
  \end{tabular}
  \caption{Dimensions of the neighbour-set function spaces of the
    heptagon alphabet with initial entries $a_{1i}$ and the dimensions
    of the full cluster-adjacent heptagon functions}
  \label{neisets}
\end{table}
}

\subsection{Integrability}

It is interesting to investigate in low weights the spaces of cluster adjacent functions without any initial entry condition. At weight two we may split the space of integrable words into those which are symmetric in the two entries of the symbol and those which are antisymmetric. The symmetric ones are trivially integrable: any word of the form $[a\otimes b]+[b\otimes a]$ is the symbol of $\log a  \, \log b$. Adjacency however constrains the possible choices of $a$ and $b$ - they must come from a common cluster, i.e. they must not correspond to distant faces on the polytope. The antisymmetric words on the other hand are not trivially integrable. However, they do automatically obey the adjacency condition, in the sense that all antisymmetric integrable weight two words are cluster adjacent, even if that condition was not imposed in constructing them. Actually they obey a stronger condition, namely that the $\mathcal{A}$-coordinates appearing in the two slots can be found in some cluster together where they are connected by an arrow.

When we investigate weight three words we find that the associated triplets of $\mathcal{A}$-coordinates are of two possible types. Each term $[a\otimes b \otimes c]$ is either of the form where $a$, $b$ and $c$ can all be found together in the same cluster or we have $c=a'$ where $a'$ is the result of mutating on $a$ in some cluster. In fact there is an even stronger condition in this latter case: if we find triplets of the form $[a\otimes b \otimes a']$ then they can always be combined so that the intermediate letter becomes the $\mathcal{X}$-coordinate associated with the mutation pair $(a,a')$. Recall that $\mathcal{X}$-coordinates are associated to one-dimensional edges of the polytope which are also associated to mutations. Moreover if there is more than one edge between the two faces labelled by $a$ and $a'$ those edges are associated to the same $\mathcal{X}$-coordinate. In other words $\mathcal{X}$-coordinates are associated to mutation pairs of $\mathcal{A}$-coordinates, hence we may denote them by $x(a,a')$.  So we have triplets of the form $[a\otimes x(a,a') \otimes a']$ or triplets $[a\otimes b \otimes c]$ where all three letters can be found together in some cluster.



\subsection{Cluster adjacency in hexagon and heptagon loop amplitudes}

We have confirmed that all the currently available results for hexagon and heptagon functions appearing in the loop expansion of MHV and NMHV amplitudes are cluster adjacent polylogarithms. That is, the functions $\mathcal{E}^{{\rm MHV}, (L)}$ and $E_{ijklm}^{(L)}$ are weight $2L$ polylogarithms whose symbols obey the cluster adjacency conditions and whose initial entries are constrained to be compatible with the physical branch cut conditions. In the hexagon case this means the initial entries are drawn from the set $\{u,v,w\}$ from (\ref{hexusual}) and in the heptagon case that they are of the form $a_{1i}$ from the heptagon alphabet given in (\ref{heptletters}).

In the MHV case the $(2L-1,1)$ coproduct of the polylogarithmic functions which appear is constrained in the the final entries are drawn only from $\mathcal{A}$-coordinates of the form $\langle i \,j-1\,j\,j+1\rangle$. This behaviour follows from an analysis of the $\bar{Q}$-equation of \cite{CaronHuot:2011kk,Bullimore:2011kg}. This has the consequence that the $(n-1,1)$ coproduct of the MHV amplitudes is heavily constrained,
\be
\mathcal{E}^{(2L-1,1)} = \sum_{i,j} [\mathcal{E}_{ij} \otimes \langle i\, j-1\, j\, j+1 \rangle]\,,
\ee
where $\mathcal{E}_{ij}$ is a neighbour set function of the $\mathcal{A}$-coordinate $\langle i\, j-1\, j\, j+1\rangle$, i.e. it is a weight $(2L-1)$ polylogarithm whose symbol's final entries are drawn from the neighbour set of $\langle i \, j-1 \,j \, j+1\rangle$.

In the NMHV case there is an interplay between the R-invariants and the final entries of the symbols of the polylogarithms which appear. We will address this point in greater detail in Sect. \ref{NMHVloops}.

\section{Cluster adjacency of tree-level BCFW recursion}

It is clear from the above discussion that cluster adjacency of polylogarithms or symbols has a non-abelian character. Two $\mathcal{A}$-coordinates $a$ and $a'$ which cannot appear next to each other are allowed to appear in the same word if they are appropriately separated by intermediate $\mathcal{A}$-coordinates. For example, if they are separated by one step only the $\mathcal{X}$-coordinate associated to the relevant mutation appears between them, as discussed above. This non-abelian behaviour is due to the fact that the symbol comes with an ordering which ultimately reflects the fact that monodromies of the associated iterated integrals do not commute with each other.

However we now discuss a setting where an abelian form of cluster adjacency holds. It is in the context of the poles of rational functions contributing to tree-level amplitudes. Here we will restrict our discussion to the cluster adjacency properties of BCFW tree-amplitudes for NMHV and N$^2$MHV helicity configurations. The superconformal and dual superconformal symmetries are known to combine into a Yangian structure \cite{Drummond:2009fd}. BCFW expansions for tree amplitudes are solved in terms of Yangian invariants. These quantities can be found as residues in the Grassmannian integral of \cite{ArkaniHamed:2009dn,Mason:2009qx}.

The pattern we find can be stated as follows: \emph{every Yangian invariant in the BCFW expansion of tree amplitudes has poles given by $\mathcal{A}$-coordinates which can be found together in a common cluster.}

Expressions for BCFW expansions may be generated directly in momentum twistor variables using the {\tt bcfw.m} package provided in \cite{Bourjaily:2010wh}. We give explicit examples showing all BCFW terms obey the cluster adjacency property up to eight points. 
As well as providing another example in which the cluster algebra structure plays a role in controlling the singularities of amplitudes, the discussion of R-invariants will be relevant later when we consider NMHV loop amplitudes.

\subsection{NMHV}
\label{NMHVBCFW}
The BCFW expansion of the $n$-point NMHV tree amplitude of $\mathcal{N}=4$ SYM (divided by the MHV tree) is given by
\begin{equation}
A_{n,1}^{\text{tree}} = \sum_{1<i<j<n} [1 i i+1 j j+1]
\label{BCFWtree}
\end{equation}
where we remind that the R-invariant $[ijklm]$ is given by
\begin{equation}
[i j k l m] = \frac{\la \la i j k l m \ra \ra }{\ab{ijkl}\ab{jklm}\ab{klmi}\ab{lmij}\ab{mijk}}\,.
\end{equation}
Here the denominator is a product of Pl\"ucker coordinates which are examples of $\mathcal{A}$-coordinates of the cluster algebra associated to ${\rm Conf}_n(\mathbb{P}^3)$. The numerator is a polynomial in momentum twistors and the Grassmann parameters $\chi_i$ encoding the supermultiplet structure, $\la \la i j k l m \ra \ra = (\chi_i \la j k l m \ra + \text{cyclic})^4$.

The R-invariants are not all independent; there are ${n-1 \choose 4}$ linearly independent ones due to identities of the form 
\begin{equation}\label{eq:Ridentity}
[abcde] - [bcdef] + [cdefa] - [defab] + [efabc] - [fabcd] = 0\,.
\end{equation}

We will now show that the $\mathcal{A}$-coordinates which describe the poles of R-invariants obey an abelian form of cluster adjacency: it is always possible to find a cluster where all the poles of an R-invariant appear together. Since the poles multiply in a commutative fashion there is no ordering to them and it is natural therefore that adjacency simply requires them all to appear together in some cluster.

\subsubsection*{Five points}
Five-points is a trivial example as there is just one R-invariant and hence the amplitude is simply
\begin{equation}\label{eq:5ptAmp}
\cA_{5,1} = [12345],
\end{equation}
also ${\rm Conf}_5(\mathbb{P}^3)$ contains just one cluster containing all frozen nodes.
\begin{figure}
	\begin{center}
		\begin{tikzpicture}[scale=0.75]
		{\footnotesize
		\node[frozenblue] (1234) at (0,0) {$\ab{1234}$};
		\node[frozenblue] (1235) at (2.5,0) {$\ab{1235}$};
		\node[frozenblue] (1245) at (5,0) {$\ab{1245}$};
		\node[frozenblue] (1345) at (7.5,0) {$\ab{1345}$};
		\node[frozenblue] (2345) at (10,0) {$\ab{2345}$};

		\draw[norm] (1234) edge (1235) (1235) edge (1245) (1245) edge (1345) (1345) edge (2345);
		}
		\end{tikzpicture}
	\end{center}
	\caption{The single ${\rm Conf}_5(\mathbb{P}^3) \sim A_0$ cluster. All nodes are frozen.}
	\label{5point}
\end{figure}
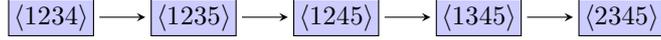
\noindent The nodes in Fig. \ref{5point} are coloured blue to indicate that they are present as poles in the R-invariant $[12345]$.
The basic R-invariant  \eqref{eq:5ptAmp} and its associated cluster will be the starting point for analysing all other NMHV R-invariants.

\subsubsection*{Six points}
At six-points there is only one type of R-invariant, $[12345]$ and its cyclic rotations, which make up the six-point, NMHV, tree given as
\begin{equation}\label{eq:6ptAmp1}
\cA_{6,1}= [12345] + [12356] + [13456] = [12346] + [12456] + [23456].
\end{equation}
Since every R-invariant at six points is a rotation of \eqref{eq:5ptAmp} in ${\rm Conf}_6(\mathbb{P}^3)$ we can identify each one with a single cluster in the polytope, one of which is
\begin{figure}
\begin{center}
\begin{tikzpicture}[scale=0.75]
{\footnotesize
\node[frozenblue] (1234) at (-2, 1.25) {$\ab{1234}$};
\node[blue] (1235) at (0, 0) {$\ab{1235}$};
\node[frozen] (1236) at (2.25, 0) {$\ab{1236}$};
\node[blue] (1245) at (0,-1.25) {$\ab{1245}$};
\node[frozen] (1256) at (2.25,-1.25) {$\ab{1256}$};
\node[blue] (1345) at (0,-2.5) {$\ab{1345}$};
\node[frozen] (1456) at (2.25,-2.5) {$\ab{1456}$};
\node[frozenblue] (2345) at (0,-3.75) {$\ab{2345}$};
\node[frozen] (3456) at (2.25,-3.75) {$\ab{3456}$};

\draw[->,shorten <=4pt, shorten >=3pt] (1234.south east) -- (1235.north west);
\draw[norm] (1235) -- (1236);
\draw[norm] (1245) -- (1256);
\draw[norm] (1345) -- (1456);
\draw[norm] (1235) -- (1245);
\draw[norm] (1245) -- (1345);
\draw[norm] (1345) -- (2345);
\draw[diag] (1256.north west) -- (1235.south east);
\draw[diag] (1456.north west) -- (1245.south east);
\draw[diag] (3456.north west) -- (1345.south east);
}
\end{tikzpicture}
\end{center}
\caption{The cluster containing the poles of $[12345]$ in ${\rm Conf}_6(\mathbb{P}^3)$.}
\end{figure}
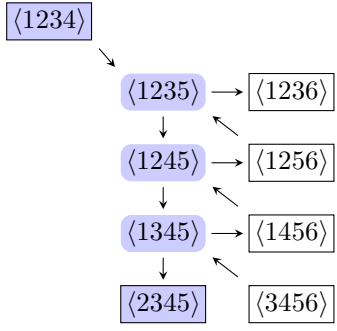
One would obtain the other five R-invariants and their associated clusters through cyclic rotations of this cluster. This can be achieved by applying the sequence of mutations illustrated in Fig. \ref{Stasheff} which generates a cyclic rotation. The clusters associated to the R-invariants are the six associated to the top and bottom corners of the square faces in Fig. \ref{Stasheff}.

Note that while the full tree amplitude (\ref{eq:6ptAmp1}) only contains physical poles of the form $\langle 1245 \rangle \sim 1/x_{25}^2 = 1/(p_2 + p_3 + p_4)^2$ and rotations, the adjacency property holds term by term in the BCFW expansion. Hence it also constrains the way in which the spurious poles at $\langle 1235 \rangle = 0$ and its cyclic rotations may appear. A consequence of the adjacency property is the well-known fact that the tree amplitude cannot have simultaneous poles in two different factorisation channels. For example, there is no term with both $\langle 1245 \rangle$ and $\langle 2356 \rangle$ in the denominator. This statement is the analogue of the fact that the Steinmann relations follow from cluster adjacency in the loop amplitudes.

\subsubsection*{Seven points and beyond}
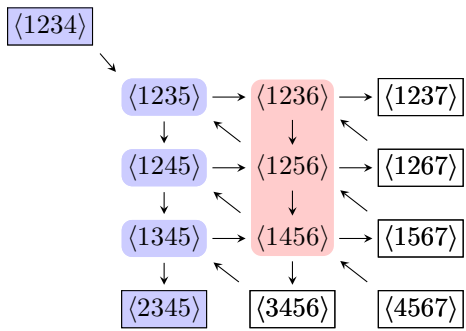
\begin{figure}
\begin{center}
\begin{tikzpicture}[scale=0.75]
{\footnotesize
\node[frozenblue] (1234) at (-2, 1.25) {$\ab{1234}$};
\node[blue] (1235) at (0, 0) {$\ab{1235}$};
\node (1236) at (2.25, 0) {$\ab{1236}$};
\node[blue] (1245) at (0,-1.25) {$\ab{1245}$};
\node (1256) at (2.25,-1.25) {$\ab{1256}$};
\node[blue] (1345) at (0,-2.5) {$\ab{1345}$};
\node (1456) at (2.25,-2.5) {$\ab{1456}$};
\node[frozenblue] (2345) at (0,-3.75) {$\ab{2345}$};
\node[frozen] (3456) at (2.25,-3.75) {$\ab{3456}$};
\node[frozen] (1237) at (4.5,0) {$\ab{1237}$};
\node[frozen] (1267) at (4.5,-1.25) {$\ab{1267}$};
\node[frozen] (1567) at (4.5,-2.5) {$\ab{1567}$};
\node[frozen] (4567) at (4.5,-3.75) {$\ab{4567}$};

\begin{scope}[blend mode=overlay,overlay]
          \node[rectangle, rounded corners, fit=(1236)(1256)(1456),fill=red!20, inner sep=0pt] {};
        \end{scope}
\node[frozenblue] (1234) at (-2, 1.25) {$\ab{1234}$};
\node[blue] (1235) at (0, 0) {$\ab{1235}$};
\node (1236) at (2.25, 0) {$\ab{1236}$};
\node[blue] (1245) at (0,-1.25) {$\ab{1245}$};
\node (1256) at (2.25,-1.25) {$\ab{1256}$};
\node[blue] (1345) at (0,-2.5) {$\ab{1345}$};
\node (1456) at (2.25,-2.5) {$\ab{1456}$};
\node[frozenblue] (2345) at (0,-3.75) {$\ab{2345}$};
\node[frozen] (3456) at (2.25,-3.75) {$\ab{3456}$};
\node[frozen] (1237) at (4.5,0) {$\ab{1237}$};
\node[frozen] (1267) at (4.5,-1.25) {$\ab{1267}$};
\node[frozen] (1567) at (4.5,-2.5) {$\ab{1567}$};
\node[frozen] (4567) at (4.5,-3.75) {$\ab{4567}$};

\draw[norm] (1235) -- (1236);
\draw[norm] (1236) -- (1237);
\draw[norm] (1245) -- (1256);
\draw[norm] (1256) -- (1267);
\draw[norm] (1345) -- (1456);
\draw[norm] (1456) -- (1567);

\draw[norm] (1235) -- (1245);
\draw[norm] (1236) -- (1256);
\draw[norm] (1245) -- (1345);
\draw[norm] (1256) -- (1456);
\draw[norm] (1345) -- (2345);
\draw[norm] (1456) -- (3456);

\draw[->,shorten <=4pt, shorten >=3pt] (1234.south east) -- (1235.north west);
\draw[diag] (1256.north west) -- (1235.south east);
\draw[diag] (1267.north west) -- (1236.south east);
\draw[diag] (1456.north west) -- (1245.south east);
\draw[diag] (1567.north west) -- (1256.south east);
\draw[diag] (3456.north west) -- (1345.south east);
\draw[diag] (4567.north west) -- (1456.south east);
}
\end{tikzpicture}
\end{center}
\caption{A cluster containing the poles of $[12345]$ in ${\rm Conf}_7(\mathbb{P}^3)$. The unfrozen nodes highlighted in red generate an $A_3$ subalgebra by repeated mutation.}
\label{fig:R67cluster}
\end{figure}

At seven points there are three types of R-invariant, 
\begin{equation}
\label{Rinvs7pts}
[12345] \text{ \& cyclic,} \quad [12346]  \text{ \& cyclic,} \quad [12356] \text{ \& cyclic.}
\end{equation}
The tree amplitude takes the form
\begin{equation}
\label{A71}
\cA_{7,1} = [12345] + [12356] + [12367] + [13456] + [13467] + [14567]\,. 
\end{equation}
As with \eqref{eq:6ptAmp1}, the BCFW representation of this amplitude is not unique due to the identity among the R-invariants \eqref{eq:Ridentity}.
At seven points multiple clusters contain the poles of a given R-invariant and hence R-invariants are associated to sub-algebras in the full ${\rm Conf}_7(\mathbb{P}^3)$ cluster algebra. For example, the initial cluster in Fig. \ref{heptinitial} contains all the poles of $[12345]$. It also contains three more unfrozen nodes in the second column. Performing all possible mutations in the second column generates an entire $A_3$ subalgebra, all of whose clusters contain the poles of $[12345]$. This is illustrated in Fig. \ref{fig:R67cluster}. The other two types of R-invariants in (\ref{Rinvs7pts}) appear respectively in $A_2$ and $A_1$ subalgebras.


One form of the eight-point NMHV tree amplitude is given by
\begin{equation}
\begin{aligned}
\cA_{8,1} = &[12345] + [12356] + [12367] + [12378] + [13456] \\ 
					+ &[13467] + [13478] + [14567] + [14578] + [15678].
\end{aligned}
\end{equation}
As we can see, more types of R-invariants begin to appear at eight points so we have presented their subalgebras in Table \ref{table:Rinvs} below along with their subalgebras at lower points.
\begin{table}
  \renewcommand{\arraystretch}{1.3}
\centering
	\begin{tabular}{@{} c  c  c  c  c @{}}
          \toprule
	$n$ & 5 & 6 & 7 & 8 \\ \hline
	$[12345]$ & $A_0$ & $A_0$ & $A_3$ & $E_6$ \\ 
	$[12356]$ & $\--$ & $A_0$ & $A_1$ & $A_4$ \\ 
	$[12346]$ & $\--$ & $A_0$ & $A_2$ & $A_5$ \\ 
	$[13467]$ & $\--$ & $\--$ & $A_1$ & $A_2 \times A_1 \times A_1$ \\ 
          $[12357]$ & $\--$ & $\--$ & $A_2$ & $A_4$ \\ 
          \bottomrule
	\end{tabular}
	\caption{Various R-invariants and their subalgebras in ${\rm Conf}_n(\mathbb{P}^3)$ at different multiplicities $n$.}
	\label{table:Rinvs}
\end{table}
\noindent The notation $A_0$ in Table \ref{table:Rinvs} indicates that a single cluster is associated to that R-invariant.
The last R-invariant $[12357]$ does not appear in the BCFW expansion of any tree in formula (\ref{BCFWtree}) amplitude but we can nevertheless associate a sub-algebra to this Yangian invariant object.

As described in Sect. \ref{gencyc} above, one can rotate the nodes in an initial-type cluster by mutating up all consecutive columns. Using this we can show that one can obtain any R-invariant by starting with the initial cluster, which we associate to $[12345]$, and mutating in different ${\rm Conf}_n(\mathbb{P}^3)$ sub-algebras. We illustrate this procedure with the following eight-point example: we will find a cluster in ${\rm Conf}_8(\mathbb{P}^3)$ which contains the poles of $[13467]$. 

\begin{figure}
\centering
\begin{tikzpicture}[scale=0.75]
{\footnotesize
\node[frozen] (1234) at (-2, 1.25) {$\ab{1234}$};
\node (1235) at (0, 0) {$\ab{1235}$};
\node (1236) at (2.25, 0) {$\ab{1236}$};
\node (1245) at (0,-1.25) {$\ab{1245}$};
\node (1256) at (2.25,-1.25) {$\ab{1256}$};
\node (1345) at (0,-2.5) {$\ab{1345}$};
\node (1456) at (2.25,-2.5) {$\ab{1456}$};
\node[frozen] (2345) at (0,-3.75) {$\ab{2345}$};
\node[frozen] (3456) at (2.25,-3.75) {$\ab{3456}$};
\node (1237) at (4.5,0) {$\ab{1237}$};
\node (1267) at (4.5,-1.25) {$\ab{1267}$};
\node (1567) at (4.5,-2.5) {$\ab{1567}$};
\node[frozen] (1238) at (6.75,0) {$\ab{1238}$};
\node[frozen] (1278) at (6.75,-1.25) {$\ab{1278}$};
\node[frozen] (1678) at (6.75,-2.5) {$\ab{1678}$};
\node[frozen] (5678) at (6.75,-3.75) {$\ab{5678}$};
\node[frozen] (4567) at (4.5,-3.75) {$\ab{4567}$};
        
\draw[norm] (1235) -- (1236);
\draw[norm] (1236) -- (1237);
\draw[norm] (1245) -- (1256);
\draw[norm] (1256) -- (1267);
\draw[norm] (1345) -- (1456);
\draw[norm] (1456) -- (1567);
\draw[norm] (1237) -- (1238);
\draw[norm] (1267) -- (1278);
\draw[norm] (1567) -- (1678);

\draw[norm] (1235) -- (1245);
\draw[norm] (1236) -- (1256);
\draw[norm] (1245) -- (1345);
\draw[norm] (1256) -- (1456);
\draw[norm] (1345) -- (2345);
\draw[norm] (1456) -- (3456);
\draw[norm] (1237) -- (1267);
\draw[norm] (1267) -- (1567);
\draw[norm] (1567) -- (4567);

\draw[->,shorten <=4pt, shorten >=3pt] (1234.south east) -- (1235.north west);
\draw[diag] (1256.north west) -- (1235.south east);
\draw[diag] (1267.north west) -- (1236.south east);
\draw[diag] (1456.north west) -- (1245.south east);
\draw[diag] (1567.north west) -- (1256.south east);
\draw[diag] (3456.north west) -- (1345.south east);
\draw[diag] (4567.north west) -- (1456.south east);
\draw[diag] (1278.north west) -- (1237.south east);
\draw[diag] (1678.north west) -- (1267.south east);
\draw[diag] (5678.north west) -- (1567.south east);
}
\end{tikzpicture}

\caption{A cluster containing the poles of $[12345]$ in ${\rm Conf}_8(\mathbb{P}^3)$.}
\label{fig:8ptinitcluster}
\end{figure}
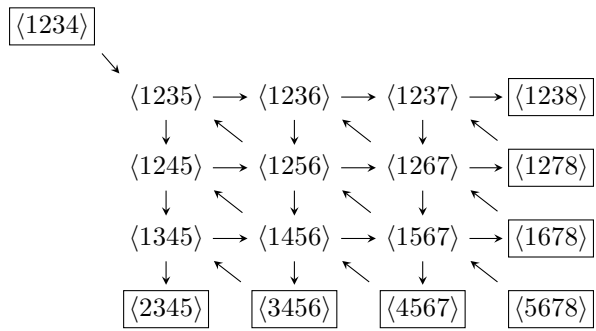

\begin{figure}
\centering
\begin{tikzpicture}[scale=0.75]
{\footnotesize
\node (1236) at (0,0) {$\ab{1236}$};
\node (2367) at (0,1.25) {$\ab{2367}$};
\node (1367) at (0,2.5) {$\ab{1367}$};
\node (2346) at (2.25,0) {$\ab{2346}$};
\node (3467) at (2.25,1.25) {$\ab{3467}$};
\node (1467) at (2.25,2.5) {$\ab{1467}$};
\node[frozen] (1678) at (0,3.75) {$\ab{1678}$};
\node (1567) at (2.25,3.75) {$\ab{1567}$};
\node (1267) at (-2.25,2.5) {$\ab{1267}$};
\node (1237) at (-2.25,1.25) {$\ab{1237}$};
\node[frozen] (1234) at (2.25,-1.25) {$\ab{1234}$};
\node[frozen] (2345) at (4.5,-1.25) {$\ab{2345}$};
\node[frozen] (1278) at (-4.5,2.5) {$\ab{1278}$};
\node[frozen] (1238) at (-4.5,1.25) {$\ab{1238}$};
\node[frozen] (3456) at (4.5,0) {$\ab{3456}$};
\node[frozen] (4567) at (4.5,2.5) {$\ab{4567}$};
\node[frozen] (5678) at (4.5,3.75) {$\ab{5678}$};

\begin{scope}[blend mode=overlay,overlay]
          \node[rectangle,fill=green!20, rounded corners, fit=(1367)(1467)(2367)(3467)(1236)(2346), inner sep=0.3pt] {};
        \end{scope}
\node (1236) at (0,0) {$\ab{1236}$};
\node (2367) at (0,1.25) {$\ab{2367}$};
\node (1367) at (0,2.5) {$\ab{1367}$};
\node (2346) at (2.25,0) {$\ab{2346}$};
\node (3467) at (2.25,1.25) {$\ab{3467}$};
\node (1467) at (2.25,2.5) {$\ab{1467}$};
\node[frozen] (1678) at (0,3.75) {$\ab{1678}$};
\node (1567) at (2.25,3.75) {$\ab{1567}$};
\node (1267) at (-2.25,2.5) {$\ab{1267}$};
\node (1237) at (-2.25,1.25) {$\ab{1237}$};
\node[frozen] (1234) at (2.25,-1.25) {$\ab{1234}$};
\node[frozen] (2345) at (4.5,-1.25) {$\ab{2345}$};
\node[frozen] (1278) at (-4.5,2.5) {$\ab{1278}$};
\node[frozen] (1238) at (-4.5,1.25) {$\ab{1238}$};
\node[frozen] (3456) at (4.5,0) {$\ab{3456}$};
\node[frozen] (4567) at (4.5,2.5) {$\ab{4567}$};
\node[frozen] (5678) at (4.5,3.75) {$\ab{5678}$};

\draw[norm] (1236) -- (2346);
\draw[norm] (2367) -- (3467);
\draw[norm] (1367) -- (1467);
\draw[norm] (4567) -- (1467);
\draw[norm] (2346) -- (3456);
\draw[norm] (1567) -- (1678);
\draw[norm] (5678) -- (1567);
\draw[norm] (1267) -- (1367);
\draw[norm] (1267) -- (1278);
\draw[norm] (1237) -- (1238);

\draw[norm] (1367) -- (2367);
\draw[norm] (2367) -- (1236);
\draw[norm] (1467) -- (3467);
\draw[norm] (3467) -- (2346);
\draw[norm] (2346) -- (1234);
\draw[norm] (1237) -- (1267);
\draw[norm] (1467) -- (1567);

\draw[diag] (2346.north west) -- (2367.south east);
\draw[diag] (3467.north west) -- (1367.south east);
\draw[diag] (1234.north west) -- (1236.south east);
\draw[diag] (2345.north west) -- (2346.south east);
\draw[diag] (3456.north west) -- (3467.south east);
\draw[diag] (1236.north west) -- (1237.south east);
\draw[diag] (3467.north east) -- (4567.south west);
\draw[diag] (1278.south east) -- (1237.north west);
\draw[diag] (1567.south east) -- (4567.north west);
\draw[diag] (1678.south west) -- (1267.north east);
}
\end{tikzpicture}
\caption{The cluster obtained after five cyclic mutations of Fig. \ref{fig:8ptinitcluster} in the first two columns.}
\label{intermediate}
\end{figure}
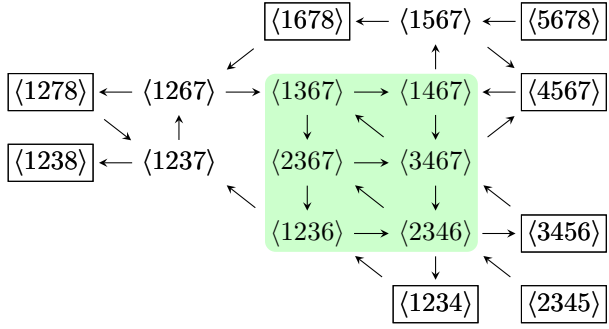

\begin{figure}
\centering
\begin{tikzpicture}[scale=0.75]
{\footnotesize
\node[blue] (1346) at (0,0) {$\ab{1346}$};
\node[blue] (1347) at (0,1.25) {$\ab{1347}$};
\node[blue] (1367) at (0,2.5) {$\ab{1367}$};
\node[red] (2346) at (2.25,-1.25) {$\ab{2346}$};
\node[blue] (3467) at (2.25,1.25) {$\ab{3467}$};
\node[blue] (1467) at (2.25,2.5) {$\ab{1467}$};
\node[frozen] (1678) at (0,3.75) {$\ab{1678}$};
\node[red] (1567) at (2.25,3.75) {$\ab{1567}$};
\node (1267) at (-2.25,2.5) {$\ab{1267}$};
\node (1237) at (-2.25,1.25) {$\ab{1237}$};
\node[frozen] (1234) at (-2.25,-1.25) {$\ab{1234}$};
\node[frozen] (2345) at (4.5,-1.25) {$\ab{2345}$};
\node[frozen] (1278) at (-4.5,2.5) {$\ab{1278}$};
\node[frozen] (1238) at (-4.5,1.25) {$\ab{1238}$};
\node[frozen] (3456) at (4.5,0) {$\ab{3456}$};
\node[frozen] (4567) at (4.5,2.5) {$\ab{4567}$};
\node[frozen] (5678) at (4.5,3.75) {$\ab{5678}$};

\begin{scope}[blend mode=overlay,overlay]
          \node[rectangle, rounded corners, fit=(1267)(1237),fill=red!20, inner sep=0pt] {};
        \end{scope}
\node[blue] (1346) at (0,0) {$\ab{1346}$};
\node[blue] (1347) at (0,1.25) {$\ab{1347}$};
\node[blue] (1367) at (0,2.5) {$\ab{1367}$};
\node[red] (2346) at (2.25,-1.25) {$\ab{2346}$};
\node[blue] (3467) at (2.25,1.25) {$\ab{3467}$};
\node[blue] (1467) at (2.25,2.5) {$\ab{1467}$};
\node[frozen] (1678) at (0,3.75) {$\ab{1678}$};
\node[red] (1567) at (2.25,3.75) {$\ab{1567}$};
\node (1267) at (-2.25,2.5) {$\ab{1267}$};
\node (1237) at (-2.25,1.25) {$\ab{1237}$};
\node[frozen] (1234) at (-2.25,-1.25) {$\ab{1234}$};
\node[frozen] (2345) at (4.5,-1.25) {$\ab{2345}$};
\node[frozen] (1278) at (-4.5,2.5) {$\ab{1278}$};
\node[frozen] (1238) at (-4.5,1.25) {$\ab{1238}$};
\node[frozen] (3456) at (4.5,0) {$\ab{3456}$};
\node[frozen] (4567) at (4.5,2.5) {$\ab{4567}$};
\node[frozen] (5678) at (4.5,3.75) {$\ab{5678}$};

\draw[norm] (1234) -- (1237);
\draw[norm] (2346) -- (1234);
\draw[norm] (1237) -- (1267);
\draw[norm] (1237) -- (1347);
\draw[norm] (1237) -- (1267);
\draw[norm] (1237) -- (1238);
\draw[norm] (1267) -- (1278);
\draw[norm] (1267) -- (1367);
\draw[norm] (1367) -- (1467);
\draw[norm] (1467) -- (1567);
\draw[norm] (1467) -- (3467);
\draw[norm] (1567) -- (1678);
\draw[norm] (5678) -- (1567);
\draw[norm] (2345) -- (2346);
\draw[norm] (1346) -- (1347);
\draw[norm] (1347) -- (1367);
\draw[norm] (4567) -- (1467);

\draw[->,shorten >=2pt] (1347.south west) -- (1234.35);
\draw[->,shorten <=3pt] (1234.north east) -- (1346.south west);
\draw[diag] (1346.south east) -- (2346.north west);
\draw[diag] (2346.north east) -- (3456.south west);
\draw[diag] (3456.north west) -- (3467.south east);
\draw[diag] (3467.south west) -- (1346.north east);
\draw[diag] (3467.north east) -- (4567.south west);
\draw[diag] (1678.south west) -- (1267.north east);
\draw[diag] (1278.south east) -- (1237.north west);
\draw[diag] (1367.south west) -- (1237.north east);
\draw[diag] (1567.south east) -- (4567.north west);
}
\end{tikzpicture}
\caption{A cluster containing the poles of the R-invariant $[13467]$.}
\label{final}
\end{figure}
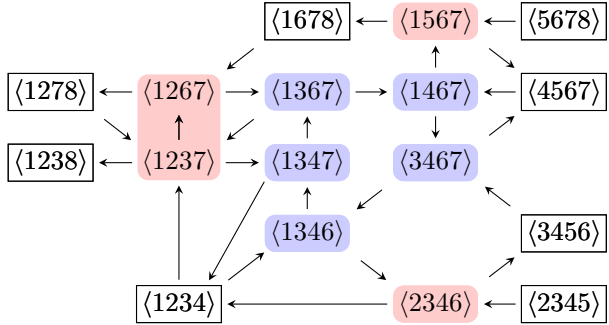

Starting from $[12345]$, the sequence of rotations to get $[13467]$ is
\begin{equation}
[12345] \xrightarrow{+4} [12356] \xrightarrow{+5} [13467]
\end{equation}
where the rotations are in ${\rm Conf}_6(\mathbb{P}^3)$ and ${\rm Conf}_7(\mathbb{P}^3)$ respectively. To find a cluster in ${\rm Conf}_8(\mathbb{P}^3)$ with all the $\mathcal{A}$-coordinates we need we start from the initial cluster (shown in Fig. \ref{fig:8ptinitcluster}) and mutate in the ${\rm Conf}_7(\mathbb{P}^3)$ subalgebra (the first two columns) such that its nodes rotate by five to arrive at the cluster shown in Fig. \ref{intermediate}. Then we mutate in the ${\rm Conf}_6(\mathbb{P}^3)$ subalgebra (the first column only) such that its nodes rotate by four.
Beginning with the ${\rm Conf}_8(\mathbb{P}^3)$ initial cluster we employ our mutation prescription by mutating up the first column, followed by the second column, repeating this another four times which results in the cluster shown in Fig. \ref{intermediate} where the unchanged topology of the ${\rm Conf}_7(\mathbb{P}^3)$ subalgebra is given in green.
 We now mutate up the first column in the green section four times, resulting in the final cluster shown in Fig. \ref{final} where the poles of $[13467]$ are in blue and the $A_2 \times A_1 \times A_1$ subalgebra is in red in agreement with Table \ref{table:Rinvs}.
Using this procedure one can locate a cluster which contains the poles of any R-invariant for an arbitrary number of points.

\subsection{Beyond NMHV}
Beyond NMHV, terms in BCFW tree amplitudes are more complicated than simple R-invariants so it is less obvious that one could associate subalgebras of ${\rm Conf}_n(\mathbb{P}^3)$ cluster algebras to individual terms. We show, up to eight points, that one can do this in much the same way as for NMHV.

\subsubsection*{Six points}
At six points the N$^2$MHV amplitude is equivalent to the $\overline{\text{MHV}}$ amplitude. It is given by
\begin{equation} \label{eq:6ptAmp}
\cA_{6,2} = \frac{\dab{123456}}{\ab{1234}\ab{1236}\ab{1256}\ab{1456}\ab{2345}\ab{3456}}
\end{equation}
where
\begin{equation}
\dab{ijklmn} = \frac{\dab{ijkmn} \dab{jklmn}}{\ab{jkmn}^4}
\end{equation}
is cyclically invariant and polynomial although not manifestly so in this form.

Identifying a cluster with \eqref{eq:6ptAmp} is trivial since every pole is an adjacent bracket and hence appears in every cluster in ${\rm Conf}_6(\mathbb{P}^3)$ i.e. one can associate this amplitude with the entire $A_3$ cluster algebra.
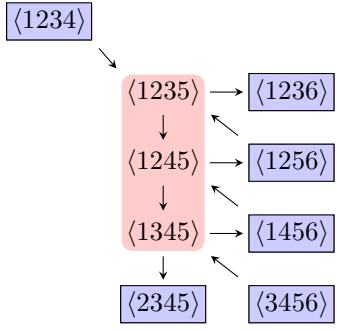
\begin{figure}
\begin{center}
\begin{tikzpicture}[scale=0.75]
{\footnotesize
\node[frozenblue] (1234) at (-2, 1.25) {$\ab{1234}$};
\node (1235) at (0, 0) {$\ab{1235}$};
\node[frozenblue] (1236) at (2.25, 0) {$\ab{1236}$};
\node (1245) at (0,-1.25) {$\ab{1245}$};
\node[frozenblue] (1256) at (2.25,-1.25) {$\ab{1256}$};
\node (1345) at (0,-2.5) {$\ab{1345}$};
\node[frozenblue] (1456) at (2.25,-2.5) {$\ab{1456}$};
\node[frozenblue] (2345) at (0,-3.75) {$\ab{2345}$};
\node[frozenblue] (3456) at (2.25,-3.75) {$\ab{3456}$};

\begin{scope}[blend mode=overlay,overlay]
          \node[rectangle, rounded corners, fit=(1235)(1245)(1345),fill=red!20, inner sep=0pt] {};
        \end{scope}

\node[frozenblue] (1234) at (-2, 1.25) {$\ab{1234}$};
\node (1235) at (0, 0) {$\ab{1235}$};
\node[frozenblue] (1236) at (2.25, 0) {$\ab{1236}$};
\node (1245) at (0,-1.25) {$\ab{1245}$};
\node[frozenblue] (1256) at (2.25,-1.25) {$\ab{1256}$};
\node (1345) at (0,-2.5) {$\ab{1345}$};
\node[frozenblue] (1456) at (2.25,-2.5) {$\ab{1456}$};
\node[frozenblue] (2345) at (0,-3.75) {$\ab{2345}$};
\node[frozenblue] (3456) at (2.25,-3.75) {$\ab{3456}$};

\draw[->,shorten <=4pt, shorten >=3pt] (1234.south east) -- (1235.north west);
\draw[norm] (1235) -- (1236);
\draw[norm] (1245) -- (1256);
\draw[norm] (1345) -- (1456);
\draw[norm] (1235) -- (1245);
\draw[norm] (1245) -- (1345);
\draw[norm] (1345) -- (2345);
\draw[diag] (1256.north west) -- (1235.south east);
\draw[diag] (1456.north west) -- (1245.south east);
\draw[diag] (3456.north west) -- (1345.south east);
}
\end{tikzpicture}
\end{center}
\caption{A cluster in $A_3$ corresponding to the six-point N$^{2}$MHV amplitude.}
\end{figure}
The blue nodes correspond to poles in the amplitude and the nodes highlighted in red correspond to the full $A_3$ algebra in which the amplitude lives.
\subsubsection*{Seven points}
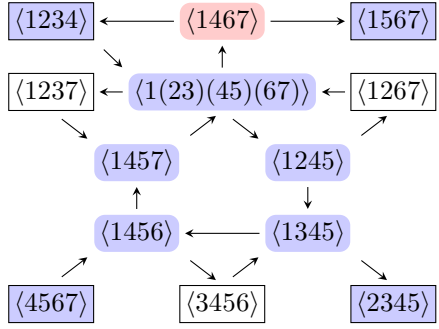
\begin{figure}
\begin{center}
\begin{tikzpicture}[scale=0.75]
{\footnotesize
\node[frozen] (3456) at (0,0) {$\ab{3456}$};
\node[frozenblue] (4567) at (-3,0) {$\ab{4567}$};
\node[frozenblue] (2345) at (3,0) {$\ab{2345}$};
\node[blue] (1456) at (-1.5,1.25) {$\ab{1456}$};
\node[blue] (1345) at (1.5,1.25) {$\ab{1345}$};
\node[blue] (1457) at (-1.5,2.5) {$\ab{1457}$};
\node[blue] (1245) at (1.5,2.5) {$\ab{1245}$};
\node[frozen] (1237) at (-3,3.75) {$\ab{1237}$};
\node[frozen] (1267) at (3,3.75) {$\ab{1267}$};
\node[blue] (s1) at (0,3.75) {$\la 1 (23) (45) (67) \ra$};
\node[red] (1467) at (0, 5) {$\ab{1467}$};
\node[frozenblue] (1234) at (-3,5) {$\ab{1234}$};
\node[frozenblue] (1567) at (3,5) {$\ab{1567}$};

\draw[norm] (1345) -- (1456);
\draw[norm] (s1) -- (1237);
\draw[norm] (1267) -- (s1);
\draw[norm] (1467) -- (1234);
\draw[norm] (1467) -- (1567);

\draw[norm] (s1) -- (1467);
\draw[norm] (1456) -- (1457);
\draw[norm] (1245) -- (1345);

\draw[diag] (4567.north) -- (1456.south west);
\draw[diag] (1456.south east) -- (3456.north);
\draw[diag] (3456.north) -- (1345.south west);
\draw[diag] (1345.south east) -- (2345.north);
\draw[diag] (1457.north east) -- (s1.south);
\draw[diag] (s1.south) -- (1245.north west);
\draw[diag] (1237.south) -- (1457.north west);
\draw[diag] (1245.north east) -- (1267.south);
\draw[diag] (1234.south east) -- (s1.north west);
}
\end{tikzpicture}
\end{center}
\caption{A cluster corresponding to the $4^{\text{th}}$ term in $\cA_{7,2}$.}
\label{fig:7ptClusterTerm4}
\end{figure}

\begin{figure}
\begin{center}
\begin{tikzpicture}[scale=0.75]
{\footnotesize
\node[red] (2367) at (0,0) {$\ab{2367}$};
\node[frozenblue] (1237) at (-3,0) {$\ab{1237}$};
\node[frozenblue] (4567) at (3,0) {$\ab{4567}$};
\node[red] (s3) at (0,1.25) {$\la 7 (23) (45) (16) \ra $};
\node[frozenblue] (1567) at (-6,1.25) {$\ab{1567}$};
\node[frozenblue] (1267) at (6,1.25) {$\ab{1267}$};
\node[blue] (1467) at (-6, 2.5) {$\ab{1467}$};
\node[blue] (s1) at (-2.5, 2.5) {$\la 1 (23) (45) (67) \ra $};
\node[blue] (s2) at (2.5, 2.5) {$\la 6 (23) (45) (17) \ra $};
\node[blue] (1367) at (6, 2.5) {$\ab{1367}$};
\node[frozen] (1234) at (-4.65,3.75) {$\ab{1234}$};
\node[frozen] (2345) at (0,3.75) {$\ab{2345}$};
\node[frozen] (3456) at (4.65,3.75) {$\ab{3456}$};

\draw[norm] (s1) -- (1467);
\draw[norm] (1367) -- (s2);
\draw[norm] (2367) -- (1237);
\draw[norm] (2367) -- (4567);

\draw[norm] (1467) -- (1567);
\draw[norm] (1367) -- (1267);
\draw[norm] (s3) -- (2367);
\draw[norm] (s1.210) -- (1237);
\draw[norm] (4567) -- (s2.330);

\draw[->, shorten <=3pt] (1237.north east) -- (s3.south west);
\draw[->, shorten >=3pt] (s3.north west) -- (s1.south);
\draw[->, shorten <=3pt] (s2.south) -- (s3.north east);
\draw[diag] (1467.north) -- (1234.south west);
\draw[diag] (1234.south east) -- (s1.160);
\draw[diag] (s1.20) -- (2345.south west);
\draw[diag] (2345.south east) -- (s2.160);
\draw[diag] (s2.20) -- (3456.south west);
\draw[diag] (3456.south east) -- (1367.north);

\begin{scope}[on background layer]
\draw[fill=red!20, thin, red!20] (s3.240) rectangle (2367.60);
\end{scope}
}
\end{tikzpicture}
\end{center}
\caption{A cluster corresponding to the $6^{\text{th}}$ term in $\cA_{7,2}$.}
\label{n2mhvterm6}
\end{figure}
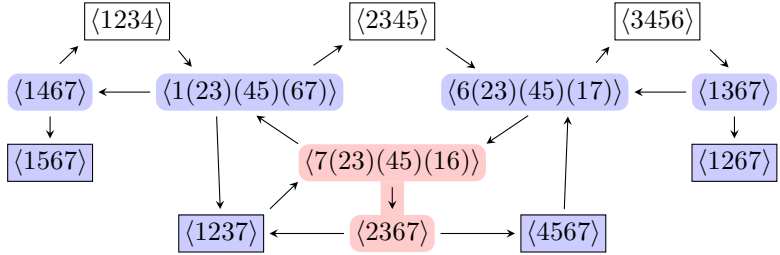

The seven-point, N$^2$MHV, tree-amplitude is equivalent to the $\overline{\text{NMHV}}$ amplitude
\begin{equation} \label{eq:7ptAmp}
\begin{aligned}
\cA_{7,2} &= \cA_{6,2} \\ &+ \frac{\dab{134567}}{\ab{1345}\ab{1347}\ab{1367}\ab{1567}\ab{3456}\ab{4567}} \\ 
&+ \frac{\dab{123467}}{\ab{1234}\ab{1237}\ab{1267}\ab{1467}\ab{2346}\ab{3467}} \\ 
&+ \frac{\dab{12345} \dab{14567}}{\ab{1234}\ab{1245}\ab{1345}\ab{1456}\ab{1457}\ab{1567}\ab{2345}\ab{4567}\la 1 (23)(45)(67) \ra} \\ 
&+ \frac{\dab{12367} \dab{23456}}{\ab{1236}\ab{1237}\ab{1267}\ab{2345}\ab{2346}\ab{2356}\ab{2367}\ab{3456}\la 6 (23)(45)(17) \ra} \\ 
&+ \frac{\dab{12367} \dab{14567}}{\ab{1237}\ab{1267}\ab{1367}\ab{1467}\ab{1567}\ab{4567}\la 1 (23)(45)(67) \ra \la 6 (23)(45)(17) \ra}.
\end{aligned}
\end{equation}
The first term is equal to the expression (\ref{eq:6ptAmp}) for the six-point amplitude. It is now in ${\rm Conf}_7(\mathbb{P}^3) \sim E_6$ therefore some of the poles are now unfrozen and the $A_3$ algebra is now a subalgebra of the full $E_6$ algebra, as shown in Fig. \ref{fig:7ptInitialCluster}.
\begin{figure}
\begin{center}
\begin{tikzpicture}[scale=0.75]
{\footnotesize
\node[frozenblue] (1234) at (-2, 1.25) {$\ab{1234}$};
\node (1235) at (0, 0) {$\ab{1235}$};
\node[blue] (1236) at (2.25, 0) {$\ab{1236}$};
\node (1245) at (0,-1.25) {$\ab{1245}$};
\node[blue] (1256) at (2.25,-1.25) {$\ab{1256}$};
\node (1345) at (0,-2.5) {$\ab{1345}$};
\node[blue] (1456) at (2.25,-2.5) {$\ab{1456}$};
\node[frozenblue] (2345) at (0,-3.75) {$\ab{2345}$};
\node[frozenblue] (3456) at (2.25,-3.75) {$\ab{3456}$};
\node[frozen] (1237) at (4.5,0) {$\ab{1237}$};
\node[frozen] (1267) at (4.5,-1.25) {$\ab{1267}$};
\node[frozen] (1567) at (4.5,-2.5) {$\ab{1567}$};
\node[frozen] (4567) at (4.5,-3.75) {$\ab{4567}$};

\begin{scope}[blend mode=overlay,overlay]
          \node[rectangle, rounded corners, fit=(1235)(1245)(1345),fill=red!20, inner sep=0pt] {};
        \end{scope}

\node[frozenblue] (1234) at (-2, 1.25) {$\ab{1234}$};
\node (1235) at (0, 0) {$\ab{1235}$};
\node[blue] (1236) at (2.25, 0) {$\ab{1236}$};
\node (1245) at (0,-1.25) {$\ab{1245}$};
\node[blue] (1256) at (2.25,-1.25) {$\ab{1256}$};
\node (1345) at (0,-2.5) {$\ab{1345}$};
\node[blue] (1456) at (2.25,-2.5) {$\ab{1456}$};
\node[frozenblue] (2345) at (0,-3.75) {$\ab{2345}$};
\node[frozenblue] (3456) at (2.25,-3.75) {$\ab{3456}$};
\node[frozen] (1237) at (4.5,0) {$\ab{1237}$};
\node[frozen] (1267) at (4.5,-1.25) {$\ab{1267}$};
\node[frozen] (1567) at (4.5,-2.5) {$\ab{1567}$};
\node[frozen] (4567) at (4.5,-3.75) {$\ab{4567}$};

\draw[norm] (1235) -- (1236);
\draw[norm] (1236) -- (1237);
\draw[norm] (1245) -- (1256);
\draw[norm] (1256) -- (1267);
\draw[norm] (1345) -- (1456);
\draw[norm] (1456) -- (1567);

\draw[norm] (1235) -- (1245);
\draw[norm] (1236) -- (1256);
\draw[norm] (1245) -- (1345);
\draw[norm] (1256) -- (1456);
\draw[norm] (1345) -- (2345);
\draw[norm] (1456) -- (3456);

\draw[->,shorten <=4pt, shorten >=3pt] (1234.south east) -- (1235.north west);
\draw[diag] (1256.north west) -- (1235.south east);
\draw[diag] (1267.north west) -- (1236.south east);
\draw[diag] (1456.north west) -- (1245.south east);
\draw[diag] (1567.north west) -- (1256.south east);
\draw[diag] (3456.north west) -- (1345.south east);
\draw[diag] (4567.north west) -- (1456.south east);
}
\end{tikzpicture}
\end{center}

\caption{A cluster containing the poles of $\cA_{6,2}$ in ${\rm Conf}_7(\mathbb{P}^3)$.}
\label{fig:7ptInitialCluster}
\end{figure}
As before, the blue nodes correspond to poles in the term while the nodes highlighted in red correspond to an $A_3$ subalgebra inside the full $E_6$ algebra in which all the poles of \eqref{eq:6ptAmp} can be found. The second and third terms of \eqref{eq:7ptAmp} can be obtained by rotating the momentum twistors in \eqref{eq:6ptAmp} by two and five units respectively and hence one can obtain clusters containing their poles by rotating Fig. \ref{fig:7ptInitialCluster} by the same amounts. We can associate the fourth term of \eqref{eq:7ptAmp} with an $A_1$ subalgebra as shown in Fig. \ref{fig:7ptClusterTerm4}. 
\noindent One can obtain the fifth term by rotating the fourth term by five units hence it also lives in an $A_1$ subalgebra found by rotating Fig. \ref{fig:7ptClusterTerm4} by five units.
Finally, the sixth term can be associated to an $A_2$ subalgebra as illustrated in Fig. \ref{n2mhvterm6}.

\subsubsection*{Eight points}
The eight-point N$^{2}$MHV amplitude is the first true N$^{2}$MHV amplitude in that it is not equivalent to the parity conjugate of another N$^{k<2}$MHV amplitude. Explicitly it is given by

{\footnotesize
\begin{align}
\label{eq:8ptAmp}
\!\!\!\cA_{8,2} &=  \cA_{7,2} \notag \\
&+ \frac{\dab{123478}}{\ab{1234}\ab{1238}\ab{1278}\ab{1478}\ab{2347}\ab{3478}} \notag \\
&+ \frac{\dab{134578}}{\ab{1345}\ab{1348}\ab{1378}\ab{1578}\ab{3457}\ab{4578}} \notag \\
&+ \frac{\dab{145678}}{\ab{1456}\ab{1458}\ab{1478}\ab{1678}\ab{4567}\ab{5678}} \notag \\
&+ \frac{\dab{12345} \dab{15678}}{\ab{1234}\ab{1235}\ab{1245}\ab{1345}\ab{1567}\ab{1568}\ab{1578}\ab{1678}\ab{2345}\ab{5678}} \notag \\
&- \frac{\dab{12378} \dab{23456}}{\ab{1237}\ab{1238}\ab{1278}\ab{2345}\ab{2346}\ab{2356}\ab{2378}\ab{3456}\la 2 3 \bar{5} \cap \bar{8} \ra } \notag \\
&+ \frac{\dab{12345} \dab{14578}}{\ab{1234}\ab{1245}\ab{1345}\ab{1457}\ab{1458}\ab{1578}\ab{2345}\ab{4578}\quadd{1}{23}{45}{78}} \notag \\
&+ \frac{\dab{12356} \dab{15678}}{\ab{1235}\ab{1256}\ab{1356}\ab{1567}\ab{1568}\ab{1678}\ab{2356}\ab{5678}\quadd{1}{23}{56}{78}} \notag \\
&+ \frac{\dab{13456} \dab{15678}}{\ab{1345}\ab{1356}\ab{1456}\ab{1567}\ab{1568}\ab{1678}\ab{3456}\ab{5678}\quadd{1}{34}{56}{78}} \notag \\
&+ \frac{\dab{12378} \dab{23467}}{\ab{1237}\ab{1238}\ab{1278}\ab{2346}\ab{2347}\ab{2367}\ab{2378}\ab{3467}\quadd{7}{23}{46}{18}} \notag \\
&+ \frac{\dab{13478} \dab{34567}}{\ab{1347}\ab{1348}\ab{1378}\ab{3456}\ab{3457}\ab{3467}\ab{3478}\ab{4567}\quadd{7}{34}{56}{18}} \notag \\
&+ \frac{\dab{12378} \dab{14578}}{\ab{1238}\ab{1278}\ab{1378}\ab{1478}\ab{1578}\ab{4578}\quadd{1}{23}{45}{78}\quadd{7}{23}{45}{18}} \notag \\
&+ \frac{\dab{12378} \dab{15678}}{\ab{1238}\ab{1278}\ab{1378}\ab{1578}\ab{1678}\ab{5678}\quadd{1}{23}{56}{78}\quadd{7}{23}{56}{18}} \notag \\
&+ \frac{\dab{13478} \dab{15678}}{\ab{1348}\ab{1378}\ab{1478}\ab{1578}\ab{1678}\ab{5678}\quadd{1}{34}{56}{78}\quadd{7}{34}{56}{18}} \notag \\
&+ \frac{\dab{12378} \Delta}{\ab{1237}\ab{1238}\ab{1378}\ab{2378}\ab{4567}\la 2 3 \bar{5} \cap \bar{8} \ra \quadd{7}{23}{45}{18}\quadd{7}{23}{46}{18}\quadd{7}{23}{56}{18}}
\end{align}
}
where in the last term we have the quantity $\Delta^{0|4} = \delta^{0|4}(\chi_2 \ab{1378}\ab{4567} - \chi_3 \ab{1278}\ab{4567} - \chi_4 \quadd{7}{23}{56}{18} + \chi_5 \quadd{7}{23}{46}{18} - \chi_6 \quadd{7}{23}{45}{18} - \chi_7 \la 2 3 \bar{5} \cap \bar{8} \ra)$.

At eight points, ${\rm Conf}_8(\mathbb{P}^3)$ is an infinite cluster algebra, however we can still associate finite subalgebras to each of the 20 terms in the amplitude. These subalgebras are displayed in Table \ref{n2mhv-tab} where terms 1-6 are those in \eqref{eq:7ptAmp}.
\begin{table}
\centering
	{
	\begin{tabular}{@{} c  c  c  c  c  c  c  c @{}}
          \toprule
	Term & Sub-Algebra & Term & Sub-Algebra & Term & Sub-Algebra & Term & Sub-Algebra \\ \midrule
	$1$ & $A_3 \times A_3$ & $6$ & $A_2$ & $11$ & $A_3 \times A_1$ & $16$ & $A_1 \times A_1$ \\ 
	$2$ & $A_3 \times A_2$ & $7$ & $A_3 \times A_3$ & $12$ & $A_1 \times A_1$ & $17$ & $A_2 \times A_1$ \\ 
	$3$ & $A_3 \times A_1$ & $8$ & $A_3$ & $13$ & $A_1 \times A_1$ & $18$ & $A_3 \times A_2$ \\ 
	$4$ & $A_2 \times A_1$ & $9$ & $A_3 \times A_3$ & $14$ & $A_2 \times A_1$ & $19$ & $A_2 \times A_1$ \\ 
          $5$ & $A_1 \times A_1$ & $10$ & $A_3£$ & $15$ & $A_1 \times A_1$ & $20$ & $A_2$ \\
          \bottomrule
	\end{tabular}
	}
	\caption{Subalgebras associated to terms in $\cA_{8,2}$.}
	\label{n2mhv-tab}
\end{table}
Although the subalgebras shown in Table \ref{n2mhv-tab} are all finite, at higher points they may become infinite. For example, the subalgebra associated to \eqref{eq:6ptAmp} at ten points will be $A_3 \times {\rm Conf}_8(\mathbb{P}^3)$ which is infinite as ${\rm Conf}_8(\mathbb{P}^3)$ is infinite. 

The tenth term is a new type of term of the form
\begin{equation}
[12345][56781]\,,
\end{equation}
to which we can associate an $A_3$ subalgebra, a cluster belonging to which takes the form shown in Fig. \ref{fig:RProductCluster} below. The left and right columns of blue nodes in Fig. \ref{fig:RProductCluster} correspond to the poles of $[12345]$ and $[56781]$ respectively while the red column signifies the $A_3$ subalgebra to which we associate this term.
\begin{figure}
\centering
\begin{tikzpicture}[scale=0.75]
{\footnotesize
\node[frozenblue] (1234) at (-2, 1.25) {$\ab{1234}$};
\node[frozen] (1238) at (1.25,1.25) {$\ab{1238}$};
\node[frozen] (1278) at (3.75,1.25) {$\ab{1278}$};
\node[frozenblue] (1678) at (7, 1.25) {$\ab{1678}$};
\node[blue] (1235) at (0, 0) {$\ab{1235}$};
\node (1258) at (2.5, 0) {$\ab{1258}$};
\node[blue] (1245) at (0,-1.25) {$\ab{1245}$};
\node (1256) at (2.5,-1.25) {$\ab{1256}$};
\node[blue] (1345) at (0,-2.5) {$\ab{1345}$};
\node (1456) at (2.5,-2.5) {$\ab{1456}$};
\node[frozenblue] (2345) at (-2,-3.75) {$\ab{2345}$};
\node[frozen] (3456) at (1.25,-3.75) {$\ab{3456}$};
\node[blue] (1578) at (5,0) {$\ab{1578}$};
\node[blue] (1568) at (5,-1.25) {$\ab{1568}$};
\node[blue] (1567) at (5,-2.5) {$\ab{1567}$};
\node[frozen] (4567) at (3.75,-3.75) {$\ab{4567}$};
\node[frozenblue] (5678) at (7, -3.75) {$\ab{5678}$};

\begin{scope}[blend mode=overlay,overlay]
          \node[rectangle, rounded corners, fit=(1258)(1256)(1456),fill=red!20, inner sep=0pt] {};
        \end{scope}
\node[frozenblue] (1234) at (-2, 1.25) {$\ab{1234}$};
\node[frozen] (1238) at (1.25,1.25) {$\ab{1238}$};
\node[frozen] (1278) at (3.75,1.25) {$\ab{1278}$};
\node[frozenblue] (1678) at (7, 1.25) {$\ab{1678}$};
\node[blue] (1235) at (0, 0) {$\ab{1235}$};
\node (1258) at (2.5, 0) {$\ab{1258}$};
\node[blue] (1245) at (0,-1.25) {$\ab{1245}$};
\node (1256) at (2.5,-1.25) {$\ab{1256}$};
\node[blue] (1345) at (0,-2.5) {$\ab{1345}$};
\node (1456) at (2.5,-2.5) {$\ab{1456}$};
\node[frozenblue] (2345) at (-2,-3.75) {$\ab{2345}$};
\node[frozen] (3456) at (1.25,-3.75) {$\ab{3456}$};
\node[blue] (1578) at (5,0) {$\ab{1578}$};
\node[blue] (1568) at (5,-1.25) {$\ab{1568}$};
\node[blue] (1567) at (5,-2.5) {$\ab{1567}$};
\node[frozen] (4567) at (3.75,-3.75) {$\ab{4567}$};
\node[frozenblue] (5678) at (7, -3.75) {$\ab{5678}$};
        
\draw[norm] (1578) -- (1258);
\draw[norm] (1245) -- (1256);
\draw[norm] (1568) -- (1256);
\draw[norm] (1345) -- (1456);
\draw[norm] (1456) -- (1567);
\draw[norm] (1258) -- (1235);

\draw[norm] (1235) -- (1245);
\draw[norm] (1256) -- (1258);
\draw[norm] (1245) -- (1345);
\draw[norm] (1256) -- (1456);
\draw[norm] (1568) -- (1578);
\draw[norm] (1567) -- (1568);

\draw[->,shorten <=4pt, shorten >=3pt] (1234.south east) -- (1235.north west);
\draw[->,shorten <=4pt, shorten >=3pt] (1345.south west) -- (2345.north east);
\draw[->,shorten <=4pt, shorten >=3pt] (1578.north east) -- (1678.south west);
\draw[->,shorten <=4pt, shorten >=3pt] (5678.north west) -- (1567.south east);
\draw[diag] (1258.south east) -- (1568.north west);
\draw[diag] (1456.north west) -- (1245.south east);
\draw[norm] (3456) -- (1345);
\draw[norm] (4567) -- (1456);
\draw[norm] (1456) -- (3456);
\draw[norm] (1567) -- (4567);
\draw[norm] (1235) -- (1238);
\draw[norm] (1238) -- (1258);
\draw[norm] (1258) -- (1278);
\draw[norm] (1278) -- (1578);
}
\end{tikzpicture}
\caption{A cluster containing the poles of $[12345][56781]$ in ${\rm Conf}_8(\mathbb{P}^3)$.}
\label{fig:RProductCluster}
\end{figure}
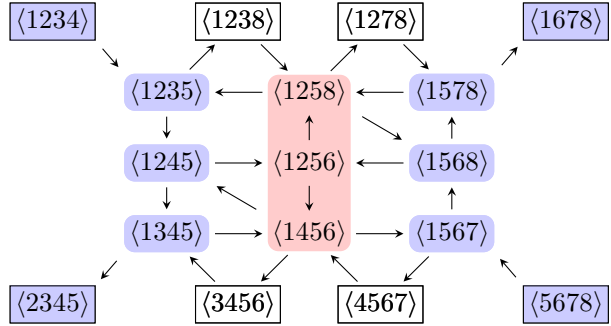


\subsection{Discussion}

We have shown that all NMHV R-invariants obey the cluster adjacency property in that their poles can all be found together in some cluster. We have also shown that the BCFW terms in the expansion of N$^2$MHV trees also obey cluster adjacency for six, seven, and eight points. To each term is associated some subalgebra in the full polytope where every cluster contains all of the poles. Similar structures have emerged in the study of the Grassmannian integrals of \cite{ArkaniHamed:2009dn,Mason:2009qx} and on-shell diagrams \cite{ArkaniHamed:2012nw}. The difference here is that the properties we observe between poles (both physical and spurious) are phrased in the same language that we have found relates the branch cuts (symbol entries) of the integrated amplitudes.

The results for tree-level NMHV and N$^2$MHV are highly suggestive that there should exist a general relation between the singularities of the Yangian invariant leading singularities and the cluster algebras associated to ${\rm Conf}_n(\mathbb{P}^3)$.  A natural question is whether an extension of the notion of cluster adjacency holds for all Yangian invariants. This would lead us to consider quantities which go beyond $\mathcal{A}$-coordinates for ${\rm Conf}_n(\mathbb{P}^3)$ such as the four-mass box leading singularity which exhibits square root branch cuts in momentum twistor variables. Studying such quantities should lead to insight on what cluster adjacency has to say beyond rational $\mathcal{A}$-coordinates and should have implications for understanding the boundary structure of higher polytopes and the type of transcendental functions which appear beyond seven-point amplitudes.


Certain operations can also be performed on Yangian invariants \cite{ArkaniHamed:2010kv}, e.g. the `fusing' of two Yangian invariants is also a Yangian invariant. Could one find a cluster interpretation of such an operation? The cluster shown in Fig. \ref{fig:RProductCluster} contains the poles of the product of two Yangian invariants and could also be indicative of the amalgamation procedure \cite{ArkaniHamed:2012nw} whereby two clusters can be joined together to produce a cluster in a larger algebra.

\section{NMHV loop amplitudes}
\label{NMHVloops}

Now we are in a position to relate the cluster adjacency properties described in the two previous sections. The first amplitudes which exhibit both poles and cuts non-trivially are the NMHV loop amplitudes. 




\subsection{Hexagons}

The BDS-like subtracted NMHV hexagon is often written in terms of a parity even function $E(u,v,w)=E(Z_1,\ldots,Z_6)$ and a parity odd function\footnote{Sometimes $\tilde{E}(y_u,y_v,y_w)$ denoted simply as $\tilde{E}(u,v,w)$, in which case one should in addition take care to remember its odd parity.} $\tilde{E}(y_u,y_v,y_w)=\tilde{E}(Z_1,\ldots,Z_6)$, where we have drawn attention to their dependence on the twistor variables. Here we will adopt a shorthand notation which makes reference to the which of the cyclically ordered twistors $Z_i$ sits in the first argument,
\begin{equation}
\begin{aligned}
E_1 &= E(u,v,w)\,, \quad &E_2 &= E(v,w,u)\,,\quad &E_3 &= E(w,u,v)\,, \\
\tilde{E}_1 &= E(y_u,y_v,y_w)\,, \quad &\tilde{E}_2 &= -\tilde{E}(y_v,y_w,y_u)\,,\quad &\tilde{E}_3 &= E(y_w,y_u,y_v)\,.
\end{aligned}
\end{equation}
The parity properties of $E$ and $\tilde{E}$ imply
\be
E_4 = E_1\,, \qquad \tilde{E}_4 = - \tilde{E}_1\,.
\ee
With this notation the hexagon NMHV amplitude takes the form
\begin{align}
\label{NMHV6pt}
\mathcal{E}_{6,{\rm NMHV}}   = &E_{1}[(1)+(4)] + E_{2}[(2)+(5)]+E_{3}[(3)+(6)] \notag\\
+ &\tilde{E}_{1}[(1)-(4)] + \tilde{E}_{2}[(2)-(5)] + \tilde{E}_{3}[(3)-(6)] \,.
\end{align}
Here we have adopted a common shorthand notation for the R-invariants: we write $(1) = [23456]$ and cyclically related formulae. The function $\tilde{E}$ is taken to obey
\begin{equation}
\tilde{E}_{1} - \tilde{E}_{2} + \tilde{E}_{3}=0\,.
\end{equation}
We may equivalently write $\mathcal{E}_6^{\rm NMHV}$ as follows,
\begin{equation}
\label{NMHV6}
\mathcal{E}_{6,{\rm NMHV}} = (1) F_{1} + \text{ cyc.} \qquad F_1=E_1+\tilde{E}_1\,.
\end{equation}
In (\ref{NMHV6}) the notation `cyc' refers to all cyclic rotations of the momentum twistors. At $L$ loops the functions $E$ and $\tilde{E}$ are weight $2L$ polylogarithms.

To discuss the cluster adjacency properties of the hexagon NMHV amplitudes we should consider the $(2L-1,1)$ coproduct of $\mathcal{E}_{6,{\rm NMHV}}$,
\be
\label{dFNMHV}
\mathcal{E}_{6,{\rm NMHV}}^{(2L-1,1)} = (1) \sum_{i<j<k<l} [F_{1}^{\langle ijkl \rangle} \otimes \langle ijkl \rangle]  + \text{ cyc.}
\ee
Cluster adjacency manifest itself in two ways in the above expression. Firstly the $F^{\langle ijkl \rangle}$ are neighbour set functions for $\langle ijkl \rangle$. This is the statement that $F$ and hence $E$ and $\tilde{E}$ cluster adjacent polylogarithms in the sense described in Sect. \ref{CApolysdef}. Secondly we find that the different functions $F_1^{\langle ijkl \rangle}$ appearing in (\ref{dFNMHV}) are constrained by the fact that $F_1$ appears with the R-invariant $(1)$ in (\ref{NMHV6}).

In order to reveal the additional constraints that cluster adjacency places on the form of $F$ we exploit the fact that the R-invariants obey the identity
\begin{equation}
(1)-(2)+(3)-(4)+(5)-(6)=0\,.
\end{equation}
This allows us to modify the presentation of $\mathcal{E}_{6,{\rm NMHV}}^{(2L-1,1)}$ by adding to it a vanishing term of the form
\begin{equation}
\label{addzero}
[(1)-(2)+(3)-(4)+(5)-(6)] Z_{1} \,,
\end{equation}
where $Z$ is given by
\be
Z_{1} = \sum_{i<j<k<l} [Z^{\langle ijkl \rangle}_{1}\otimes \langle ijkl \rangle]\,.
\ee
Here (by cyclically symmetrising (\ref{addzero}) if necessary) we can require that $Z$ is anti-cyclic,
\begin{equation}
\label{cycZ}
Z_{2}= - Z_{1}\,.
\end{equation}
This means that the presentation of $\mathcal{E}_{6,{\rm NMHV}}^{(2L-1,1)}$ is still manifestly cyclic,
\begin{equation}
\label{E6copmod}
\mathcal{E}_{6,{\rm NMHV}}^{(2L-1,1)} = (1) \sum_{i<j<k<l} [(F_{1}^{\langle ijkl \rangle} + Z_1^{\langle ijkl \rangle}) \otimes \langle ijkl \rangle]  + \text{ cyc.}
\end{equation}


We find the following additional cluster adjacency property of all hexagon NMHV loop amplitudes:
\emph{there exists a $Z$ such that the only $\mathcal{A}$-coordinates $\langle ijkl \rangle$ appearing in (\ref{E6copmod}) are in the neighbour set of every $\mathcal{A}$-coordinate in the denominator of the R-invariant $(1)$}. 

As we have discussed in Sect. \ref{NMHVBCFW}, the R-invariant $(1)=[23456]$ is associated to a single cluster in ${\rm Conf}_6(\mathbb{P}^3)$ (in fact it is the one whose triangulation involves all the chords of the form $(1i)$). It follows that the only unfrozen $\mathcal{A}$-coordinates allowed in the final entries are the ones of that cluster, namely $\langle 2346 \rangle = (15)$, $\langle 2356 \rangle = (14)$ and $\langle 2456\rangle=(13)$.  The following unfrozen $\mathcal{A}$-coordinates,
\begin{equation}
\label{absentAs}
\{\langle 1235 \rangle,\langle 1245 \rangle,\langle 1246 \rangle,\langle 1345 \rangle,\langle 1346 \rangle,\langle 1356 \rangle\}\,,
\end{equation}
are therefore forbidden in the sum in (\ref{E6copmod}) above.

Note that since $Z$ is multiplied by zero in (\ref{addzero}) we do not need to require that it is integrable, nor even that it is homogeneous.  Nevertheless, the fact that it exists and obeys (\ref{cycZ}) has the following implications for the final entries (or $(n-1,1)$ coproduct) of $F$,
\begin{align}
F_{1}^{\langle 1235 \rangle}  &= - Z_{1}^{\langle 1235 \rangle}\,,\notag\\
F_{1}^{\langle 1246 \rangle}  &= - Z_{1}^{\langle 1246 \rangle}\,,\notag\\
F_{1}^{\langle 1345 \rangle}  &= - Z_{1}^{\langle 1345 \rangle}\,,\notag\\
F_{1}^{\langle 1356 \rangle}  &= - Z_{1}^{\langle 1356 \rangle}\,,\notag\\
F_{1}^{\langle 1245 \rangle}  &= - Z_{1}^{\langle 1245 \rangle}\,, \notag\\
F_{1}^{\langle 1346 \rangle}  &= - Z_{1}^{\langle 1346 \rangle}\,.
\end{align}
The anti-cyclicity of $Z$ implies\footnote{We remind the reader that the subscripts refer to the arguments of functions. For example, $Z_6^{\langle 1235 \rangle}$ means $Z_1^{\langle 1235 \rangle}|_{Z_i \rightarrow Z_{i-1}}$ and not the $\langle1235 \rangle$ coproduct element of $Z_6$.}
\begin{align}
Z^{\langle 1246 \rangle}_{1} &= - Z^{\langle 1235 \rangle}_{6}\,,\notag\\
Z^{\langle 1345 \rangle}_{1} &= + Z^{\langle 1235 \rangle}_{3}\,,\notag\\
Z^{\langle 1356 \rangle}_{1} &= + Z^{\langle 1235 \rangle}_{5}\,,\notag \\
Z^{\langle 1346 \rangle}_{1} &= + Z^{\langle 1245 \rangle}_{3}\,.
\end{align}
Combining the above two sets of relations we deduce that adjacency implies the following relations among the coproducts of $F$\,,
\begin{align}
F_{1}^{\langle 1246 \rangle}  &= -F_{6}^{\langle 1235 \rangle} \,,\notag\\
F_{1}^{\langle 1345 \rangle}  &=+F_{3}^{\langle 1235 \rangle} \,, \notag\\
F_{1}^{\langle 1356 \rangle}  &=+ F_{5}^{\langle 1235 \rangle} \,, \notag\\
F_{1}^{\langle 1346 \rangle}  &=+ F_{3}^{\langle 1245 \rangle}  \,.
\label{Fcoproductrels}
\end{align}
The equations (\ref{Fcoproductrels}) are the consequences of cluster adjacency between the final entries of the coproduct of $F=E+\tilde{E}$ and the R-invariants.

As discussed in \cite{Dixon:2015iva}, similar coproduct relations follow from the $\bar{Q}$-equation of \cite{CaronHuot:2011kk,Bullimore:2011kg}. We may ask how the $\bar{Q}$ conditions are related to the adjacency ones. To do this it is simplest to count how many homogeneous (final entry)$\otimes$(R-invariant) combinations are allowed by cluster adjacency. To do this one may choose five independent R-invariants, say $(1),(2),(3),(4),(5)$, and nine $d \log$'s of multiplicatively independent homogeneous letters and make an arbitrary linear combination of all 45 possible products. We expand the resulting expression into the $d \log \langle ijkl \rangle$ and eliminate all pairs $(m)\,d\log \langle ijkl \rangle $ which obey adjacency (taking care to remember that some $\mathcal{A}$-coordinates are compatible with the R-invariant $(6) = (1)-(2)+(3)-(4)+(5)$) and require the resulting combination to vanish. This yields 27 conditions, leaving 18 linearly independent homogeneous (final entry)$\otimes$(R-invariant) combinations. This is exactly the same number of linearly independent combinations which are compatible with the $\bar{Q}$ final entry conditions described in \cite{Dixon:2015iva}. 

We conclude that for the NMHV hexagon, the cluster adjacency property is equivalent to the $\bar{Q}$ final entry conditions. One should nevertheless stress that the $\bar{Q}$ equation itself is stronger than just the final entry conditions as it expresses the $(2L-1,1)$ coproduct entries in terms of and integral over a limit of certain heptagon amplitudes.
We find it remarkable that cluster adjacency property in its various forms encompasses both the (extended) Steinmann conditions as well as some of the implications of dual superconformal symmetry.

\subsection{Heptagons}

In the case of heptagons it is possible to write down 21 R-invariants,
\begin{equation}
  [34567]=(12), \,\,\,
  [24567]=(13), \,\,\,
  [23567]=(14)\,\,\,\text{\& cyclic}\,.
\end{equation}
They satisfy seven six-term identities of the form
\begin{equation}
\label{eq:sixterm}
(12) - (13) + (14) - (15) + (16) - (17) = 0 \quad \text{\& cyclic} \,.
\end{equation}
Only six of these identities are linearly independent and the number
of independent R-invariants is therefore 15. In a canonical basis (as
used in \cite{Dixon:2016nkn,CaronHuot:2011kk}) which comprises the
tree amplitude and 14 other R-invariants, the BDS-like-normalised
amplitude is expressed as follows:
\begin{equation}
  \label{eq:hepnonred}
 \mathcal{E}_{7,{\rm NHMV}}
  =
  \cA_{7,1} E_0
  +\bigl[(12)\, E_{12}+\text{cyclic} + (14)\, E_{14}+\text{cyclic}\bigr]\, ,
\end{equation}
where $\cA_{7,1}$ is equal to the NMHV tree amplitude, given in (\ref{A71}).

The property of cluster adjacency again manifests itself in the heptagon NMHV amplitudes. It is possible to find a representation of the $(2L-1,1)$ coproduct of the form
\be
\label{heptcoproduct}
\mathcal{E}_{7,{\rm NMHV}}^{(2L-1,1)} = \sum_{a\in\mathcal{A}} \bigl[[(12) e_{12}^{a} + (13)e_{13}^{a} + (14)e_{14}^{a}]\otimes a \bigr] + \text{ cyc.}
\ee
Here the sum is over the heptagon alphabet (\ref{heptletters}). As in the hexagon case, adjacency manifests itself in two ways in (\ref{heptcoproduct}). Firstly each of the $e_{ij}^a$ is a weight $(2L-1)$ heptagon neighbour set function for the letter $a$. This implies that the functions $E_0$ and $E_{ij}$ in (\ref{eq:hepnonred}) are cluster adjacent polylogarithms. Secondly, only some of the $e_{ij}^a$ are non-zero: the ones where the letter $a$ is cluster adjacent to all of the poles of the R-invariant $(ij)$. For example, the
R-invariant $(12)$ contains three poles that are non-frozen cluster
${\cal A}$ coordinates, namely $\langle 3567 \rangle \sim a_{34}$,
$ \langle 3467 \rangle \sim a_{15}$, and $\langle 3457 \rangle \sim a_{26}$:
\begin{equation}
  (12) =
  \frac{\bigl(\langle 3456 \rangle \chi_7 + \text{cyclic}\bigr)^4}
  {\langle 4567 \rangle \langle 3567 \rangle \langle 3467 \rangle \langle 3457 \rangle \langle 3456 \rangle}\,.
\end{equation}
The intersection of the homogeneous neighbour sets of these coordinates defines the neighbour set of the R-invariant $(12)$, and similarly for the other R-invariants:
\begin{equation}
  \begin{aligned}[t]
    \hns{(12)} &= \hns{a_{34}} \, \cap \hns{a_{15}}\, \cap \hns{a_{26}}\\
    &= \{a_{11}, a_{12}, a_{15}, a_{21},a_{22},a_{26},a_{31}, a_{32}, a_{34}, a_{53}, a_{55}, a_{57}\}\,,\\
     \hns{(13)}&= \hns{a_{21}} \, \cap \hns{a_{33}}\, \cap \hns{a_{41}}\, \cap \hns{a_{43}}\\
     &= \{a_{11}, a_{13}, a_{21}, a_{23},a_{31},a_{33},a_{41}, a_{43}, a_{62}\}\,,\\
     \hns{(14)}&= \hns{a_{11}} \, \cap \hns{a_{14}}\, \cap \hns{a_{21}}\, \cap \hns{a_{34}}\, \cap \hns{a_{46}}\\
     &= \{a_{11}, a_{14}, a_{21}, a_{24},a_{31},a_{34},a_{46}\}\,.
  \end{aligned}
  \label{heptNMHVns}
\end{equation}
Only the (final entry)$\otimes$(R-invariant) combinations compatible with the above and their cyclic rotations are allowed by cluster adjacency.

Note that the representation (\ref{heptcoproduct}) employs the full redundant set of R-invariants. Upon elimination of the redundant R-invariants, the coproducts of the functions $E_0$ and $E_{ij}$ in (\ref{eq:hepnonred}) above are seen to be related to the quantities $e_{ij}$ via
\be
E_0^a = \sum_i e^a_{i,i+2}\,, \qquad E^a_{12} = e^a_{12} -e^a_{16} - e^a_{24} - e^a_{46}\,, \qquad E^a_{14} = e^a_{14} - e^a_{16} - e^a_{46}\,.
\ee
As in the hexagon case, we do not require that the combinations $\sum_a [e^a_{ij} \otimes a]$ are integrable; only $\sum_a [E_0^a \otimes a]$ and $\sum_a [E_{ij}^a \otimes a]$ are integrable. Nevertheless, just as in the hexagon case, the existence and adjacency properties of the $e^a_{ij}$ imply relations on the coproducts of the functions $E_0$ and $E_{ij}$.

Out of the $7\times(7+9+12) = 196$ cluster adjacent (final entry)$\otimes$(R-invariant) combinations allowed by (\ref{heptNMHVns}). The following linear combinations of cluster adjacent (final entry)$\otimes$(R-invariant) products vanish due to identities,
\begin{align}
[(12) - (13) + (14) - (15) + (16) - (17)]\otimes\{a_{11},a_{21},a_{31}\}
\end{align}
as do their cyclic rotations. This allows us to eliminate 21 such combinations leaving 175 independent cluster adjacent combinations.

The 175 combinations form a 
larger set than the more restricted set of 147 NMHV (final entry)$\otimes$(R-invariant) combinations derived by Caron-Huot which are
compatible with the $\bar Q$ equation. These 147 combinations are listed in \cite{Dixon:2016nkn}. Using the
identities (\ref{eq:sixterm}), these NMHV final entries can be
rewritten in the following manifestly cluster-adjacent way in which
the final entries of the function multiplying the R-invariant $(ij)$
are in the set $\hnsqbar{(ij)}$ where:
\label{sec:clust-adjc-nmhv}
    \label{eq:qbarca}
\begin{align}
  \hnsqbar{(12)} &= \{ a_{15}, a_{21}, a_{26}, a_{32}, a_{34}, a_{53}, a_{57}\}
                  \subset \hns{(12)}\notag \\
  \hnsqbar{(13)} &= \{ a_{21}, a_{23}, a_{31}, a_{33}, a_{41}, a_{43}, a_{62}\}
                  \subset \hns{(13)\notag }\\
  \hnsqbar{(14)} &= \{ a_{11}, a_{14}, a_{21}, a_{24}, a_{31}, a_{34}, a_{46}\}
                  \subset \hns{(14)}\qquad  \text{\& cyclic}\,.
\end{align}
The above set of $7\times3\times7 = 147$ (final entry)$\otimes$(R-invariant) pairs are equivalent up to using identities to the set presented in \cite{Dixon:2016nkn}. In contrast to the form presented in \cite{Dixon:2016nkn}, the
$\bar Q$-compatible final entries are monomials in the letters, which
makes it trivial to verify cluster adjacency properties. Note that
the list of (final entry)$\otimes$(R-invariant) pairs (\ref{eq:qbarca}) is not
unique since it is possible to trade some combinations with others
using the six-term identities (\ref{eq:sixterm}).

In \cite{NMHV4} we will make use of the above cluster adjacent form for the NMHV heptagon amplitude to allow for an efficient implementation of the bootstrap programme at four loops.

  \section{Conclusions}

We have explored and extended the role of cluster algebras and their relation to the appearance of singularities in scattering amplitudes in $\mathcal{N}=4$ super Yang-Mills theory. The picture which emerges is very geometric in nature, the boundary structure of the cluster polytope controls the way in which both poles and branch cuts appear. Codimension-one faces of the cluster polytope correspond to unfrozen $\mathcal{A}$-coordinates which appear in the symbol alphabet. The branch cuts exhibit a non-abelian structure, with sequential cuts corresponding to faces which do not touch being forbidden. Poles in BCFW terms for tree amplitudes (and more conjecturally Yangian invariants) exhibit an abelianised version of adjacency; they all correspond to $\mathcal{A}$-coordinates from the same cluster. The same adjacency structure also relates the poles of R-invariants and the final entries (i.e. derivatives) of the polylogarithms which appear in the NMHV amplitudes.

The structures we have uncovered naturally lead to many further questions.
\begin{itemize}
\item Can we use adjacency to construct integrable words without
  having to apply the bootstrap techniques? This question is even of
  interest if we do not insist on the physical initial entry
  conditions and indeed one can ask it for all finite cluster
  algebras, not just the cases of physical interest described here. A
  hint that this might be possible comes from the observation that
  mutation pairs $\{a,a'\}$ appearing in a triple always appear in the
  form $[a \otimes x(a,a') \otimes a']$ where $x(a,a')$ is the
  $\mathcal{X}$-coordinate associated to any mutation which takes $a$
  to $a'$.

\item Can we extend our results to general N${}^k$MHV BCFW terms or
  more generally Yangian invariants? Going beyond BCFW terms will lead
  to expressions which involve quantities more complicated that
  $\mathcal{A}$-coordinates. Perhaps we will learn something about how
  such singularities interact with the known ones and how they relate
  to adjacency.

\item When considering loop amplitudes, to what extent does the
  structure seen here extend to the octagon and beyond? There are
  several issues at stake here. Firstly, we would ideally like to
  define a Steinmann IR finite quantity for all $n$, while the
  BDS-like subtraction only exists for $n\neq0 \text{ mod } 4$. We
  also know that at eight points and beyond we will have to deal with
  letters which involve square roots and are not just rational
  functions of Pl\"ucker coordinates. It is important to understand
  what role adjacency plays when these are included - it will
  necessarily go beyond the definition of adjacency in terms of
  $\mathcal{A}$-coordinates we have used here. This question is
  connected to the question of whether we can understand the boundary
  structure of the cluster polytope for ${\rm Conf}_n(\mathcal{P}^3)$
  for $n\geq8$. Further, there will be the question of whether
  adjacency can be extended beyond the polylogarithmic case to include
  the elliptic functions appearing in e.g. the ten-point two-loop
  N${}^3$MHV amplitude.

\item To what extent do adjacency constraints arise beyond planar
  $\mathcal{N}=4$ amplitudes? For sufficiently many external legs
  there will always be Steinmann constraints on scattering
  amplitudes. A natural question is whether these extend to further
  constraints between pairs of singularities which are not both simple
  unitarity cuts of amplitudes. The geometrical picture of the
  relations between singularities described here suggests that it is
  important to understand the relevant geometry and its boundary
  structure in the more general setting. This geometry is necessarily
  more complicated in the general case of massless scattering where
  dual conformal symmetry is broken.

\end{itemize}

It will be fascinating to explore the above questions. Ultimately we
might hope to be able to give a simple geometric or algebraic
construction of physical scattering amplitudes.

\section*{Acknowledgments}
JMD, JAF and \"OCG are supported by ERC consolidator grant 648630 IQFT.

  \bibliographystyle{JHEP}
  
  \bibliography{biblio}

\end{document}